\documentclass[aps,prd,floatfix,amsmath,amssymb,onecolumn,groupedaddress,11pt]{revtex4-2}
\usepackage{graphicx}
\usepackage{dcolumn}
\usepackage{bm}
\usepackage{amsfonts,mathtools}
\usepackage{physics}
\usepackage{braket}
\usepackage[caption=false]{subfig}
\DeclareUnicodeCharacter{2212}{-}
\usepackage{float}
\usepackage{comment}
\usepackage{subfig}
\usepackage{hhline}

\begin{document}

\title{Flavored Gauge-Mediated Supersymmetry Breaking Models with Discrete Non-Abelian Symmetries}
\author{Shu Tian Eu}
\author{Lisa L.~Everett}
\author{Todd Garon}
\author{Neil Leonard}
\affiliation{ \textit{ Department of Physics, University of Wisconsin-Madison, Madison, WI, 53706}}

\begin{abstract}
We investigate flavored gauge mediation models in which the Higgs and messenger doublets are embedded in multiplets of the discrete non-Abelian symmetry $\mathcal{S}_3$. 
In these theories, the $\mathcal{S}_3$ symmetry correlates the flavor structure of the quark and lepton Yukawa couplings with the structure of the messenger Yukawa couplings that contribute to the soft supersymmetry breaking mass parameters.  We provide a systematic exploration of possible scenarios within this framework that can accommodate hierarchical quark and charged lepton masses, and examine the resulting phenomenological implications in each case. We find a heavier spectrum for the superpartner masses compared to flavored gauge mediation models controlled by Abelian symmetries, which can be directly traced back to the need in our scenarios for two vectorlike pairs of messenger fields for viable electroweak symmetry breaking. 
\end{abstract}

\maketitle
\section{Introduction}
In the LHC era, the search for physics beyond the Standard Model (SM) has proven elusive, and standard frameworks for TeV-scale new physics are highly constrained. For the well-studied case of extensions of the Standard Model to include softly broken $\mathcal{N}=1$ supersymmetry, such as the minimal supersymmetric standard model (MSSM), the LHC bounds indicate that if softly broken supersymmetry does indeed play a role in any new physics at the next rung of the energy ladder, its implementation is necessarily more complicated and ostensibly fine-tuned than originally anticipated. In this context, given the vast nature of the parameter space associated with the soft supersymmetry breaking sector, frameworks such as the MSSM can remain viable. However, patterns of possibly viable MSSM parameter regions would then be indicated, perhaps pointing to a specific organizing principle at higher energies.

One such example is within the context of gauge-mediated supersymmetry breaking (GMSB). In its minimal implementation, its distinctive phenomenology is characterized by a superpartner mass spectrum with a sizable splitting between the $SU(3)_c$-charged superpartners (squarks and gluinos) and the superpartners charged only under the electroweak symmetry (sleptons and electroweakinos), with the splitting governed by the messenger mass scale and the number of messenger pairs  (taken to be $\mathbf{5}$, $\overline{\bf 5}$ with respect to $SU(5)$). However, the minimal implementation does not easily allow for a 125 GeV Higgs mass, requiring very high messenger scales and subsequent squark and gluino masses that are far out of reach of the LHC \cite{Dine:1981za,ALVAREZGAUME198296,DIMOPOULOS1981353,DIMOPOULOS1983479,Dine:1981za,DINE1982227,Nappi:1982hm,Dine_1993,Dine_1995,Dine_1996,Dine_1997,Giudice_1999,Draper_2012,Arbey_2012,Adeel_Ajaib_2012,Fischler_2014,Calibbi_2014}.  As such, nonminimal implementations of gauge mediation, such as general gauge mediation \cite{Meade:2008wd}, or scenarios in which the MSSM fields and the messenger fields interact directly via renormalizable superpotential couplings, have now long been explored \cite{Dine_1997,Giudice_1999,Chacko_2002,Shadmi_2012,Evans_2011,evans2011probing,Evans_2012,Kang_2012,Craig_2013,Albaid_2013,Abdullah_2013,P_rez_2013,Byakti_2013,Evans_2013,Calibbi_2013,evans2015chiral,Galon_2013,Fischler_2014,Calibbi_2014,Joaquim:2006mn,Joaquim:2006uz}. 

Of the many intriguing options for direct couplings between the messenger and matter sectors, the {\it flavored gauge mediation} framework, which exploits the fact that the electroweak Higgs fields can mix with the doublet components of the messenger pairs, has been of particular interest in the literature \cite{Abdullah_2013,Shadmi_2012,Ierushalmi_2016,P_rez_2013,Byakti_2013,Evans_2013,Calibbi_2013,evans2015chiral,Galon_2013,Jeli_ski_2015,Everett_2018,Everett_2019,Ahmed_2017}. In flavored gauge mediation (FGM) models, Higgs-messenger mixing leads to the generation of messenger Yukawa couplings, which affect the prediction of the soft supersymmetry breaking mass parameters at the input (messenger mass) scale. The messenger Yukawa contributions not only affect the superpartner mass spectrum, but also can generically can lead to the nontrivial possibility of flavor mixing in the soft terms. In viable FGM scenarios, therefore, the messenger Yukawa couplings are controlled by additional symmetries, and their forms are also intimately connected to the generation of the MSSM Yukawa couplings of the quarks and leptons. The case of $U(1)$ symmetries, as explored extensively for example in \cite{Calibbi_2014}, allows for great flexibility in constructing viable models with one or more vectorlike pairs of messengers. In addition, it was shown in \cite{Ierushalmi_2016} that flavor-mixing contributions to the soft terms in such scenarios are much smaller than naive expectations might suggest, and can be consistent with stringent bounds from flavor-changing processes, depending on the model in question.

Instead of using Abelian symmetries to control the messenger Yukawa couplings, an alternative is to build models based on discrete non-Abelian symmetries. Such symmetries have been extensively used as governing principles for the generation of viable SM fermion masses and mixing parameters \cite{Xing_1997,Fritzsch_1994,Fritzsch_2000}. In flavored gauge mediation, this possibility was first explored in detail in \cite{P_rez_2013}, where the authors constructed a two-family scenario based on the discrete non-Abelian symmetry $\mathcal{S}_3$, with the Higgs and messenger fields connected within $\mathcal{S}_3$ doublets.  This idea was then extended to incorporate three families \cite{Everett_2018,Everett_2019,Everett_2020}. Most notably, it was realized in \cite{Everett_2018} that to avoid a severe $\mu/B_\mu$ problem, the Higgs-messenger sector should be extended to include $\mathcal{S}_3$ singlet representations as well as doublet representations. This leads to scenarios with a minimal number $N=2$ of messenger pairs (in contrast to the $U(1)$ cases, which allow for one messenger pair), which enhances the splitting of the squark and gluino masses compared to the slepton and electroweakino masses. Further embedding of the MSSM fields in $\mathcal{S}_3$ representations allows for the possibility that $\mathcal{S}_3$ can play a role as part of the family symmetry that governs the SM fermion masses and mixings. A specific implementation of this idea was explored in  \cite{Everett_2019}, as well as in \cite{Everett_2020}, in which the Higgs-messenger singlets play a dominant role in generating the third family SM fermion masses.

The purpose of this paper is to provide a comprehensive analysis of the FGM $\mathcal{S}_3$ scenario, summarizing and extending our previous work. The aim is to  explore other viable corners of parameter space of these theories and the subsequent effects of including nonleading corrections to the fermion masses.  We identify several viable parameter regions, describe their phenomenological consequences, and compare them to the $U(1)$ FGM benchmark scenarios in the literature. We will see that quite generally, it is not easy to generate viable fermion masses while maintaining flavor-diagonal soft terms, and we will characterize the extent to which such flavor nondiagonal terms are constrained in these theories. The examples studied here all feature very heavy squarks and gluinos, very heavy Higgs fields, and lighter sleptons, charginos, and neutralinos. As such, they provide working examples of currently allowed MSSM parameter space that will continue to be constrained at the LHC and future colliders.

This paper is structured as follows. We begin with a brief overview of the flavored gauge mediation framework studied here, and describe various options for obtaining hierarchical quark and charged lepton masses. Next, we describe several concrete models, and analyze their mass spectra in detail. Finally, we present our summary and conclusions.

\section{Theoretical Background}
As described in \cite{Everett_2018}, the FGM $\mathcal{S}_3$ scenario studied here assumes a specific set of Higgs-messenger fields and supersymmetry-breaking fields. The quantum numbers of these fields with respect to $\mathcal{S}_3$ are given in Table~{\ref{tab:11}}.
\begin{table}[htbp]
   \centering
    \begin{tabular}{c|cccccc|cc}
     & $\mathcal{H}_u^{(2)}$&$\mathcal{H}_u^{(1)}$ & $\mathcal{H}_d^{(2)}$& $\mathcal{H}_d^{(1)}$ & $T_{\bar{d}k}$ & $T_{dk}$   &$X_H$ & $X_T$\\
    \hline
    $\mathcal{S}_3$ &$\mathbf 2 $& $\mathbf 1$& $\mathbf 2 $& $\mathbf 1 $ & $\mathbf 1 $ & $\mathbf 1 $ &\textbf 2 & $\mathbf 1 $\\
   \end{tabular}
   \caption{The field content and $\mathcal{S}_3$ charges for the messenger and supersymmetry breaking sectors.}
   \label{tab:11}
\end{table} 
Here the $\mathcal{H}_{u,d}^{(2)}$ are Higgs-messenger $\mathcal{S}_3$  doublets, the $\mathcal{H}_{u,d}^{(1)}$ are Higgs-messenger $\mathcal{S}_3$  singlets, and $X_H$ is a supersymmetry breaking field that also breaks the $\mathcal{S}_3$ symmetry. The $T_{\bar{d}k,dk}$ denote the $SU(3)_c$ triplets which have the appropriate quantum numbers to complete approximate $\mathbf{5}$, $\overline{\mathbf{5}}$ multiplets with the messengers and $X_T$ is the supersymmetry breaking field that couples to these triplets \footnote{The triplet messengers and the $X_T$ field are chosen for simplicity to be singlets with respect to the $\mathcal{S}_3$ Higgs-messenger symmetry. Note that this choice is not consistent with a full embedding of this scenario into a grand unified theory. This would require more extended model-building that would also need to address the well-known doublet-triplet splitting issue in grand unified models).}. Focusing on the Higgs-messenger fields, we can write $\mathcal{H}_{u,d}^{(2)}$ and $\mathcal{H}_{u,d}^{(1)}$ as 
\begin{eqnarray}
\mathcal{H}_{u}&\equiv&  \left (\begin{array}{c} (\mathcal{H}^{(2)}_u)_1 \\ (\mathcal{H}^{(2)}_u)_2 \\ \mathcal{H}^{(1)}_u \end{array} \right )  \equiv  \left (\begin{array}{c} \mathcal{H}^{(2)}_{u1}\\ \mathcal{H}^{(2)}_{u2}\\ \mathcal{H}^{(1)}_u\end{array} \right )= \mathcal{R}_u \left (\begin{array}{c} H_u\\M_{u1} \\  M_{u2} \end{array} \right ) \nonumber \\
\mathcal{H}_{d}&\equiv& \left (\begin{array}{c} (\mathcal{H}^{(2)}_d)_1 \\ (\mathcal{H}^{(2)}_d)_2 \\ \mathcal{H}^{(1)}_d \end{array} \right ) \equiv  \left (\begin{array}{c} \mathcal{H}^{(2)}_{d1}\\ \mathcal{H}^{(2)}_{d2}\\ \mathcal{H}^{(1)}_d\end{array} \right )=\mathcal{R}_d \left (\begin{array}{c} H_d\\M_{d1}  \\ M_{d2} \end{array} \right ),
\label{higgs_s3}
\end{eqnarray}
in which $H_{u,d}$ are the electroweak Higgs fields of the MSSM, $M_{u1,d1}$ and $M_{u2,d2}$ are gauge mediation messenger doublets, and $\mathcal{R}_{u,d}$ are unitary matrices whose form is governed by the couplings of the Higgs-messenger fields to $X_H$, which obtains both a scalar and $F$-component vacuum expectation value (VEV). As shown in  \cite{Everett_2018}, consistency requirements and obtaining the needed mass hierarchy between the MSSM Higgs fields $H_{u,d}$ and the heavy messengers $M_{ui,di}$ require that $\mathcal{R}_{u,d}$ are given by
\begin{eqnarray}
\mathcal{R}_{u,d}= \left ( \begin{array}{ccc} \frac{1}{\sqrt{3}} & \mp \frac{1}{2} \left (1+\frac{1}{\sqrt{3}} \right) & \frac{1}{2} \left (1-\frac{1}{\sqrt{3}} \right) \\  \frac{1}{\sqrt{3}} & \pm \frac{1}{2} \left (1-\frac{1}{\sqrt{3}} \right) & -\frac{1}{2} \left (1+\frac{1}{\sqrt{3}} \right) 
\\  \frac{1}{\sqrt{3}} &  \pm \frac{1}{\sqrt{3}} &  \frac{1}{\sqrt{3}} \end{array} \right ).
\label{rotationmatrices}
\end{eqnarray}
We turn now to the MSSM fields and their interactions with the Higgs-messenger fields.  Although various possibilities exist, as discussed in \cite{Everett_2018}, 
we make the key assumption that the three generations of SM quarks and leptons are embedded into doublet and singlet representations of $\mathcal{S}_3$, as summarized in Table~\ref{tab:12}.
\begin{table}[htbp]
   \centering
    \begin{tabular}{c|cccc|cccccccccc|c}
  & $\mathcal{H}_u^{(2)}$&$\mathcal{H}_u^{(1)}$ & $\mathcal{H}_d^{(2)}$& $\mathcal{H}_d^{(1)}$  & $Q_{\mathbf 2}$ &$Q_{\mathbf 1}$& $\bar u_{\mathbf 2}$ & $\bar u_{\mathbf 1 }$&$\bar d_{\mathbf 2}$& $\bar d_{\mathbf 1}$ & $L_{\mathbf{2}}$ & $L_{\mathbf{1}}$ & $\bar{e}_{\mathbf{2}}$ & $\bar{e}_{\mathbf{1}}$&$X_H$\\
    \hline
    $\mathcal{S}_3$ &$\mathbf 2 $& $\mathbf 1$& $\mathbf 2 $& $\mathbf 1 $  & $\mathbf 2 $&$\mathbf 1$  & $\mathbf 2 $&$\mathbf 1$  & $\mathbf 2 $&$\mathbf 1$ & $\mathbf 2 $&$\mathbf 1$  & $\mathbf 2 $&$\mathbf 1$ &\textbf 2\\
   \end{tabular}
   \caption{Charges for an $\mathcal{S}_3$ model of the Higgs-messenger fields and the MSSM matter fields. Here the $SU(3)$ triplet messengers and the associated $X_T$ field are not displayed for simplicity.
   }
   \label{tab:12}
\end{table} 
With these $\mathcal{S}_3$ charge assignments, the superpotential couplings of the MSSM matter fields and the Higgs-messenger fields, for example for the up quarks, are given by 
\begin{eqnarray}
W^{(u)}= \tilde{y}_u\big[Q_{\mathbf 2}  \bar u_{\mathbf 2}  \mathcal{H}^{(2)}_u+\beta_{1u}Q_{\mathbf 2}  \bar u_{\mathbf 2} \mathcal{H}^{(1)}_u + \beta_{2u} Q_{\mathbf 2}  \bar u_{\mathbf 1}  \mathcal{H}^{(2)}_u +\beta_{3u} Q_{\mathbf 1}  \bar u_{\mathbf 2}  \mathcal{H}^{(2)}_u+ \beta_{4u} Q_{\mathbf 1}  \bar u_{\mathbf 1}  \mathcal{H}^{(1)}_u\big].
\label{wu}
\end{eqnarray}
In Eq.~(\ref{wu}), $\tilde{y}_u$ is a dimensionless overall factor, and the quantities $\beta_{1u}$, $\beta_{2u}$, $\beta_{3u}$, and $\beta_{4u}$ are dimensionless quantities that characterize the different couplings as allowed by $\mathcal{S}_3$. (Analogous forms hold for the down quarks and the charged leptons; we will ignore the effects of neutrino masses.) In the basis given by 
\begin{eqnarray}
Q= (Q_\mathbf{2}, Q_\mathbf{1})^T = ((Q_{\mathbf 2})_1, (Q_{\mathbf 2})_2 ,Q_{\mathbf 1})^T, \qquad \overline{u}= (\overline{u}_{\mathbf{2}}, \overline{u}_\mathbf{1})^T= ((\overline{u}_{\mathbf 2})_1, (\overline{u}_\mathbf{2})_2 ,\overline{u}_{\mathbf 1})^T,
\end{eqnarray}
the superpotential couplings of Eq.~(\ref{wu}) can be expressed in matrix form as
\begin{eqnarray}
W^{(u)}=\tilde{y}_uQ^T\left( \begin{matrix} \mathcal{H}^{(2)}_{u1}&\beta_{1u}\mathcal{H}^{(1)}_{u}&\beta_{2u} \mathcal{H}^{(2)}_{u2}\\ \beta_{1u} \mathcal{H}^{(1)}_u& \mathcal{H}^{(2)}_{u2}& \beta_{2u}\mathcal{H}^{(2)}_{u1}\\ \beta_{3u}\mathcal{H}^{(2)}_{u2}& \beta_{3u}\mathcal{H}^{(2)}_{u1}&\beta_{4u} \mathcal{H}^{(1)}_u\end{matrix}\right)\bar u. \label{UpYukawas}
\end{eqnarray}
From here, we can easily identify the MSSM Yukawa coupling $Y_u$ and the messenger Yukawa couplings $Y_{u1}'$, $Y_{u2}'$, as
\begin{eqnarray}
Y_u=  \frac{\tilde{y}_{u}}{\sqrt{3}} \left (\begin{array}{ccc} 1 & \beta_{1u} & \beta_{2u} \\ \beta_{1u} & 1 & \beta_{2u} \\ \beta_{3u} & \beta_{3u} & \beta_{4u}           \end{array} \right ),
\label{eq:yud}
\end{eqnarray}
and 
\begin{equation}
Y^\prime_{u1}=\tilde{y}_{u} \left (\begin{array}{ccc} -\frac{1}{2}-\frac{1}{2\sqrt{3}} & \frac{\beta_{1u}}{\sqrt{3}} & \;\; \frac{\beta_{2u}}{2} - \frac{\beta_{2u}}{2\sqrt{3}} \\  \frac{\beta_{1u}}{\sqrt{3}} & \;\; \frac{1}{2}-\frac{1}{2\sqrt{3}} & -\frac{\beta_{2u}}{2} - \frac{\beta_{2u}}{2\sqrt{3}} \\ \;\; \frac{\beta_{3u}}{2} - \frac{\beta_{3u}}{2\sqrt{3}} & -\frac{\beta_{3u}}{2} - \frac{\beta_{3u}}{2\sqrt{3}} & \frac{\beta_{4u}
}{\sqrt{3}}
\end{array} \right )
\end{equation}
\begin{equation}
\;\; Y^\prime_{u2}=\tilde{y}_{u} \left (\begin{array}{ccc} \;\; \frac{1}{2}-\frac{1}{2\sqrt{3}} & \frac{\beta_{1u}}{\sqrt{3}} &  -\frac{\beta_{2u}}{2} - \frac{\beta_{2u}}{2\sqrt{3}} \\  \frac{\beta_{1u}}{\sqrt{3}} & -\frac{1}{2}-\frac{1}{2\sqrt{3}} & \;\; \frac{\beta_{2u}}{2} - \frac{\beta_{2u}}{2\sqrt{3}} \\ -\frac{\beta_{3u}}{2} - \frac{\beta_{3u}}{2\sqrt{3}} & \;\; \frac{\beta_{3u}}{2} - \frac{\beta_{3u}}{2\sqrt{3}} & \frac{\beta_{4u}
}{\sqrt{3}}
\end{array} \right ).
\end{equation}
These results are for the up sector; again, analogous relations hold for the down quarks and charged leptons, with the replacements $u\rightarrow d,e$ in all parameters, respectively. 

For arbitrary values of the coefficients, Eq.~(\ref{eq:yud}) does not result in hierarchical fermion masses. It is only at special values of the couplings, corresponding to various enhanced symmetry points, that we can obtain a realistic quark mass hierarchy at leading order.  To see this, we note that we can diagonalize this system explicitly and examine parameter sets where viable eigenvalue hierarchies can be obtained. For example, in the up quark sector, we can follow standard procedures and consider the Hermitian combinations $Y_u Y_u^\dagger$ and $Y_u^\dagger Y_u$. It is straightforward to calculate following exact results for their eigenvalues (denoted by $\lambda_{1u,2u,3u})$:
\begin{eqnarray}
\lambda_{1u} = \frac{\tilde{y}_u^2}{3}(1-\beta_{1u})^2, \qquad \lambda_{2u,3u} = \frac{\tilde{y}_u^2}{6} \left ((1+\beta_{1u})^2+2(\beta_{2u}^2+\beta_{3u}^2)+ \beta_{4u}^2 \mp \sqrt{\Lambda_u} \right ),
\label{eq:eigenvaluessq}
\end{eqnarray} 
in which $\Lambda_u$ is given by
\begin{eqnarray}
\Lambda_u &=& (1+\beta_{1u})^4+4(\beta_{2u}^4+\beta_{3u}^4)+\beta_{4u}^4+4((1+\beta_{1u})^2+\beta_{4u}^2)(\beta_{2u}^2+\beta_{3u}^2)-2(1+\beta_{1u})^2\beta_{4u}^2\nonumber \\ &-&8\beta_{2u}^2\beta_{3u}^2+16(1+\beta_{1u})\beta_{2u}\beta_{3u} \beta_{4u}.
\label{eq:Lambdaudef}
\end{eqnarray}
Clearly, for arbitrary values of the parameters, the eigenvalues are not hierarchical. However, in looking for leading-order results in which only one eigenvalue is sizable, we can easily identify two general scenarios of interest, depending on the ordering of the mass eigenvalues.  One option is that $\lambda_{1u}$ is one of the small eigenvalues, which would have $\beta_{1u}\rightarrow 1$, and $\lambda_{2u}$ is the other, and hence $\lambda_{3u}$ generically has an $O(1)$ value.  Another option is that $\lambda_{1u}$ is the large eigenvalue, such that $\beta_{1u}\neq 1$, and both $\lambda_{2u,3u}$ are small. We now discuss each possibility in turn.  

In what follows, we will focus on the up quarks, but our default assumption will be that the down quarks and the charged leptons will take similar forms.  Mixing possible options for eigenvalue hierarchies in the different charged fermion sectors will not be considered here for simplicity. 

\subsection{Case 1: $\lambda_{1u,2u}\ll \lambda_{3u}$ (encompassing the ``singlet-dominated" and ``democratic" limits)} 
We begin with the situation that $\beta_{1u}\rightarrow 1$, such that $\lambda_{1u}$ is a small eigenvalue, and explore parameter regimes in which $\lambda_{2u}$ is also small.  For simplicity, we first consider the case in which both vanish, such that to this order of approximation we have one massive third-generation, and two massless generations.  It is easily verified that in this regime, both eigenvalues vanish for 
\begin{eqnarray}
\beta_{1u}=1, \qquad \beta_{2u}\beta_{3u}=\beta_{4u}.
\label{eq:betarelations}
\end{eqnarray}
This case includes what we call the {\it democratic} limit, in which all the $\beta_{iu}=1$, and thus the MSSM Yukawas take on the well-known democratic form \cite{Fritzsch_1994}. The democratic limit was originally studied at leading order in \cite{Everett_2018}, and will be studied in more detail below, including subleading corrections. This case also includes what we will call the {\it singlet-dominated} limit, which is the case in which $ \beta_{4u} \gg  \beta_{1u,2u,3u}$, as $\beta_{4u}$ is the parameter related to the strength of the superpotential coupling involving only $\mathcal{S}_3$ singlet fields.  In the singlet-dominated limit, the MSSM and messenger Yukawa couplings at leading order, in the diagonal quark mass basis, only have nonvanishing $3-3$ entries, allowing for sizable stop mixing and consequently lighter superpartner masses than the other examples we will consider (as we will see).  This limiting case was studied in some detail in \cite{Everett_2019} and \cite{Everett_2020}, and will be considered below as a benchmark scenario for purposes of comparison.

For Case 1, incorporating Eq.~(\ref{eq:betarelations}) and up to possible rephasings to ensure that the fermion masses are real and positive, the diagonalization matrices $U_{uL}$ and $U_{uR}$ take the form 
\begin{eqnarray}
U_{uL}=\left (\begin{array}{ccc} \;\;\; \frac{1}{\sqrt{2}} & -\frac{\beta_{3u}}{\sqrt{2}\sqrt{2+\beta_{3u}^2}} & \frac{1}{\sqrt{2+\beta_{3u}^2}} 
\\ -\frac{1}{\sqrt{2}} & -\frac{\beta_{3u}}{\sqrt{2}\sqrt{2+\beta_{3u}^2}} & \frac{1}{\sqrt{2+\beta_{3u}^2}}
\\ \;\;\; 0 &\frac{\sqrt{2}}{\sqrt{2+\beta_{3u}^2}} & \frac{\beta_{3u}}{\sqrt{2+\beta_{3u}^2}} 
\end{array}\right ), \qquad  
U_{uR}=\left (\begin{array}{ccc} \;\;\; \frac{1}{\sqrt{2}} & -\frac{\beta_{2u}}{\sqrt{2}\sqrt{2+\beta_{2u}^2}} & \frac{1}{\sqrt{2+\beta_{2u}^2}} \\
 -\frac{1}{\sqrt{2}} & -\frac{\beta_{2u}}{\sqrt{2}\sqrt{2+\beta_{2u}^2}} & \frac{1}{\sqrt{2+\beta_{2u}^2}}
\\ \;\;\; 0 &\frac{\sqrt{2}}{\sqrt{2+\beta_{2u}^2}} & \frac{\beta_{2u}}{\sqrt{2+\beta_{2u}^2}} 
\end{array}\right ).
\label{eq:case1diagmatrices}
\end{eqnarray}
Assuming these forms with no further rephasings, the messenger Yukawa couplings in the diagonal quark mass basis then take the form
\begin{equation}
Y_{u1}^\prime = \tilde{y}_u \left (\begin{array}{ccc} -\frac{\sqrt{3}}{2} & \frac{3 \beta_{2u}}{2\sqrt{2+\beta_{2u}^2}} & \frac{\beta_{2u}^2-1}{\sqrt{2}\sqrt{2+\beta_{2u}^2}} \\ \frac{3\beta_{3u}}{2\sqrt{2+\beta_{3u}^2}} & \frac{3\sqrt{3} \beta_{2u}\beta_{3u}}{2\sqrt{2+\beta_{2u}^2}\sqrt{2+\beta_{3u}^2}} & \frac{\sqrt{3}(\beta_{2u}^2-1)\beta_{3u}}{\sqrt{2}\sqrt{2+\beta_{2u}^2}\sqrt{2+\beta_{3u}^2}} \\ \frac{\beta_{3u}^2-1}{\sqrt{2}\sqrt{2+\beta_{3u}^2}} & \frac{\sqrt{3}(\beta_{3u}^2-1)\beta_{2u}}{\sqrt{2}\sqrt{2+\beta_{2u}^2}\sqrt{2+\beta_{3u}^2}} & \frac{(\beta_{2u}^2-1)(\beta_{3u}^2-1)}{\sqrt{3}\sqrt{2+\beta_{2u}^2}\sqrt{2+\beta_{3u}^2}}
\end{array} \right )
\end{equation}
\begin{equation}
Y_{u2}^\prime =\tilde{y}_u \left (\begin{array}{ccc} -\frac{\sqrt{3}}{2} & -\frac{3 \beta_{2u}}{2\sqrt{2+\beta_{2u}^2}} & \frac{\beta_{2u}^2-1}{\sqrt{2}\sqrt{2+\beta_{2u}^2}} \\ -\frac{3\beta_{3u}}{2\sqrt{2+\beta_{3u}^2}} & \frac{3\sqrt{3} \beta_{2u}\beta_{3u}}{2\sqrt{2+\beta_{2u}^2}\sqrt{2+\beta_{3u}^2}} & \frac{\sqrt{3}(\beta_{2u}^2-1)\beta_{3u}}{\sqrt{2}\sqrt{2+\beta_{2u}^2}\sqrt{2+\beta_{3u}^2}} \\ -\frac{\beta_{3u}^2-1}{\sqrt{2}\sqrt{2+\beta_{3u}^2}} & \frac{\sqrt{3}(\beta_{3u}^2-1)\beta_{2u}}{\sqrt{2}\sqrt{2+\beta_{2u}^2}\sqrt{2+\beta_{3u}^2}} & \frac{(\beta_{2u}^2-1)(\beta_{3u}^2-1)}{\sqrt{3}\sqrt{2+\beta_{2u}^2}\sqrt{2+\beta_{3u}^2}}
\end{array} \right ).
\label{eqn:case1yukawa}
\end{equation}
From these forms, we see that in the democratic limit, the messenger Yukawas only have nonvanishing entries in the upper $2\times 2$ block, as follows:
\begin{equation}
Y_{u1}^\prime = \tilde{y}_u \left (\begin{array}{ccc} -\frac{\sqrt{3}}{2} & \frac{\sqrt{3}}{2} & 0 \\ \frac{\sqrt{3}}{2} & \frac{\sqrt{3}}{2} & 0 \\ 0& 0&0 
\end{array} \right ), \qquad Y_{u2}^\prime = \tilde{y}_u \left (\begin{array}{ccc} -\frac{\sqrt{3}}{2} & -\frac{\sqrt{3}}{2} & 0 \\ -\frac{\sqrt{3}}{2} & \frac{\sqrt{3}}{2} & 0 \\ 0& 0&0 
\end{array} \right ).
\end{equation}
In the singlet-dominated limit, the $3-3$ entries dominate, with $Y^\prime_{u1,u2}={\rm Diag}(0,0,\tilde{y}_u\beta_{2u}\beta_{3u}/\sqrt{3})$. 

\subsection{Case 2: $\lambda_{2u,3u}\ll \lambda_{1u}$ (the ``doublet-dominated" limit)}
For this case, it is necessary that $\beta_{1u}\neq 1$ such that $\lambda_{1u}\gg \lambda_{2u,3u}$.  For concreteness, we take $\beta_{1u}\rightarrow -1$, and thus require $\beta_{2u,3u,4u}\ll 1$, as well as $\Lambda_u \rightarrow 0$.  Indeed, $\lambda_{2u,3u}=0$ is achieved for $\beta_{1u}=-1,\beta_{2u}=\beta_{3u}=\beta_{4u}=0$. To see this, we note that for $\beta_1=-1$ only, the condition for $\Lambda_u=0$ is as follows:
\begin{equation}
-8 \beta_{2u}^2 \beta_{3u}^2+4(\beta_{2u}^4+\beta_{3u}^4)+4(\beta_{2u}^2+\beta_{3u}^2)\beta_{4u}^2+\beta_{4u}^2=0,
\end{equation}
which is zero only for $\beta_{4u}=0$ and $\beta_{2u}=\beta_{3u}$. We will take $\beta_{1u} = -1$ and $\beta_{4u} = 0$, but leave $\beta_{2u}$ and $\beta_{3u}$ unconstrained at present, recalling that we will need to restrict ourselves to the case that $\beta_{2u,3u}\ll \vert \beta_{1u} \vert =1$.  This limit is the {\it doublet-dominated} limit, since now $\vert \beta_{1u} \vert \gg \beta_{2u,3u} \gg \beta_{4u}=0$, and $\beta_{1u}$ controls the superpotential coupling involving only $\mathcal{S}_3$ doublet fields. In this limit, the mass eigenvalues take the form (assuming for concreteness that $\beta_{3u}>\beta_{2u}$):
\begin{equation}
\lambda_{1u} = \frac{2\tilde{y}_u^2}{3}, \qquad \lambda_{2u}= \frac{2 \tilde{y}_u^2 \beta_{2u}^2}{3}, \qquad \lambda_{3u} =  \frac{2 \tilde{y}_u^2 \beta_{3u}^2}{3},
\end{equation}
such that $\lambda_{2u} < \lambda_{3u} < \lambda_{1u}$. (For $\beta_{3u}< \beta_{2u}$, the placement of $\beta_{2u}$ and $\beta_{3u}$ in $\lambda_{2u}$ and $\lambda_{3u}$ is reversed.) We now take $ \sqrt{3} \tilde{y}_u/2=y_t$ to identify $y_t$ as the top quark Yukawa coupling to leading order.  The diagonalization matrices $U_{uL}$ and $U_{uR}$ now take the following particularly simple forms:
\begin{equation}
U_{uL}=\left (\begin{array}{ccc} \frac{1}{\sqrt{2}} & 0 & \frac{1}{\sqrt{2}} \\ \frac{1}{\sqrt{2}} & 0 & -\frac{1}{\sqrt{2}} \\ 0& 1 & 0 
\end{array} \right ), \qquad
U_{uR}=\left (\begin{array}{ccc} 0& \frac{1}{\sqrt{2}} &  \frac{1}{\sqrt{2}} \\ 0 & \frac{1}{\sqrt{2}} &  -\frac{1}{\sqrt{2}} \\ 1 & 0 & 0 
\end{array} \right ).
\end{equation}
The messenger Yukawas in the diagonal quark basis are then given by
\begin{equation}
Y_{u1}^\prime = y_t \left (\begin{array}{ccc} -\frac{\beta_{2u}}{2\sqrt{2}}& -\frac{3}{4} & -\frac{\sqrt{3}}{4} \\ \;\;\; 0 &  -\frac{\beta_{3u}}{2\sqrt{2}} & \frac{\sqrt{3}\beta_{3u}}{2\sqrt{2}}\\ \frac{\sqrt{3}\beta_{2u}}{2\sqrt{2}} & -\frac{\sqrt{3}}{4} & \;\;\; \frac{1}{4} \end{array} \right ), \qquad
Y_{u2}^\prime = y_t\left (\begin{array}{ccc}  -\frac{\beta_{2u}}{2\sqrt{2}}& -\frac{3}{4} & \frac{\sqrt{3}}{4} \\ \;\;\; 0 &  -\frac{\beta_{3u}}{2\sqrt{2}} & -\frac{\sqrt{3}\beta_{3u}}{2\sqrt{2}}\\ -\frac{\sqrt{3}\beta_{2u}}{2\sqrt{2}} & \frac{\sqrt{3}}{4} & \;\;\; \frac{1}{4} \end{array} \right ).
\end{equation}

\section{Models}
As described in the previous subsection, we have identified two cases with hierarchical quark and charged lepton masses. The first (Case 1) satisfies Eq.~(\ref{eq:betarelations}), and includes two possible scenarios at leading order: the  singlet-dominated limit, in which it is the $\mathcal{S}_3$ singlet couplings of the MSSM fields and the Higgs-messenger fields that dominate the superpotential, and the democratic limit, in which all the couplings of $\mathcal{S}_3$ representations in the superpotential are precisely equal at leading order, resulting in an enhanced $\mathcal{S}_{3L}\times \mathcal{S}_{3R}$ symmetry. The singlet-dominated limit was explored in \cite{Everett_2019} and \cite{Everett_2020}, and the democratic limit at leading order in \cite{Everett_2018}. 
The second (Case 2) is what we call the  doublet-dominated limit, as in this case the dominant couplings are those involving only $\mathcal{S}_3$ doublets. 
In what follows, we will discuss these scenarios in greater detail.

\subsection{Case 1 models}

We begin the discussion of Case 1 models with the singlet-dominated limit, which was studied in detail in \cite{Everett_2019,Everett_2020}. In this scenario, sizable stop mixing can occur due to the FGM contributions to the third-generation soft trilinear scalar coupling. This in turn allows for the squarks and gluinos to be in the $O(5-6 \; {\rm TeV})$ range, which is relatively light compared to generic parameter choices for this class of FGM models.  A variety of subleading corrections to this limit can be considered, including the possibility of generating nontrivial masses for the second-generation fields and the possibility of viable quark mixing at the first subleading order. For the case described in \cite{Everett_2020}, the corrections to the soft terms that result from these terms have only minimal effects on the superpartner masses.
Furthermore, in this case flavor-violating contributions to the soft terms also do not result at the first subleading order in the quantities that control the lighter generation quark and lepton masses, though this is not necessarily generic.  Here we will not revisit this case in detail other than as a point of comparison for the new scenarios considered in this work.

Let us now turn to the democratic limit, for which the Yukawa coupling parameters $\beta_{1i}=\beta_{2i}= \beta_{3i}= \beta_{4i}=1$, where $i=u,d,e$.  In this case, the MSSM Yukawa matrices take the form \begin{equation}
    Y_i=\frac{\tilde{y}_i}{\sqrt{3}}\begin{pmatrix}
    1 & 1 & 1\\
    1 & 1 & 1\\
    1 & 1 & 1 
    \end{pmatrix}.
\end{equation}
This is the well-known flavor democratic mass matrix form, which exhibits an $\mathcal{S}_{3L}\times \mathcal{S}_{3R}$ symmetry. At leading order, this mass matrix has two vanishing eigenvalues, and one $O(1)$ eigenvalue, to be identified with the third-generation.  As shown in \cite{Everett_2018}, the messenger Yukawa matrices have nonzero entries only in the upper $2\times 2$ block in the diagonal quark mass basis.

We now address the generation of the first- and second-generation fermion masses and the effects on the sfermion masses through the messenger Yukawa corrections. Here we choose to break the $\mathcal{S}_{3L}\times \mathcal{S}_{3R}$ symmetry to $\mathcal{S}_{2L}\times \mathcal{S}_{2R}$ and then to $\mathcal{S}_{1L}\times \mathcal{S}_{1R}$, which generates a nonzero mass for the first- and second-generation fermions (see e.g.~\cite{Fritzsch_2000}). This can be achieved via the following terms:
\begin{equation}
    Y_i^{(\text{corr})}=\frac{\tilde{y}_i\epsilon_i}{\sqrt{3}}\begin{pmatrix}
    0 & 0 & 1\\
    0 & 0 & 1\\
    1 & 1 & 1
    \end{pmatrix}+\frac{\tilde{y}_i\sigma_i}{\sqrt{3}}\begin{pmatrix}
    1 & 0 & -1\\
    0 & -1 & 1\\
    -1 & 1 & 0
    \end{pmatrix},
    \label{eq:demcorrections}
\end{equation}
in which $\epsilon_i$ and $\sigma_i$ are real dimensionless perturbative parameters associated with symmetry breaking from $\mathcal{S}_3$ to $\mathcal{S}_2$ and $\mathcal{S}_2$ to $\mathcal{S}_1$ respectively. In our scenario, the $\epsilon$ perturbations of the up quarks (the down quarks and charged leptons have analogous structures) can be generated in superpotential at the renormalizable level by
\begin{equation}
    \epsilon_u y_u[\beta_{2u} Q_\textbf{2}\bar{u}_\textbf{1}\mathcal{H}_u^{(2)}+\beta_{3u}Q_\textbf{1}\bar{u}_\textbf{2}\mathcal{H}_u^{(2)}+\beta_{4u} Q_\textbf{1}\bar{u}_\textbf{1}\mathcal{H}_{u}^{(1)}],
\end{equation}
while the $\sigma$ perturbations can be generated via non-renormalizable operators. These superpotential terms add corrections of the form of Eq.~(\ref{eq:demcorrections}) to the Yukawa matrix for the up-type quarks, and corrections of the following form to the up-type messenger Yukawa matrices: 
\begin{equation}
\begin{split}
    Y_{u1}^{\prime(\text{corr})}&=\tilde{y}_u\epsilon_u\begin{pmatrix}
    0 & 0 & \frac{1}{2}-\frac{1}{2\sqrt{3}} \\
    0 & 0 & -\frac{1}{2}-\frac{1}{2\sqrt{3}}\\
    \frac{1}{2}-\frac{1}{2\sqrt{3}} & -\frac{1}{2}-\frac{1}{2\sqrt{3}}& \frac{1}{\sqrt{3}}
    \end{pmatrix}+\tilde{y}_u\sigma_u\left(
\begin{array}{ccc}
 \frac{1}{2}-\frac{1}{2 \sqrt{3}} & 0 & \frac{1}{2}+\frac{1}{2 \sqrt{3}} \\
 0 & \frac{1}{2}+\frac{1}{2 \sqrt{3}} & \frac{1}{2}-\frac{1}{2 \sqrt{3}} \\
 \frac{1}{2}+\frac{1}{2 \sqrt{3}} & \frac{1}{2}-\frac{1}{2 \sqrt{3}} & 0 \\
\end{array}
\right)\\
     Y_{u2}^{\prime(\text{corr})}&=\tilde{y}_u\epsilon_u\begin{pmatrix}
    0 & 0 & -\frac{1}{2}-\frac{1}{2\sqrt{3}}\\
    0 & 0 & \frac{1}{2}-\frac{1}{2\sqrt{3}}\\
    -\frac{1}{2}-\frac{1}{2\sqrt{3}} & \frac{1}{2}-\frac{1}{2\sqrt{3}} & \frac{1}{\sqrt{3}}
    \end{pmatrix}+\tilde{y}_u\sigma_u\left(
\begin{array}{ccc}
 -\frac{1}{2}-\frac{1}{2 \sqrt{3}} & 0 & -\frac{1}{2}+\frac{1}{2 \sqrt{3}} \\
 0 & -\frac{1}{2}+\frac{1}{2 \sqrt{3}} & -\frac{1}{2}-\frac{1}{2 \sqrt{3}} \\
 -\frac{1}{2}+\frac{1}{2 \sqrt{3}} & -\frac{1}{2}-\frac{1}{2 \sqrt{3}} & 0 \\
\end{array}
\right).
\end{split}
\end{equation}

Including these correction terms along with the leading order results, the eigenvalues
$\lambda_{1u, 2u,3u}$, are then found to be
\begin{equation}
\lambda_{1u} = 0, \qquad \lambda_{2u} = \frac{4 y_t^2 \epsilon_u^2}{81}+ O(\epsilon^3),  \qquad \lambda_{3u} = y_t^2+O(\epsilon_u^3).
\end{equation}
In these relations, we have identified the top quark Yukawa coupling $y_t$ through $\tilde{y}_u = (y_t/\sqrt{3})(1-5\epsilon_u/9+ O(\epsilon_u)^2$, which follows from setting $\lambda_{3u} = y_t^2$ through second order in $\epsilon_u$. 
As expected, $\epsilon_u$ controls the charm quark mass 

\footnote{Note that if we neglect the subleading $\sigma$ perturbations, this scheme is equivalent to replacing $\beta_{2u,3u,4u} = 1+\epsilon_u$ in the general form for the superpotential couplings. As such, the mixing matrices are easily obtained using the results of Eq.~(\ref{eq:case1diagmatrices}) with the appropriate substitutions.}.

In the diagonal quark mass basis, the messenger Yukawa matrices for the up-type quark sector are given  to order in $\sigma_u/\epsilon_u$ by 

\begin{equation}
    \begin{split}
    Y_{u1}^{\prime \text{(diag)}}&=y_t\left(
\begin{array}{ccc}
 -\frac{1}{2}+\frac{5 \epsilon_u }{18}-\frac{3 \sqrt{3} \sigma_u }{2 \epsilon_u } & -\frac{1}{2}-\frac{\epsilon_u }{18}+\frac{3 \sqrt{3} \sigma_u }{2 \epsilon_u } & -\frac{5 \epsilon_u
   }{9 \sqrt{2}} \\
 -\frac{1}{2}-\frac{\epsilon_u }{18}+\frac{3 \sqrt{3} \sigma_u }{2 \epsilon_u } & \frac{1}{2}+\frac{\epsilon_u }{6}+\frac{3 \sqrt{3} \sigma_u }{2 \epsilon_u } & \frac{\epsilon_u }{3
   \sqrt{2}} \\
 -\frac{5 \epsilon_u }{9 \sqrt{2}} & \frac{\epsilon_u }{3 \sqrt{2}} & -\frac{\epsilon_u }{9} \\
\end{array}
\right)\\
    Y_{u2}^{\prime \text{(diag)}}&=y_t\left(
\begin{array}{ccc}
 -\frac{1}{2}+\frac{5 \epsilon_u }{18}+\frac{3 \sqrt{3} \sigma_u }{2 \epsilon_u } & \frac{1}{2}+\frac{\epsilon_u }{18}+\frac{3 \sqrt{3} \sigma_u }{2 \epsilon_u } & \frac{5 \epsilon_u
   }{9 \sqrt{2}} \\
 \frac{1}{2}+\frac{\epsilon_u }{18}+\frac{3 \sqrt{3} \sigma_u }{2 \epsilon_u } & \frac{1}{2}+\frac{\epsilon_u }{6}-\frac{3 \sqrt{3} \sigma_u }{2 \epsilon_u } & \frac{\epsilon_u }{3
   \sqrt{2}} \\
 \frac{5 \epsilon_u }{9 \sqrt{2}} & \frac{\epsilon_u }{3 \sqrt{2}} & -\frac{\epsilon_u }{9} \\
\end{array}
\right).
    \end{split}
\end{equation}
Analogous forms are easily obtained for the MSSM and messenger Yukawa matrices for down-type quarks and leptons in the diagonal quark mass basis with the replacements $\epsilon_u \rightarrow \epsilon_{d,e}$ and $y_t \rightarrow y_{b,\tau}$.
The relative strengths of the parameters $\epsilon_{u,d,e}$ and $\sigma_{u,d,e}$ can be estimated from the fact that these parameters govern the fermion masses of the lighter generations. More precisely, up to $O(1)$ prefactors, $\epsilon_{u,d,e}$ is related to $m_{c,s,\mu}/m_{t,b,\tau}$, while $\sigma_{u,d,e}$ is constrained by $m^2_{u,d,e}/m^2_{t,b,\tau} \sim \sigma^2_{u,d,e}/\epsilon_{u,d,e}$.
From these relations, it is straightforward to obtain that $\epsilon_u\approx 3\times 10^{-2}$ and $\sigma_u\approx 1\times 10^{-3}$. Similarly, $\epsilon_d \approx 0.1$, $\sigma_d\approx 9\times 10^{-3}$, $\epsilon_e\approx 0.3$, and $\sigma_e\approx 8 \times 10^{-3}$. These parameter values also yield hierarchical quark mixing angles of the 
Cabibbo-Kobayashi-Maskawa (CKM) matrix, in which the largest angle is the Cabibbo angle, $\sin\theta_c\sim 0.17$. 
While the quark mixing angles are not fully realistic (the Cabibbo angle is clearly too small compared to its experimentally determined value), for the purposes of this study it is a reasonable starting point for the analysis.

We now find the nonvanishing corrections to the soft supersymmetry breaking terms, assuming for simplicity that the ratio of the $F$ terms to the scalar VEVs for the $X_H$ and $X_T$ terms are identical (both will be denoted as $\Lambda$). We provide the expressions for these correction terms in the Appendix. As expected, in the limit that the perturbation parameters are set to zero, the result is what was found in \cite{Everett_2018}. When the perturbations are added, the diagonal entries of the soft mass-squared terms are corrected at second order in the $\epsilon$ parameters. This generates nonzero (but small) diagonal $3-3$ entries. In addition, with nonzero perturbations, flavor off-diagonal contributions to the corrections to the soft terms are generated. More precisely, the $\epsilon_{u,d,e}$  parameters introduce nonvanishing $\delta m_{f_{23}}^2$ terms at first order in $\epsilon$, while the $\sigma_{u,d,e}$ introduce nonvanishing $\delta m_{f_{12}}^2$ and $\delta m_{f_{21}}^2$ terms. Therefore, the dominant effects are expected to be seen in the $2-3$ sfermion mixings. Further details will be discussed in the next section.

\subsection{Case 2 models}
 This case corresponds to the doublet-dominated limit. Here we need $\lambda_1\gg \lambda_{2,3}$. In the limit that $\lambda_{2,3}\rightarrow 0$, we see from Eqs.~(\ref{eq:eigenvaluessq})-(\ref{eq:Lambdaudef}) this can be achieved for $\beta_1\rightarrow -1$ and $\beta_{i=2,3,4}\ll 1$, and we need $\tilde{\Lambda} \rightarrow 0$.  For $\beta_1=-1$, the condition for $\tilde{\Lambda}=0$ is as follows:
\begin{equation}
-8 \beta_2^2 \beta_3^2+4(\beta_2^4+\beta_3^4)+4(\beta_2^2+\beta_3^2)\beta_4^2+\beta_4^2=0,
\end{equation}
which is zero only for $\beta_4=0$, $\beta_2=\beta_3$. In the up quark sector, we will now set $\lambda_1=y_t^2$, such that $\tilde{y}_u^2= (3/4) y_t^2$ (analogous relations hold for the down quark  and charged lepton sectors). The $\lambda_{2i,3i}$ are directly related to $\beta_{2i,3i}$, with the specific identification dependent on the values of the $\beta_{2i,3i}$.

\noindent $\bullet$ {\it Ordering $\beta_{3i}>\beta_{2i}$}.
Let us first consider the case in which $\beta_{3i}>\beta_{2i}$, for which we have 
\begin{equation}
U_{iL}^\dagger Y_i U_{iR}= Y_i^{(\text{diag})}=
y_{t,b,\tau}\text{Diag}\left (\frac{\beta_{2i}}{\sqrt{2}}, \frac{ \beta_{3i}}{\sqrt{2}},1 \right ),
\label{eq:order1}
\end{equation}
in which $U_{iL,iR}$ take the simple forms
\begin{equation}
U_{iL}=\left (\begin{array}{ccc} \frac{1}{\sqrt{2}} & 0 & \frac{1}{\sqrt{2}} \\ \frac{1}{\sqrt{2}} & 0 & -\frac{1}{\sqrt{2}} \\ 0& 1 & 0 
\end{array} \right ), \qquad
U_{iR}=\left (\begin{array}{ccc} 0& \frac{1}{\sqrt{2}} &  \frac{1}{\sqrt{2}} \\ 0 & \frac{1}{\sqrt{2}} &  -\frac{1}{\sqrt{2}} \\ 1 & 0 & 0 
\end{array} \right ).
\label{umatricescase2a}
\end{equation}
We see that for this ordering, the $\beta_{3i}$ control the second-generation masses and the $\beta_{2i}$ control the first-generation masses.
The messenger Yukawas in the diagonal fermion mass basis (the SCKM basis) are given by
\begin{equation}
Y_{i1}^\prime = y_{t,b,\tau} \left (\begin{array}{ccc} -\frac{\beta_{2i}}{2\sqrt{2}}& -\frac{3}{4} & -\frac{\sqrt{3}}{4} \\ \;\;\; 0 &  -\frac{\beta_{3i}}{2\sqrt{2}} & \frac{\beta_{3i}}{2}\sqrt{\frac{3}{2}}\\ \frac{\beta_{2i}}{2}\sqrt{\frac{3}{2}} & -\frac{\sqrt{3}}{4} & \;\;\; \frac{1}{4} \end{array} \right ),\qquad 
Y_{i2}^\prime = y_{t,b,\tau}\left (\begin{array}{ccc}  -\frac{\beta_{2i}}{2\sqrt{2}}& -\frac{3}{4} & +\frac{\sqrt{3}}{4} \\ \;\;\; 0 &  -\frac{\beta_{3i}}{2\sqrt{2}} & -\frac{\beta_{3i}}{2}\sqrt{\frac{3}{2}}\\ -\frac{\beta_{2i}}{2}\sqrt{\frac{3}{2}} & \frac{\sqrt{3}}{4} & \;\;\; \frac{1}{4} \end{array} \right ).
\end{equation}
Given that we can identify $\beta_{2i,3i}$ with the first and second-generation masses, respectively, we can write for example for the up-type quarks (with $y_{u,d,e}=\beta_{2u,d,} y_t/\sqrt{2}$ and $y_c=\beta_{3u} y_t/\sqrt{2}$):
\begin{equation}
Y_{u1}^\prime = \left (\begin{array}{ccc} -\frac{y_u}{2}& -\frac{3 y_t}{4}& -\frac{\sqrt{3}y_t}{4} \\ \;\;\; 0 &  -\frac{y_c}{2} & -\frac{\sqrt{3}y_c}{2}\\ -\frac{\sqrt{3} y_u}{2} & \frac{\sqrt{3}y_t}{4} & \;\;\; \frac{y_t}{4} \end{array} \right ),\qquad 
Y_{u2}^\prime = \left (\begin{array}{ccc}   -\frac{y_u}{2}&-\frac{3 y_t}{4} & -\frac{\sqrt{3}y_t}{4} \\ \;\;\; 0 &  -\frac{y_c}{2} & \frac{\sqrt{3}y_c}{2} \\ \frac{\sqrt{3} y_u}{2}& -\frac{\sqrt{3}y_t}{4} & \;\;\; \frac{y_t}{4} \end{array} \right ).
\label{eq:messyukcase2a}
\end{equation}
From the quark and charged lepton masses, we can roughly estimate (neglecting running effects) that $\beta_{2u}/\beta_{3u}\sim 2\times 10^{-3}$, $\beta_{2d}/\beta_{3d} \sim 0.05$, $\beta_{2l}/\beta_{3l}\sim 0.005$, while $\beta_{3d}/\beta_{3l} \sim 0.4$. Hence, to leading order we can neglect the effects proportional to the first-generation masses (here the $\beta_{2i}$), and treat the effects due to the second-generation masses (the $\beta_{3i}$) perturbatively.
We thus calculate the corrections to the soft supersymmetry terms in this limit.
As before, we assume for simplicity that the ratio of the $F$ terms to the scalar vevs for the $X_H$ and $X_T$ terms are identical.
The detailed forms of these soft supersymmetry breaking terms are presented in the Appendix.  

We note that in this case, there are flavor off-diagonal contributions in the $\delta m^2_{{Q,L}_{12}}$ that are proportional to the $\beta_{3i}$, and thus scale with the second-generation quark and lepton masses. This is reminiscent of the Case 1 democratic limit with perturbations, though the dominant off-diagonal contributions occurred there in the $2-3$ sector, and here they arise in the more dangerous $1-2$ sector. We will discuss their effects in the next section.

\noindent $\bullet$ {\it Ordering $\beta_{2i}>\beta_{3i}$}.
We now consider the case in which $\beta_{2i}>\beta_{3i}$, for which the roles of $\beta_{2i}$ and $\beta_{3i}$ are switched in Eq.~(\ref{eq:order1}).
We now have 
\begin{equation}
U_{iL}=\left (\begin{array}{ccc} 0& \frac{1}{\sqrt{2}} &  -\frac{1}{\sqrt{2}} \\ 0 & \frac{1}{\sqrt{2}} &  \frac{1}{\sqrt{2}} \\ 1 & 0 & 0 
\end{array} \right ), \qquad
U_{iR}=\left (\begin{array}{ccc} \frac{1}{\sqrt{2}} & 0 & -\frac{1}{\sqrt{2}} \\ \frac{1}{\sqrt{2}} & 0 & \frac{1}{\sqrt{2}} \\ 0& 1 & 0 
\end{array} \right ).
\label{umatricescase2b}
\end{equation}
The messenger Yukawa matrices in the diagonal quark mass basis are of the form
\begin{equation}
Y_{i1}^\prime = y_{t,b,\tau} \left (\begin{array}{ccc} -\frac{\beta_{3i}}{2\sqrt{2}}& 0& -\frac{\sqrt{3}\beta_{3i}}{2\sqrt{2}} \\ \;\;\; -\frac{3}{4} &   -\frac{\beta_{2i}}{2\sqrt{2}}&\frac{\sqrt{3}}{4} \\ \frac{\sqrt{3}}{4} & -\frac{\beta_{2i}}{2}\sqrt{\frac{3}{2}} & \;\;\; \frac{1}{4} \end{array} \right ),\qquad 
Y_{i2}^\prime = y_{t,b,\tau}\left (\begin{array}{ccc}  -\frac{\beta_{2i}}{2\sqrt{2}}& 0 & \frac{\sqrt{3}\beta_{3i}}{2\sqrt{2}}\\ \;\;\; -\frac{3}{4}&  -\frac{\beta_{2i}}{2\sqrt{2}} & -\frac{\sqrt{3}}{4}\\ -\frac{\sqrt{3}}{4} & \frac{\beta_{2i}}{2}\sqrt{\frac{3}{2}}   & \;\;\; \frac{1}{4} \end{array} \right ).
\end{equation}
As in the previous section, we can ignore effects that scale with the first-generation fermion masses and keep leading contributions involving the second-generation fermion masses. Thus, we now neglect the terms proportional to $\beta_{3i}$ and keep leading-order terms proportional to the $\beta_{2i}$. We can again calculate the soft supersymmetry breaking terms, subject to the same assumptions as given for the alternate ordering. The detailed forms are included in the Appendix. 

One interesting feature of this mass ordering ($\beta_{2i}>\beta_{3i}$) is the corrections to the soft supersymmetry breaking mass terms are flavor-diagonal if we neglect effects proportional to the first-generation fermion masses. As in the alternate ordering, here we obtain contributions to $\delta m^2_{{Q,L}_{12}}$ that are proportional to the $\beta_{3i}$, but now these quantities must be much smaller since they govern the masses of the first-generation. Given the high degree of suppression of the flavor off-diagonal elements, in this case the model is clearly safe from flavor-changing neutral current constraints.

\section{Results and Discussion}
\label{resultssection}
In this section, we analyze the mass spectra of these scenarios and their phenomenological implications. We start with Case 1, focusing solely on the democratic limit with symmetry breaking effects, and then study Case 2, the doublet-dominated limit, with both orderings of the $\beta_{2i}$ and $\beta_{3i}$. The model parameters are $M_\text{mess}$, $\Lambda$, $\tan{\beta}=\langle H_u\rangle / \langle H_d \rangle$, the sign of $\mu$ (sgn($\mu$), taken here to be $+1$), and the relevant perturbation parameters, which depend on the scenario in question. Here we have followed standard procedures and replaced $\vert \mu\vert$ and $b$ with $\tan\beta$ and the $Z$ boson mass. 
The renormalization group equations are run using SoftSUSY 4.1.4 \cite{Allanach:2001kg}. 
\subsection{Case 1 models}
We start with the flavor democratic limit, which was explored in \cite{Everett_2018} for the case of third-generation masses only, i.e.~in the absence of the small perturbations that break the $\mathcal{S}_{3L}\times \mathcal{S}_{3R}$ symmetry. It was shown in \cite{Everett_2018} that this scenario leads to heavy superpartner masses, which can be traced to the absence of large stop mixing in this limit. In the presence of nonvanishing perturbations, this picture generically continues except for specific small regions of parameter space where the Higgs mass constraint can be satisfied without being bolstered by very heavy squarks.

\begin{figure}[h]
    \centering
    \includegraphics[scale=0.8]{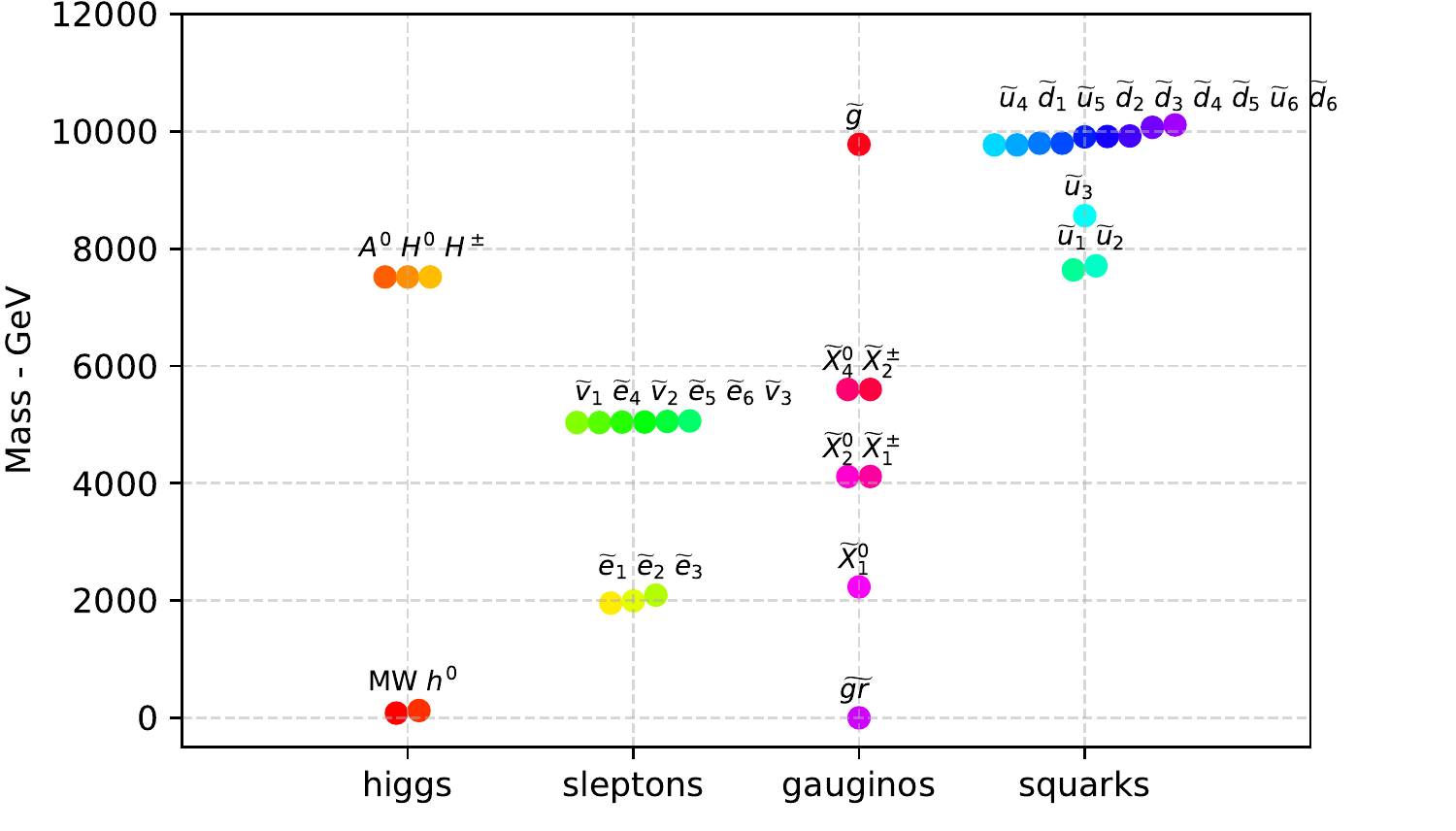}
    \caption{The sfermion mass spectrum in the Case 1 democratic limit, with $M_\text{mess}=10^{12}$ GeV, $\Lambda=8.1\times 10^5$ GeV, $\tan{\beta}=10$, $\epsilon_u=0.033$, $\epsilon_d=0.108$, $\epsilon_e=0.281$, $\sigma_u=0.001$, $\sigma_d=0.009$, and $\sigma_e=0.008$. }
    \label{fig:democraticspectrum1}
\end{figure}

In Figure~\ref{fig:democraticspectrum1}, we show a representative mass spectrum for an intermediate messenger mass scale of $M_{\rm mess}=10^{12}$ GeV and $\tan\beta=10$, where $\Lambda$ is chosen to satisfy the Higgs mass constraint \cite{Zyla:2020zbs}. As seen, the heavy Higgs particles are nearly 8 TeV, the gluino is approximately 10 TeV, and the squarks fall into three groupings: a lightest set that is close in mass to the heavy Higgs particles, a set in between, and a heavier set that is similar to the gluino mass. The sleptons are close in mass to the lightest neutralino, and the next-to-lightest superpartner (NLSP) is the lightest slepton.
The effects of nonzero $\sigma_{u,d,e}$ lead to small ($O(1\; {\rm GeV})$) splittings in the masses of $\tilde{d}_1$ and $\tilde{d}_2$, and $\tilde{u}_4$ and $\tilde{u}_5$, which are each originally identical up to order $10^{-2}$ GeV. The effect of $\epsilon_{u,d,e}$ is larger, which is expected as these have larger numerical values. For nonzero $\epsilon_u$, there is a splitting of order $\sim70$ GeV in the masses of $\tilde{u}_1$ and $\tilde{u}_2$, which are also identical up to order of $10^{-2}$ GeV in the $\mathcal{S}_{3L}\times \mathcal{S}_{3R}$ limit. Similar features are seen for $\tilde{u}_4$ and $\tilde{u}_5$. The $\epsilon_u$ corrections also introduce a small ($\sim25$ GeV) mass splitting for $\tilde{d}_1$ and $\tilde{d}_2$, which is a sign of the symmetry breaking from $\mathcal{S}_{3L}\times\mathcal{S}_{3R}$ to $\mathcal{S}_{2L}\times\mathcal{S}_{2R}$.

As noted previously, this scenario has flavor off-diagonal contributions to the corrections to the soft supersymmetry breaking terms, with the dominant contributions in the $2-3$ sector. To get an estimate of the potential sizes of these effects, we employ the standard mass insertion approximation (MIA) method, in which the quantities of interest for the quarks are
$(\delta_f^{IJ})_{XY}=(\Delta_f^{IJ})_{XY}/((m_{fI})_{XX}(m_{fJ})_{YY})$, 
where $f$ denotes the relevant matter superfield, $I,J$ are flavor indices, $X,Y$ are chirality labels, and $(\Delta_f^{IJ})_{XY}$ is an off-diagonal contribution to the sfermion soft terms \footnote{Note that MIA is a good approximation in this scenario although we have non-degenerate squark masses, since the squark masses are not strongly hierarchical~\cite{Raz_2002}.}.  We expect rather mild constraints due to the heavy sfermion and gluino masses and the suppression factors in the off-diagonal contributions to the soft terms. For the set of model parameters in Fig.~\ref{fig:democraticspectrum1}, we obtain $2-3$ squark and slepton mass insertion parameters of the order $\vert (\delta_u^{23})_{LL}\vert \sim 5\times 10^{-3}$, $\vert (\delta_u^{23})_{RR}\vert \sim 10^{-2}$, $\vert (\delta_d^{23})_{LL}\vert \sim 5\times 10^{-3}$, $\vert (\delta_d^{23})_{RR}\vert \sim 7\times 10^{-4}$, $\vert (\delta_l^{23})_{LL}\vert \sim 2\times 10^{-3}$, and $\vert (\delta_l^{23})_{RR}\vert \sim 3\times 10^{-2}$, as well as small contributions to $LR$ mixings in the $2-3$ sector (ranging from $10^{-4}$ to $10^{-7}$.  The $1-3$ and $1-2$ mass insertions are parametrically smaller, with limits that range from $10^{-4}$ to $10^{-12}$, except for $\vert (\delta_l^{12})_{RR}\vert \sim 3\times 10^{-3}$.   The resulting effects are small and within the allowed ranges (see e.g.~\cite{Misiak_1998}).

\begin{figure}[h]
\centering
\subfloat{{\includegraphics[scale=0.9]{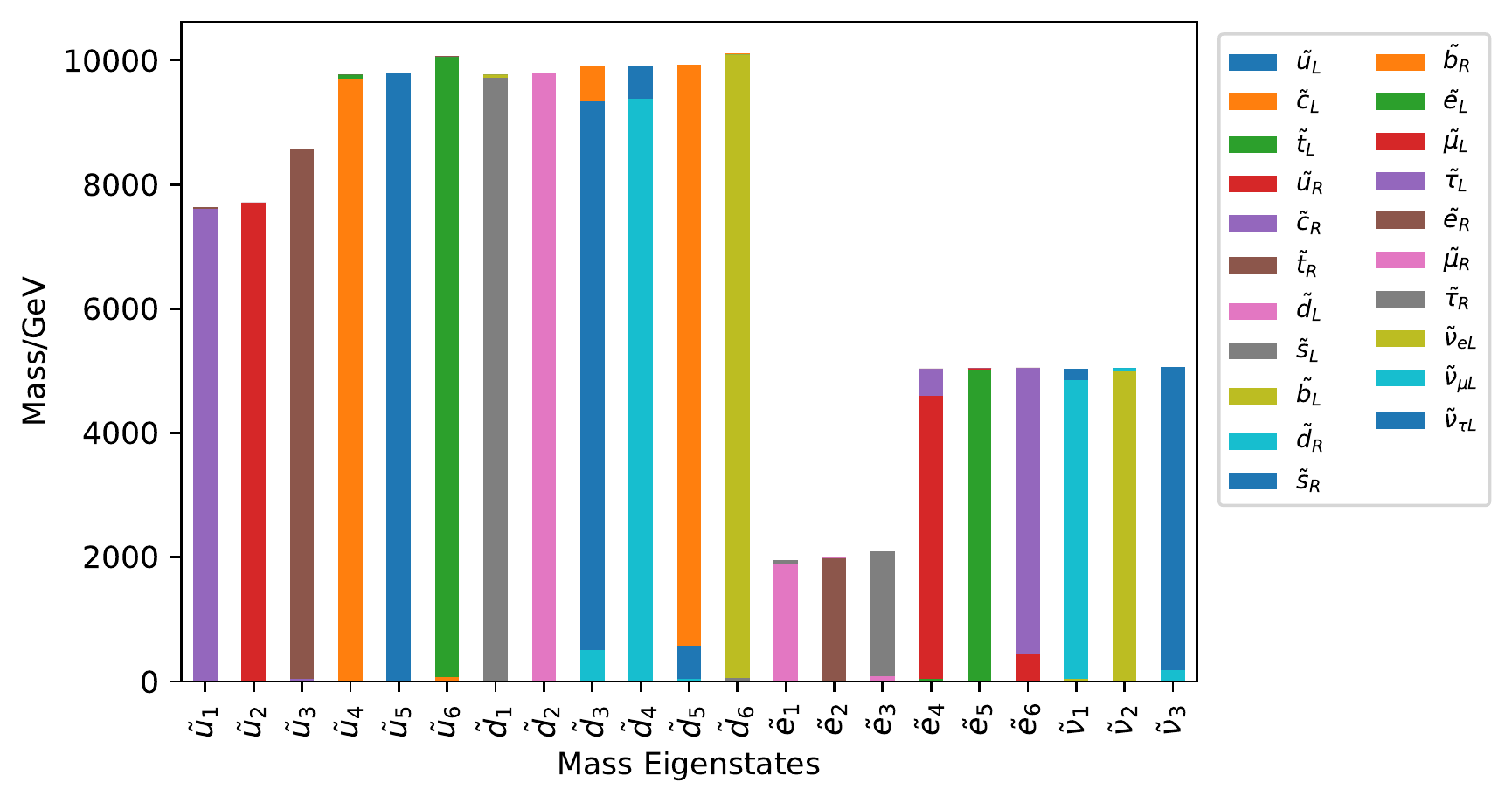}}}
\caption{
The sfermion mass eigenstates in the democratic limit with $\epsilon_u=0.033$, $\epsilon_d=0.108$, $\epsilon_e=0.281$,  $\sigma_u=0.001$, $\sigma_d=0.009$, and $\sigma_e=0.008$, with $M_\text{mess}=10^{12}$ GeV, $\Lambda=8.1\times 10^5$ GeV, and $\tan{\beta}=10$. }
\label{fig2}
\end{figure}

The composition of the mass eigenstates of the sfermions is shown in Fig.~\ref{fig2}. Without perturbations, there is almost no mixing between different flavor eigenstates. The lightest $SU(3)_c$ charged particles are the first and second-generation right-handed squarks and the lightest sleptons are the first and second-generation right-handed sleptons. In Fig.~\ref{fig2}, the results are shown for $\epsilon_u=0.033$, $\epsilon_d=0.108$, $\epsilon_e=0.281$, $\sigma_u=0.001$, $\sigma_d=0.009$, and $\sigma_e=0.008$. The lighter squarks $\tilde{u_1}$ and $\tilde{u}_2$ are again the right-handed scharm and sup. Mixing between the second and third-generations for the left handed sparticles is observed in $\tilde{u}_4$ and $\tilde{u}_6$. There is also small but nonvanishing  2-3 generational mixing among right-handed up-type squarks. For the down sector, apart from the 1-2 and 2-3 generational mixing which are larger compared to the up sector, there is also a small but nonvanishing  left-right mixing between $\tilde{b}_L$ and $\tilde{b}_R$ observed in $\tilde{d}_1$. For the sleptons, we again observe small mixing between the second and third-generation sleptons with the same handedness.

It is illustrative to compare this scenario with the singlet-dominated limit \cite{Everett_2019,Everett_2020}. In this case, the dominant contributions to the soft terms arise in the diagonal third-generation ($33$) entries, rendering this case similar to flavored gauge mediation models in which the Higgs-messenger mixing is controlled by Abelian symmetries. 
Generally this case has a light spectrum, with masses below 6 TeV. Unlike the democratic case, the heavy Higgs particles are heavier than or comparable to the $SU(3)$-charged superpartners, with masses at the $5-6$ TeV range. 
The large stop mixing due to the nonvanishing $A$ term for the third-generation fields at the messenger mass scale allows for a viable Higgs mass at smaller values of $\Lambda$ compared to the democratic limit, in which the $A$ terms vanish in the absence of the small symmetry breaking perturbations. Adding nonrenormalizable corrections as in \cite{Everett_2020} to generate the light quark and charged lepton masses does not alter this feature and generically leads to very small ($O(10^{-1} \; {\rm GeV})$) 

\subsection{Case 2 models}
We now turn to the Case 2 models, for which the superpotential couplings only involving the $\mathcal{S}_3$-doublets dominate. As described in the previous section, in this case there are two sub-categories, depending on whether the $\beta_{3i}$ or the $\beta_{2i}$ parameters control the second-generation quark and charged lepton masses.  Here we will label the mass ordering $\beta_{3i}>\beta_{2i}$ by Case 2a, and the alternate mass ordering $\beta_{2i}>\beta_{3i}$ as Case 2b.  The soft supersymmetry breaking terms for Case 2a are given in Eq.~(\ref{eq:deltamB2-1}), and the analogous quantities for Case 2b are given in Eq.~(\ref{eq:deltamB2-3}).

\begin{figure}[h!]
\centering
\subfloat{{\includegraphics[scale=0.6]{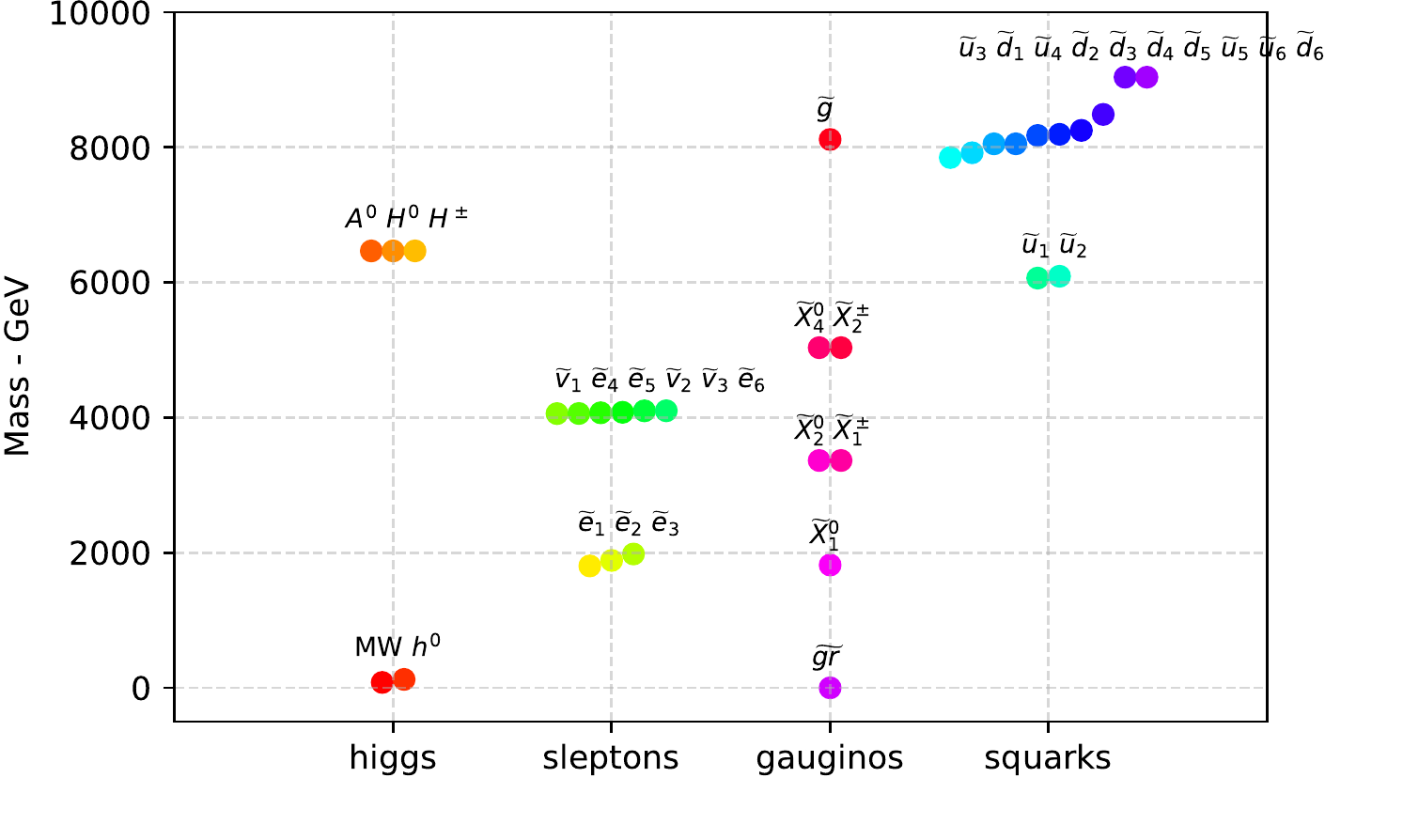}}}
\subfloat{{\includegraphics[scale=0.6]{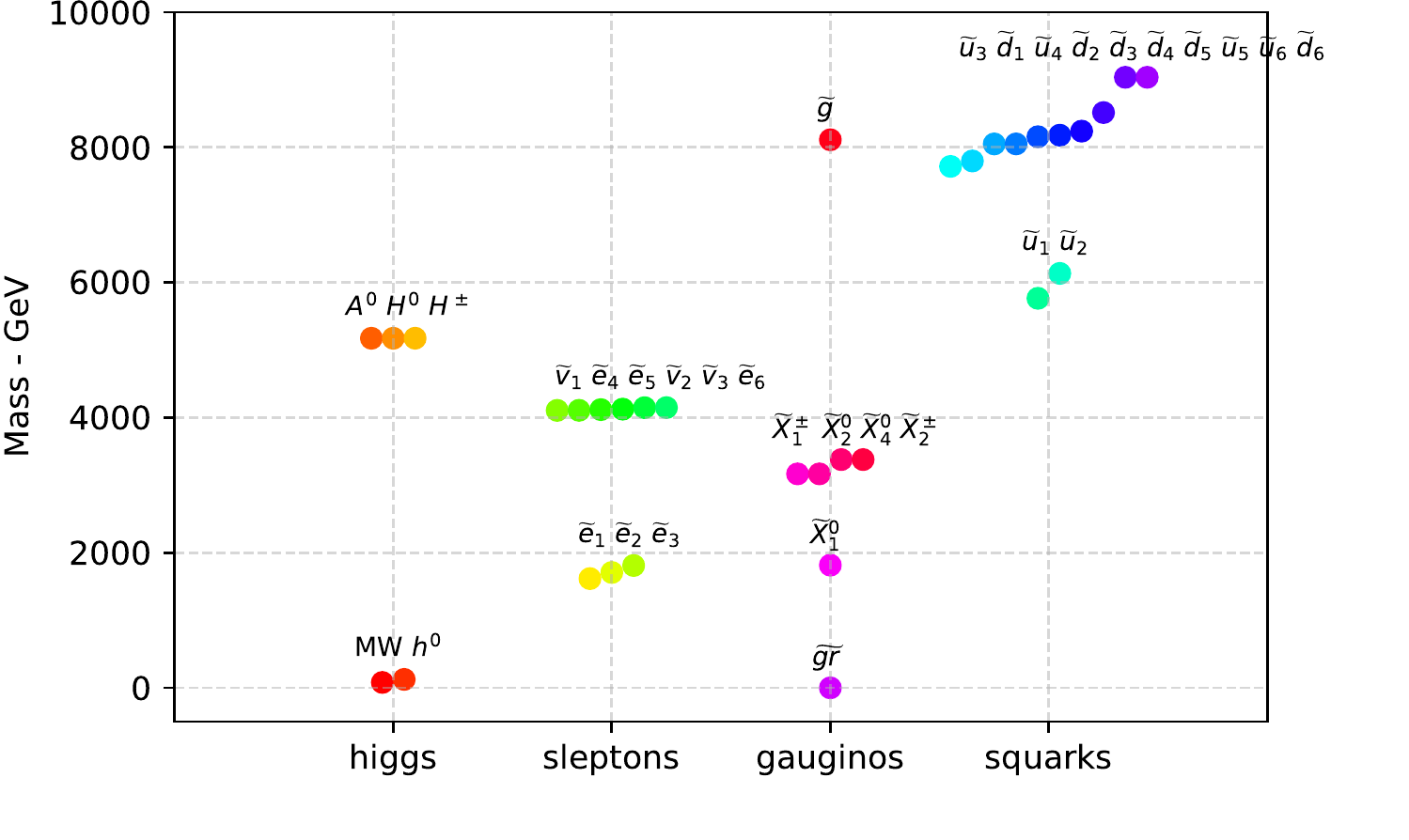}}}
\caption{The sfermion mass spectra in the doublet-dominated (Case 2) limit, with $M_\text{mess}=10^{12}$ GeV, $\Lambda=6.6\times 10^5$ GeV, for (a) Case 2a  $\beta_{3d}=0.03$, $\beta_{3u}=0.01$, $\beta_{3l}=0.08$, $\beta_{2u}=\beta_{2d}=\beta_{2l}=0$ (left), and (b) Case 2b with $\beta_{2d}=0.03$, $\beta_{2u}=0.01$, $\beta_{2l}=0.08$, $\beta_{3u}=\beta_{3d}=\beta_{3l}=0$ (right).}
\label{fig:b2spectra}
\end{figure}

In Case 2 models, there is a nonvanishing trilinear scalar parameter $A_t$ that is present in the absence of the first and second-generation quark and charged lepton masses, in contrast to the Case 1 democratic limit.  Hence, the Higgs and superpartner masses are lighter than their Case 1 democratic counterparts, though not as light as in the Case 1 singlet-dominated limit. 
In Fig.~\ref{fig:b2spectra}, we show characteristic mass spectra for $M_{\rm mess}=10^{12} \; {\rm GeV}$ and $\Lambda = 6.6\times 10^5$ GeV. Here we have included nonvanishing values for the parameters that fix the second-generation quark and charged lepton masses ($\beta_{3i}$ for Case 2a, $\beta_{2i}$ for Case 2b), and neglected the effects of the first-generation masses. The values of the perturbation parameters are chosen to yield appropriate values for the SM fermion mass values. We note here that if these quantities are taken to zero, the mass spectra are almost unchanged, with small changes that are at most $O(10^{-1}\; {\rm GeV})$, primarily in the slepton sector due to the relatively large value of the corresponding $\beta_{3l,2l}$ parameter.  

We see that in both Case 2a and Case 2b, the gluino and squark masses are similar, with the gluino at about 8 TeV and the squarks ranging from approximately $8-10$ TeV. Unlike the Case 1 singlet-dominated limit as in Fig.~\ref{fig:democraticspectrum1} in which the squark masses are generally comparable to heavy Higgses, in this case the squarks are always much heavier than the heavy Higgs bosons. The slepton masses fall into two different ranges, with the NLSP as the lightest selectron $\tilde{e}_1$. The three lightest slectrons have their masses below 2 TeV, while the other sleptons have their masses between 3$-$4 TeV. The lightest charginos and neutralinos are gaugino-dominated, with a binolike lightest neutralino, while the heavier set is higgsino-dominated.

\begin{figure}[h!]
\centering
\includegraphics[scale=0.9]{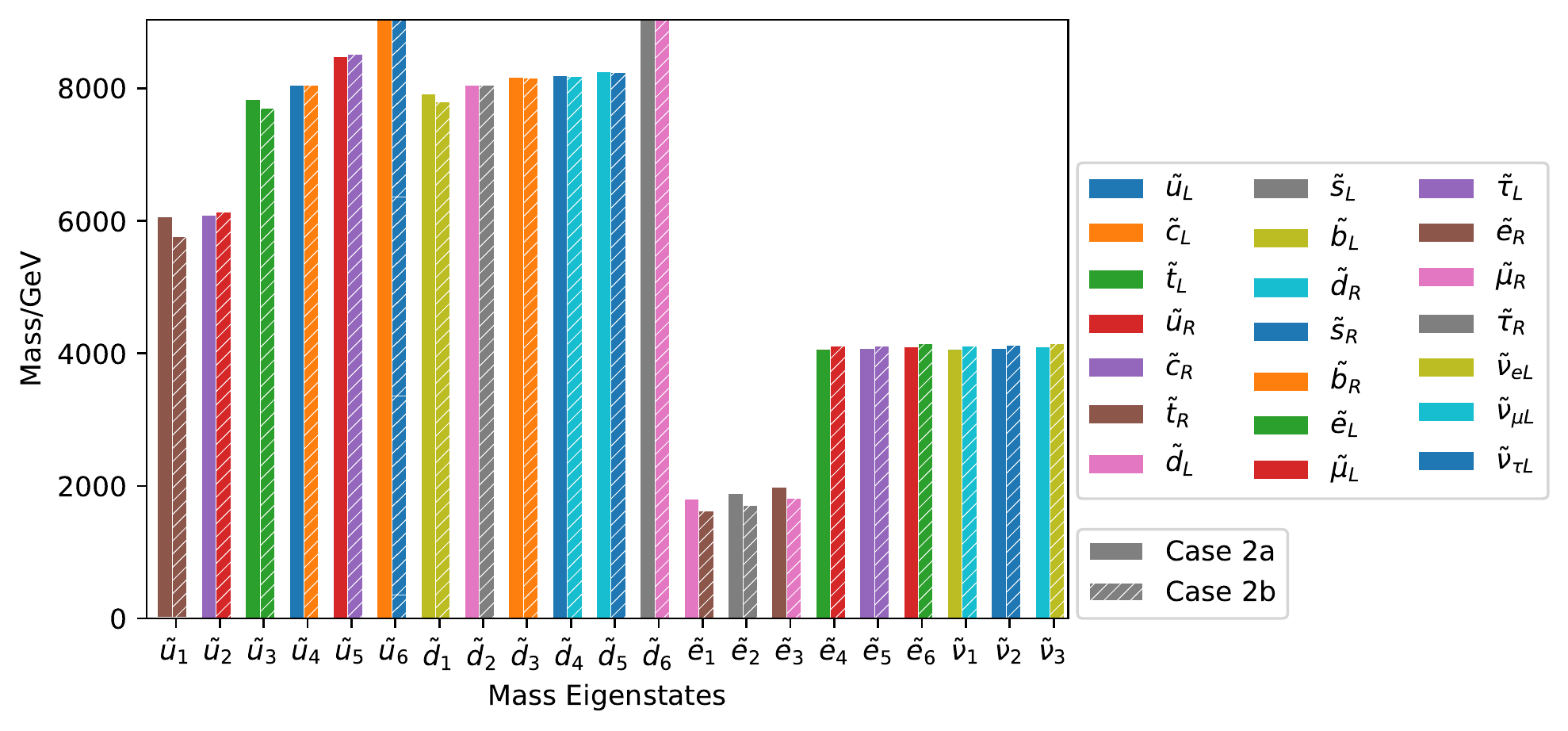}
\caption{The sfermion mass spectrum in the doublet-dominated scenarios, with ordering $\beta_{3i}>\beta_{2i}$ (Case 2a) and $\beta_{2i}>\beta_{3i}$ (Case 2b), respectively, with $M_\text{mess}=10^{12}$ GeV and $\tan{\beta}=10$. For Case 2a, $\beta_{3d}=0.03$, $\beta_{3u}=0.01$, $\beta_{3l}=0.08$. For Case 2b, $\beta_{2d}=0.03$, $\beta_{2u}=0.01$, $\beta_{2l}=0.08$. }
\label{fig8}
\end{figure}

An intriguing difference between Case 2a and Case 2b is that in Case 2b, the heavy Higgs states and the heavy charginos and neutralinos are lighter than they are in Case 2a.  For the model parameters as given in Fig.~\ref{fig:b2spectra}, we see that in Case 2b the heavy Higgs masses are in the 5-6 TeV range, while they are over 6 TeV in Case 2a, and the heavy charginos/neutralinos are also reduced by approximately 1 TeV in Case 2b compared to Case 2a. This indicates that in Case 2b, smaller values of the $\mu$ and $b$ parameters are needed for successful electroweak symmetry breaking. 

Another significant difference between Case 2a and Case 2b is that Case 2a has nonvanishing  off-diagonal contributions to squark mixing, as discussed in the previous subsection. The most significant off-diagonal sfermion mixing in Case 2a is given by $\vert (\delta_u^{12})_{LL}\vert \sim 1\times 10^{-4}$. These effects are small because the flavor off-diagonal contributions are proportional to the small quantities that govern the second-generation SM quark and charged lepton masses. In both cases, as shown in Fig.~\ref{fig8}, sfermion mixing is not significant due to the small size of the perturbation parameters.  For larger values of the messenger mass scale, nontrivial left-right mixing is observed for the third-generation down-type squarks (left-right mixing in the other sfermion sectors is negligible for all values of the messenger mass scale).

\subsection{Discussion}
In comparing the mass spectra of these scenarios (Case 1: democratic, and Cases 2a and 2b: doublet-dominated, as well as the Case 1: singlet-dominated limit as studied in \cite{Everett_2018}), there are several features of interest.
For fixed $M_\text{mess}$, the mass spectra are more compressed for larger values of $\tan{\beta}$ ($\tan{\beta}>10$) because the contributions from the bottom and tau Yukawa couplings are more significant than in the low $\tan{\beta}$ regime.  For smaller $\tan{\beta}$ values, the sparticle masses are heavier as the tree-level contribution to the light Higgs mass has decreased, requiring larger radiative corrections to boost its mass to its experimentally allowed range. The superpartner masses in this limit are thus highly split, with heavy squarks and gluinos, and lighter sleptons. For fixed $\tan\beta$ (here taken to be $\tan\beta=10$), lower values of the messenger mass scale generally lead to heavier spectra, as larger values of $\Lambda$ are needed to satisfy the light Higgs mass constraint. For higher messenger scales, due to increased renormalization group running effects, the $\mu$ and $b/\mu$ terms needed to satisfy the electroweak symmetry breaking constraints are smaller, and thus the heavy charginos and neutralinos become lighter.

\begin{figure}[t]
\centering
\subfloat{{\includegraphics[scale=0.45]{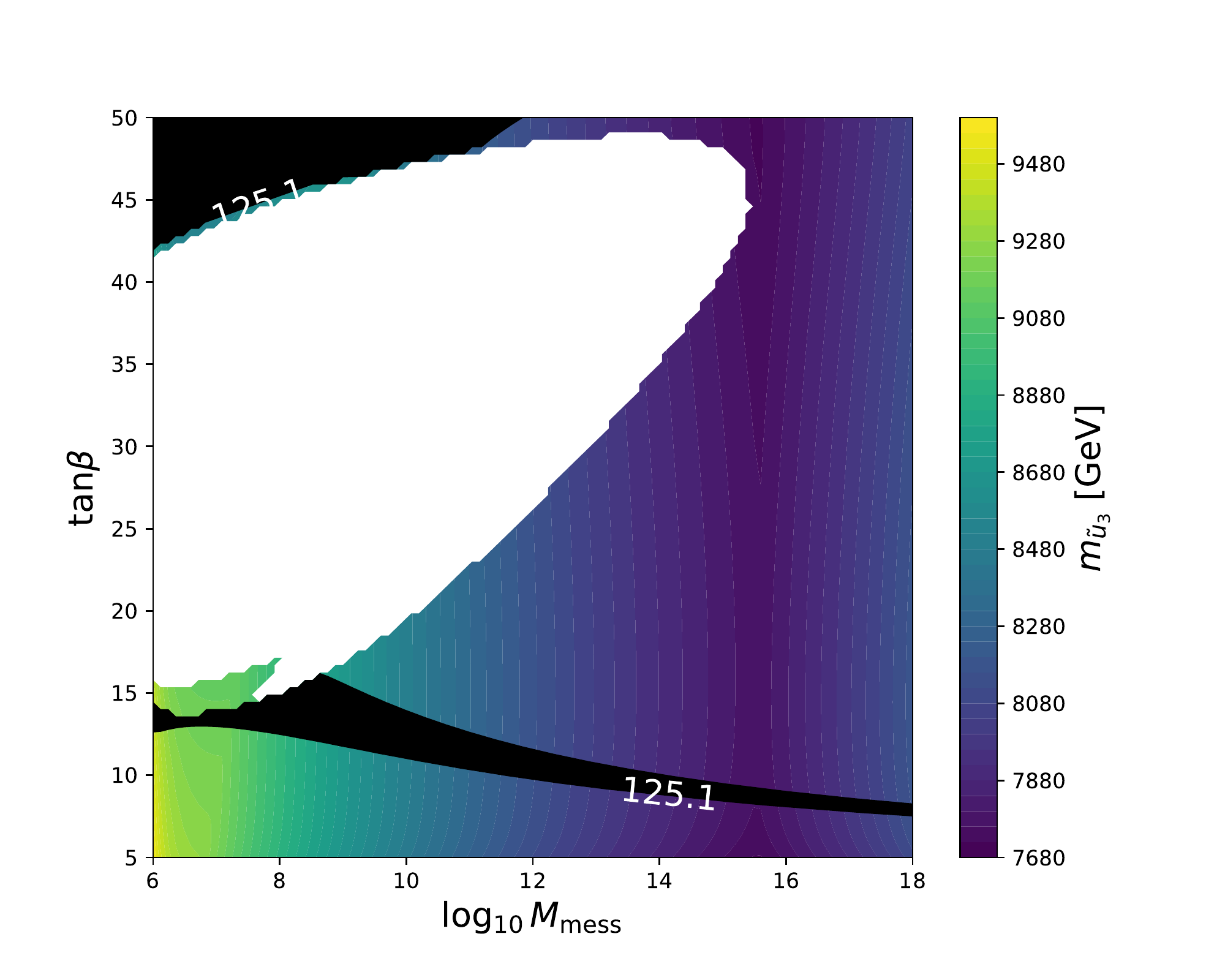}}}
\subfloat{{\includegraphics[scale=0.45]{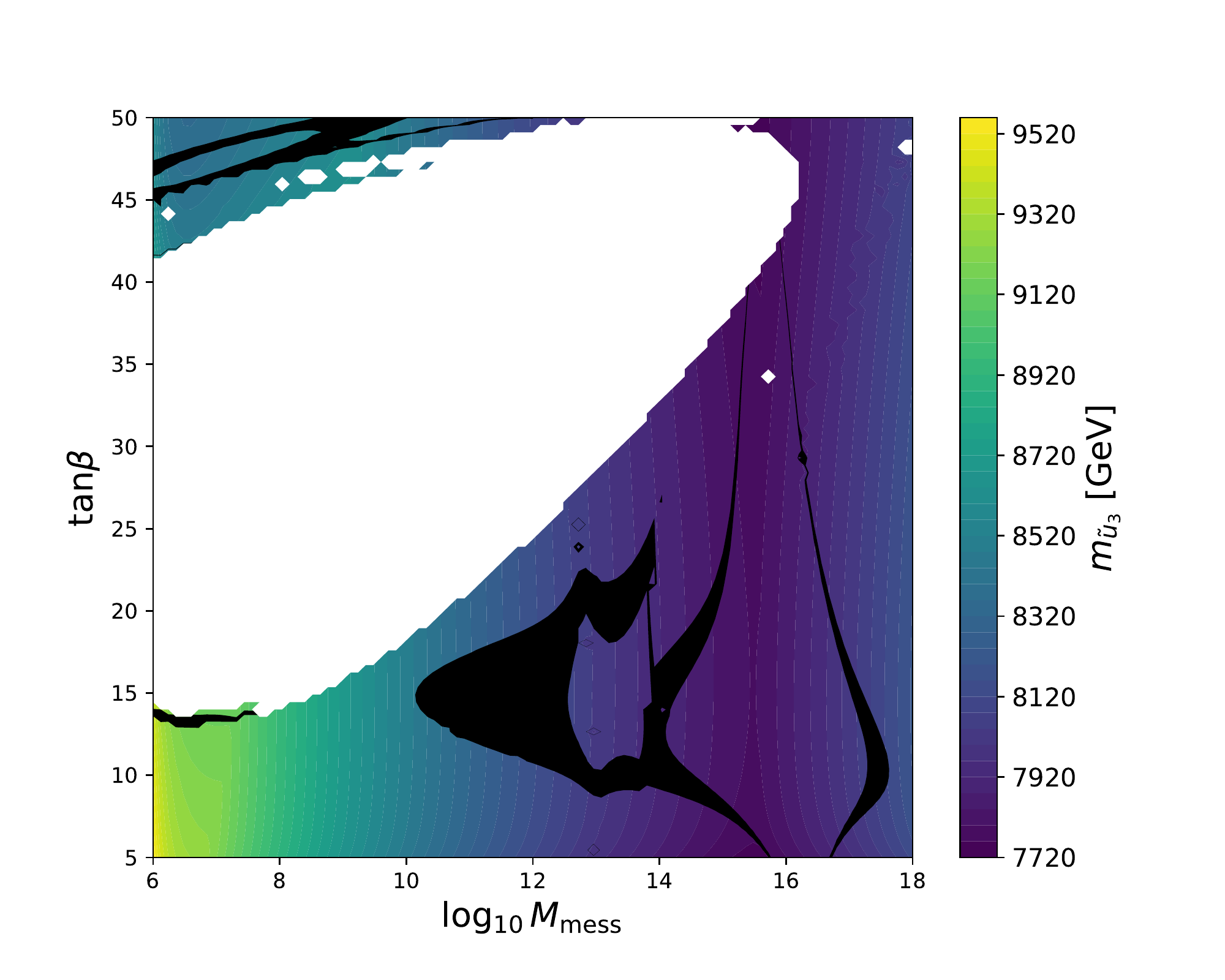}}}
\caption{(a) The Higgs mass (black band) and $\tilde{u}_3$ squark mass (color shading) for Case 1 in the democratic limit without perturbations, with 
$\Lambda=7.7\times10^5$ GeV (left). (b) The same as (a), but with $\epsilon_u=0.033$, $\epsilon_d=0.108$, $\epsilon_e=0.281$, $\sigma_u=0.001$, $\sigma_d=0.009$, $\sigma_e=0.008$ (right).  }
\label{figparameterspace}
\end{figure}

\begin{figure}[h!]
\centering
\subfloat{{\includegraphics[scale=0.35]{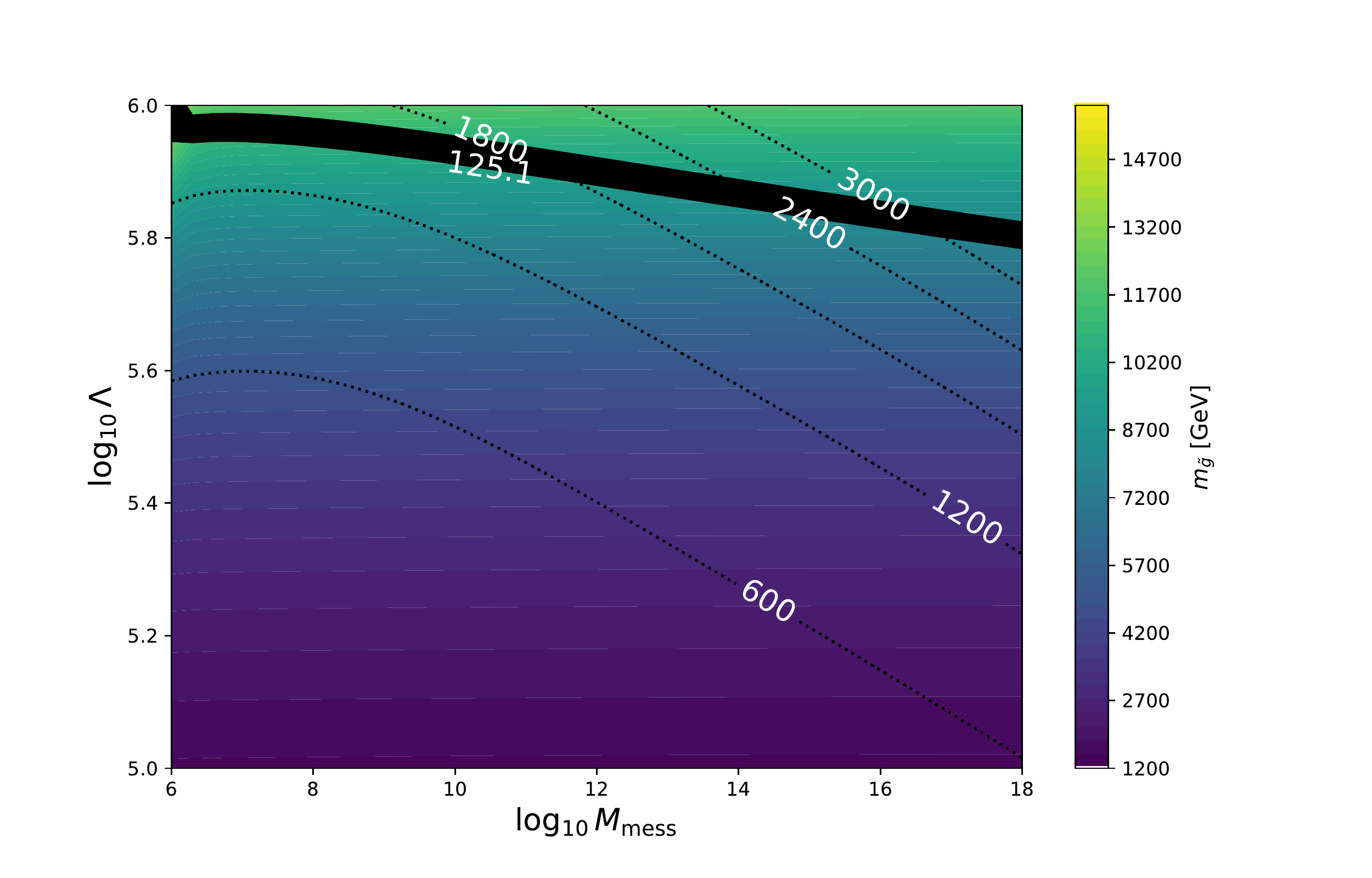}}}
\subfloat{{\includegraphics[scale=0.35]{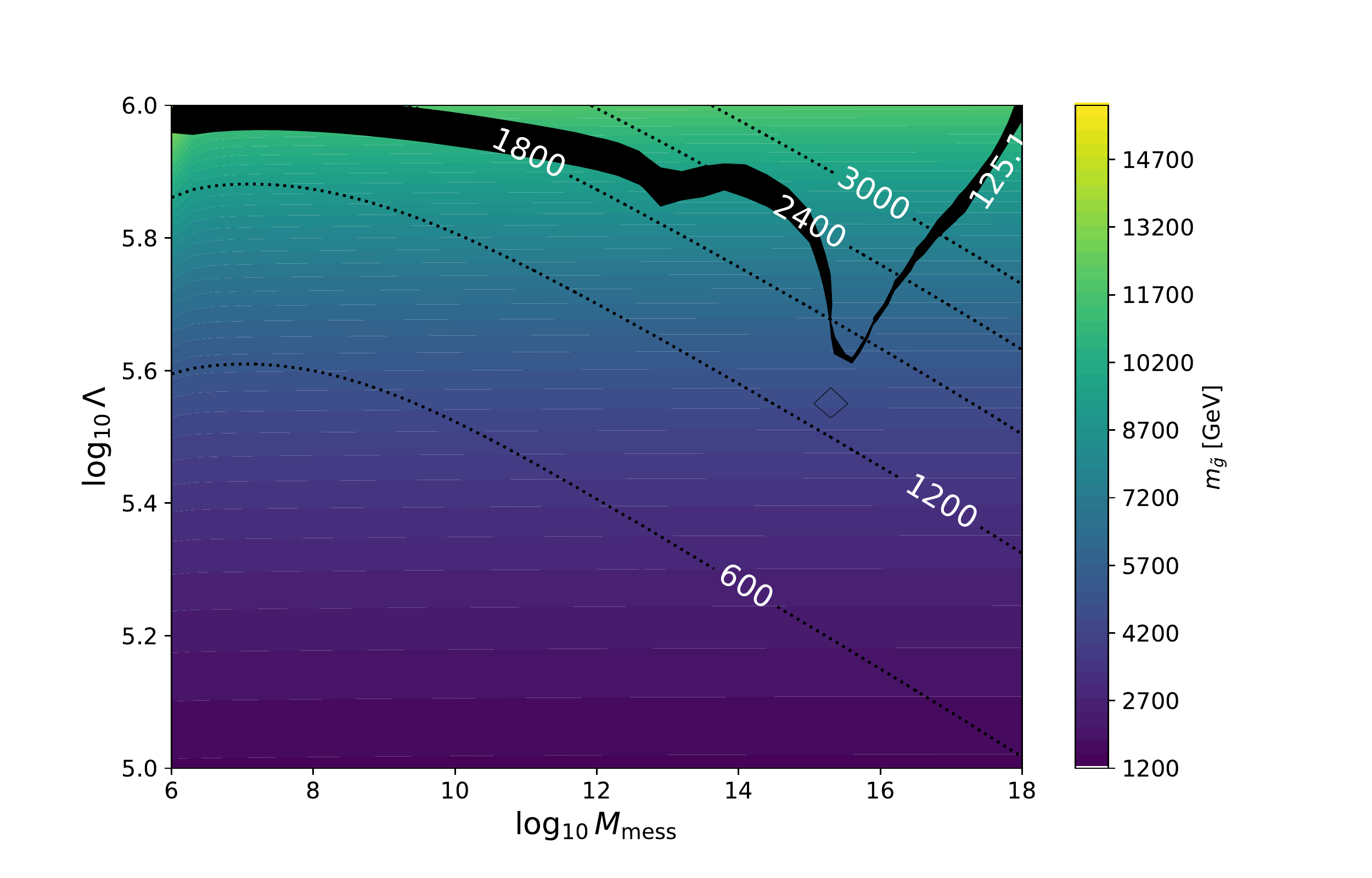}}}
\caption{(a) The Higgs mass (black band), gluino mass (color shading) and $\tilde{e}_1$ mass (dotted curves) as a function of $\Lambda$ and $M_\text{mess}$, for Case 1 in the democratic limit with $\tan\beta=10$, for (a) no perturbations (left), and (b) nonzero perturbations, with $\epsilon_u=0.033$, $\epsilon_d=0.108$, $\epsilon_e=0.281$, $\sigma_u=0.001$, $\sigma_d=0.009$, $\sigma_e=0.008$.}
\label{figparameterlambdammess}
\end{figure}
To further investigate the dependence of the mass spectra on $M_\text{mess}$ and $\tan{\beta}$, we show the Higgs mass curve for fixed $\Lambda$, with the color representing the mass of the lightest squark. We show these results in Fig.~\ref{figparameterspace} for the Case 1 democratic scenario, which show several excluded regions. 
When $\epsilon_u=\epsilon_d=\epsilon_e=0$, the central big ``hole" appears because the mass-squared of the lightest slepton is negative.
When the $\epsilon$ parameters are nonzero, the size of the holes increases, and there are also small holes that appear above the central void because $A^0$ becomes tachyonic in those regions. 
Quite generally, we see that the parameters that satisfy the Higgs mass constraint could be very different in these two cases. 
The lightest value for the mass of the $\tilde{u}_3$ squark is in the region with high $\tan\beta$ and high messenger scales. 
In Fig.~\ref{figparameterlambdammess}, we show the gluino and lightest slepton (NLSP) masses, both without perturbations (left panel) and with perturbations (right panel). The introduction of the perturbations pushes the slepton mass down to smaller values. The change in the shape of the viable Higgs mass region is even more apparent here. For low values of $M_{\rm mess}$, a higher value of $\Lambda$ is needed to satisfy the light Higgs mass constraint. For higher messenger scales  $M_\text{mess}\sim 10^{14}-10^{16}$ GeV, there is a sharp drop in the Higgs mass region that is observed. In that region, there is generally larger left-right mixing in the sbottom sector as well larger scharm-stop mixing, which result in nontrivial contributions to the Higgs mass. However, there are potential numerical instabilities related to the challenges of the Higgs mass calculation in this parameter region. A detailed resolution of these issues is beyond the scope of this paper, and is deferred to future study.

\begin{figure}[h]
\centering
\subfloat{{\includegraphics[scale=0.45]{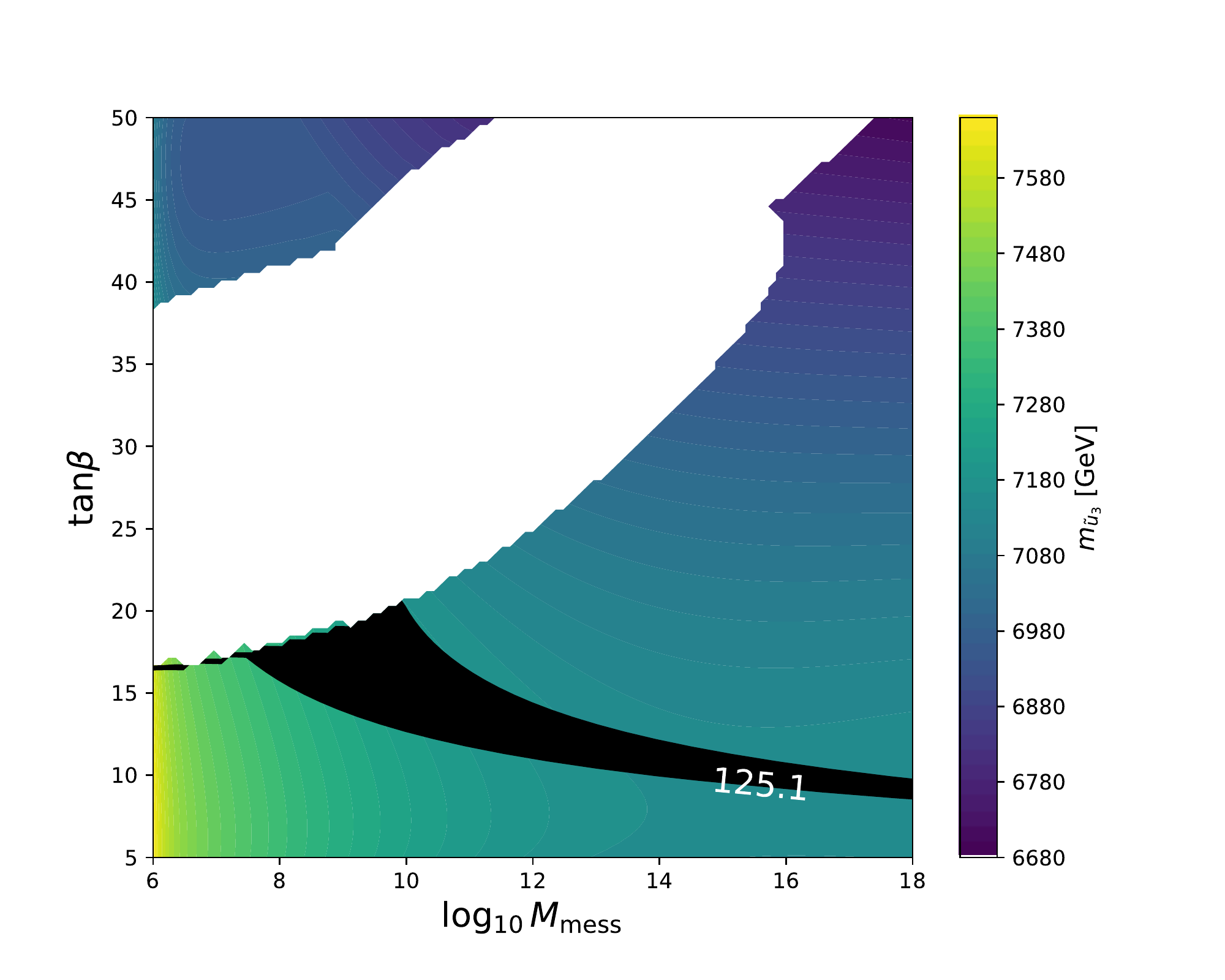}}}
\subfloat{{\includegraphics[scale=0.45]{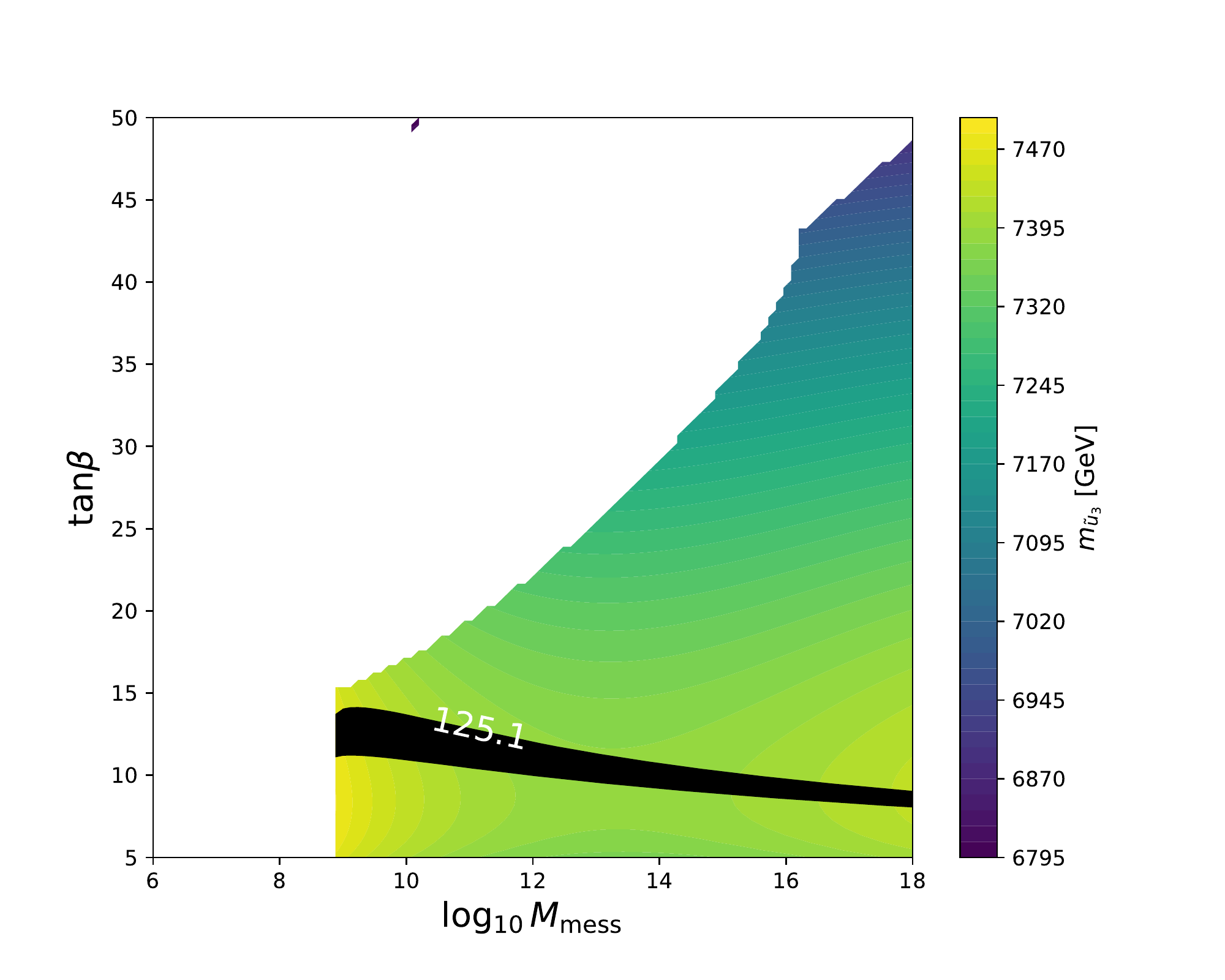}}}
\caption{The mass of the Higgs (black band) and the mass of the lightest squark $\tilde{u}_3$ (color shading) for (a) Case 2a with fixed $\Lambda=6\times10^5$ GeV, 
$\beta_{3d}=0.03$, $\beta_{3u}=0.01$, $\beta_{3l}=0.08$, $\beta_{2u}=\beta_{2d}=\beta_{2l}=0$ (left),
and (b) Case 2b with fixed $\Lambda=6.3\times10^5$ GeV, 
$\beta_{2d}=0.03$, $\beta_{2u}=0.01$, $\beta_{2l}=0.08$, $\beta_{3u}=\beta_{3d}=\beta_{3l}=0$ (right).}
\label{figb2par1}
\end{figure}

\begin{figure}[h!]
\centering
\subfloat{{\includegraphics[scale=0.35]{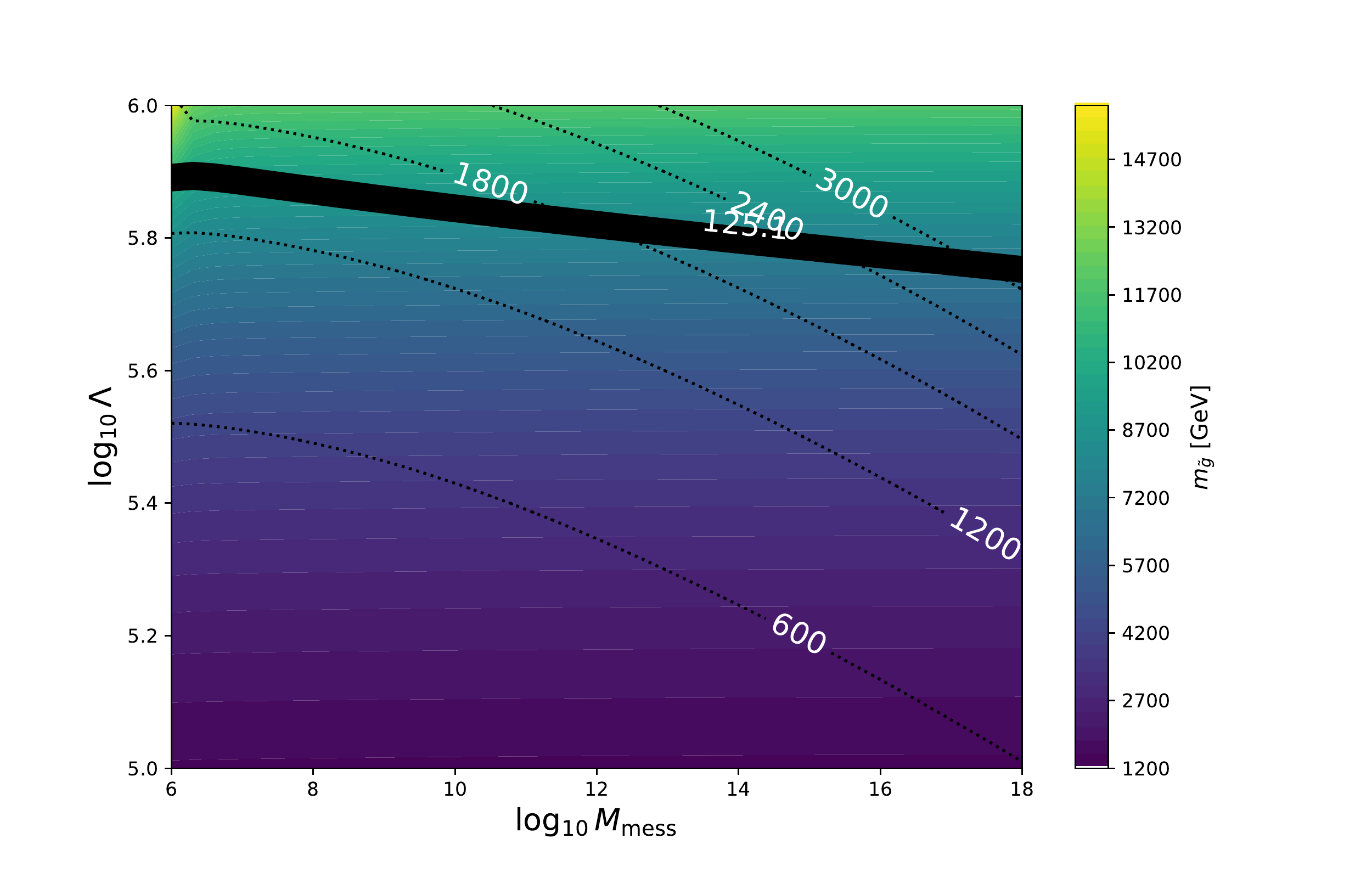}}}
\subfloat{{\includegraphics[scale=0.35]{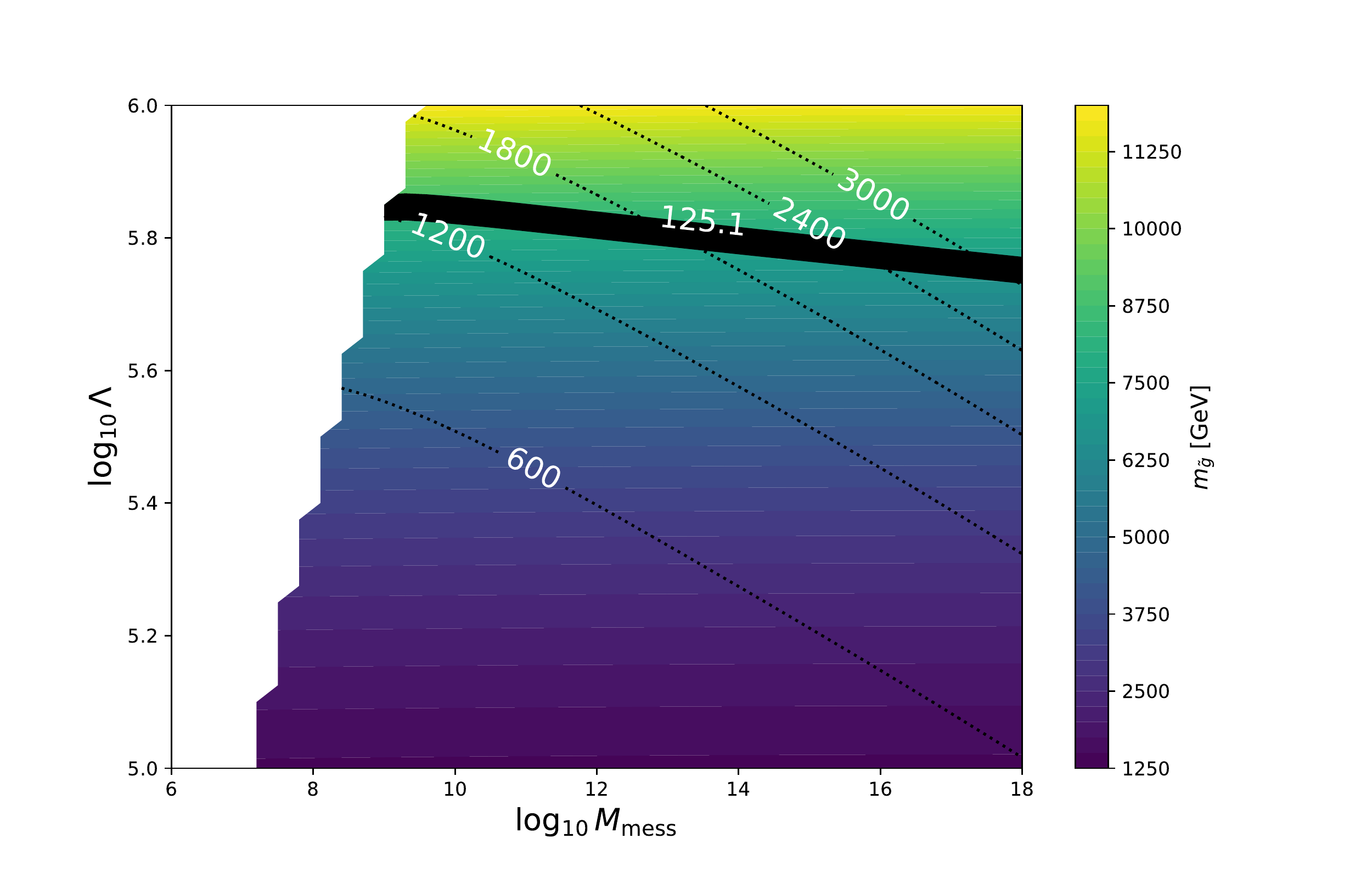}}}
\caption{The light Higgs mass (black band), gluino mass (color shading), and $\tilde{e}_1$ mass (dotted curves) with $\Lambda=6\times 10^5$ GeV, for (a) Case 2a  $\beta_{3d}=0.03$, $\beta_{3u}=0.01$, $\beta_{3l}=0.08$, $\beta_{2u}=\beta_{2d}=\beta_{2l}=0$ (left), and (b) Case 2b with $\beta_{2d}=0.03$, $\beta_{2u}=0.01$, $\beta_{2l}=0.08$, $\beta_{3u}=\beta_{3d}=\beta_{3l}=0$ (right).}
\label{figb2par2}
\end{figure}

For the Case 2 models, we also see excluded regions in the parameter scan in Fig.~\ref{figb2par1}. Here, for both Case 2a and 2b, $\Lambda$ is chosen to maximize viable parameter regions, and the perturbations have a minimal effect on the size of the void.  In Case 2a, the void appears due to tachyonic slepton masses, and the  phenomenologically viable parameter region generally lies between $\tan{\beta}\approx 5-15$, $M_\text{mess}=10^6-10^{18}$ GeV, with a fixed choice of $\Lambda=6\times10^5$ GeV. For Case 2b, apart from the central hole where the lightest slepton becomes tachyonic, the region on the left of the spectrum, which is from $M_\text{mess}\sim10^6-10^9$ GeV and $\tan{\beta}\sim 5-50$, is ruled out because the desired electroweak minimum is not present. We also see that in both cases, the viable Higgs region does not generally intersect with the region where $\tilde{e}_1$ is lighter. In Fig~\ref{figb2par2}, we fix $\tan{\beta}=10$ and show both Case 2a and Case 2b with nonzero perturbations. Note that the effects of the perturbations only slightly shift the mass curve of the NLSP upward, and have almost no effect on the viable Higgs mass region and the gluino masses, in contrast to what we have seen in the democratic limit. 

We close this section by commenting on further phenomenological aspects of this set of models. For all cases described here (both democratic and doublet-dominated models), the superpartner masses are generally heavy and split, in a way that is reminiscent of minimal gauge mediation with $N=2$.  As previously discussed, the constraints of the non-Abelian Higgs-messenger symmetry have led us to include at least two messenger pairs to avoid a catastrophic $\mu/B_\mu$ problem. Ultimately, this means that the scenarios studied in this paper have heavier and more split spectra than what can be obtained in Abelian flavored gauge mediation models (where a judicious choice of $U(1)$ charges can be made to avoid the $\mu/B_\mu$ issue seen here, without increasing the number of messenger pairs), such as in \cite{Ierushalmi_2016}, or in general gauge mediation scenarios \cite{Meade:2008wd}. We recall that in our scenario in the singlet-dominated limit as studied in \cite{Everett_2019,Everett_2020}, it is also possible to minimize the splitting of the mass spectra, though not to the extent that is possible in the Abelian flavored gauge mediation models.

As a result, the discovery potential for the scenarios studied here either via direct LHC searches or indirect constraints is not as promising as it can be in Abelian flavored gauge mediation models, or even in the singlet-dominated non-Abelian scenario. For example, it is straightforward to see that the supersymmetric contribution to muon anomalous magnetic moment (MDM) in the democratic and doublet-dominated non-Abelian flavored gauge mediation scenarios studied here is generically about two orders smaller than the current experimental value \cite{PhysRevLett.126.141801}. 
This is both due to the heavy superpartner masses as described above, and that we are generally precluded from having large values of $\tan\beta$ in these scenarios, which usually provide the largest enhancement to the MDM. Therefore, if new physics is required to resolve any future confirmed discrepancy between the SM prediction and the measured value of the muon anomalous magnetic moment, this set of flavored gauge mediation models would need to be extended to accommodate the experimental result. 

One notable difference in the non-Abelian flavored gauge mediation scenarios studied here compared to minimal gauge mediation with $N=2$ as well as the non-Abelian singlet-dominated flavored gauge mediation scenario is in regards to the NLSP composition. Here, for messenger mass scales of $10^{12}$ GeV as displayed in Fig.~\ref{fig:democraticspectrum1} and Fig.~\ref{fig:b2spectra}, the NLSP is the lightest slepton, which is a right-handed smuon. This is different from the minimal GMSB scenario and the singlet-dominated non-Abelian FGM scenario in which either staus or binolike neutralinos are the NLSP. In the scenarios studied in this paper, for this intermediate to high messenger scale, the smuon NLSP has a lifetime of $\mathcal{O}(0.001\,{\rm s})$, and the NLSP mass is generically close to about 2 TeV. This currently lies above the limits from direct production searches at $\sqrt{s}=13$ TeV \cite{CMS:2018eqb}. 
For lower values of the messenger scale ($\sim 10^6$ GeV), the lightest slepton is still the NLSP, which has a very rapid decay to the gravitino due to the lower supersymmetry breaking scale, while for very high messenger scales ($\sim 10^{14}$ GeV), the NLSP is now a long-lived binolike neutralino. We also note that in the non-Abelian flavored gauge mediation scenarios studied here, there is no significant co-NLSP behavior, both in the low messenger scale and the intermediate to high messenger scale cases. This is in contrast to minimal $N=2$ GMSB for low messenger scales ($\sim 10^6$ GeV), for which there is appreciable co-NLSP behavior among the lighter sleptons for the binolike neutralino NLSP.

We also note that in both minimal $N=2$ GMSB and our non-Abelian flavored gauge-mediation scenarios, for messenger scales of $M_\text{mess}=10^{12}$ GeV, the gravitino has a mass of $\mathcal{O}(0.1\, \text{GeV})$, and the NLSP is not long-lived enough to decay during or after Big Bang nucleosynthesis (BBN). Therefore,  the successful predictions of BBN will not be spoiled (see e.g.~\cite{Steffen:2006hw,Feng:2004zu,ASAKA2000136}). For gravitinos of this mass range, there are well-known mechanisms to ensure the desired reheating temperatures and late entropy production to avoid having the gravitinos overclose the universe, so that gravitinos can then be a plausible dark matter candidate. For lower values of the messenger scale, the situation is further improved, as the gravitinos are lighter (with masses of the order of tenths of keV for $M_{\rm mess}= 10^6$ GeV) and the NLSP decays to gravitinos much more rapidly than in the higher messenger scale case, thus avoiding the need for gravitino dilution.

\section{Conclusions}

In this paper, we have explored MSSM flavored gauge mediation models in which the Higgs-messenger mixing is controlled by a discrete non-Abelian symmetry, here taken for simplicity to be $\mathcal{S}_3$.  Building on previous analyses \cite{Everett_2018} which showed that viable models can be constructed for an extended Higgs-messenger sector that includes both $\mathcal{S}_3$ doublet and singlet fields that mix to yield one light MSSM Higgs pair and two messenger pairs, we studied various possibilities for generating plausible SM quark and charged lepton masses in the case in which the MSSM matter fields also carry $\mathcal{S}_3$ quantum numbers. While additional relations beyond $\mathcal{S}_3$ are generically needed to obtain the desired hierarchical SM fermion masses, we have identified two general categories of solutions that we broadly categorized as Case 1 and Case 2 models.  The Case 1 models obey Eq.~(\ref{eq:betarelations}), and encompass two regimes of interest: (i) the singlet-dominated limit, in which the Yukawa couplings involving only the $\mathcal{S}_3$ singlets dominate, and (ii) the democratic limit, in which the Yukawa superpotential for the MSSM fields has an enhanced $\mathcal{S}_{3L}\times \mathcal{S}_{3R}$ symmetry.  The Case 2 models, in contrast, include the doublet-dominated limit, in which the Yukawa couplings involving only the $\mathcal{S}_3$ doublet fields dominate.  The singlet-dominated limit was previously investigated in \cite{Everett_2019,Everett_2020} and served here as a point of comparison for a general analysis of the Case 1 democratic limit and the Case 2 doublet-dominated models.  We include corrections to obtain nonvanishing masses for one or both of the lighter families, as well as for the third family.  In certain cases such corrections lead to off-diagonal corrections to the soft supersymmetry breaking mass terms, but these corrections are relatively mild (a feature that is known in the literature for flavored gauge mediation models of this general type) and as a result, do not immediately lead to insurmountable problems with flavor-changing neutral current constraints.

Within Case 1 models, our analysis shows that while the singlet-dominated limit allows for examples with optimized parameter sets that yield gluino and squark masses in the 4-5 TeV range, the Case 1 democratic limit generically has significantly heavier squark and gluino masses. The Case 2 models generally also yield heavier superpartner masses, with the heavier squarks and gluino in the 7 TeV mass range.  Ultimately, the fact that the squark and gluino masses cannot be made lighter than 4-5 TeV even in the singlet-dominated limit is related to the fact that this non-Abelian Higgs-messenger mixing scenario requires at the minimum two vectorlike messenger pairs that contribute to the loop diagrams that generate the corrections to the soft terms, to tune the $\mu$ and $b$ terms independently. This should be contrasted with Abelian models, which can have just one messenger pair, and as a result can lead to benchmark scenarios in flavored gauge mediation with lighter $SU(3)$-charged superpartners that are more accessible for searches for supersymmetry at present and future colliders. 

While the spectra in all our examples remain quite heavy, and while we have not constructed fully realistic models of the SM fermion masses and mixing angles (including CP violating effects, not included here for simplicity), we nonetheless find it encouraging that this class of non-Abelian flavored gauge mediation models can include examples that survive this next level of model-building scrutiny. More work is of course needed to see if such scenarios (or plausible extensions of such scenarios) can be embedded into a more complete high-energy model. In the meantime, however, analyses such as this one can serve as a reminder of the rich framework of TeV-scale $\mathcal{N}=1$ supersymmetry, and the many ways in which it might still be hiding at or just above TeV energies.  As the Terascale continues to be explored in this data-rich era for high energy physics, hopefully we will know relatively soon if TeV-scale supersymmetry is indeed part of our physical world.

\acknowledgments
This work is supported by the U.S. Department of Energy under the contract number DE-SC0017647. 

\clearpage 
\section*{Appendix}
\subsection{Case 1 models}
We present the corrections to the soft supersymmetry breaking terms in the Case 1 democratic limit. All relevant terms with magnitudes larger than the smallest perturbation parameter $\sigma_u$ are included. In this case all terms with coefficients of order $O(10^{-3})$GeV are taken into account. For notational simplicity, we define the following quantities:
\begin{equation}
    \begin{split}
        \tilde{g}_u^2&=\frac{16}{3}g_3^2+3g_2^2+\frac{13}{15}g_1^2,\quad \tilde{g}_d^2=\frac{16}{3}g_3^2+3g_2^2+\frac{7}{15}g_1^2,\quad \tilde{g}_e^2=3g_2^2+\frac{9}{5}g_1^2,\\
        \delta_Q&=6y_t^4+6y_b^4+2y_b^2y_t^2+y_b^2y_\tau^2-\Tilde{g_u}^2y_t^2-\Tilde{g_d}^2y_b^2,\\
     \delta_{\epsilon_{u}}&=8y_t^4+y_b^2y_t^2-\tilde{g_u}^2y_t^2, \quad 
      \delta_{\epsilon_d}=8y_b^4+y_b^2y_t^2+y_b^2y_\tau^2-\tilde{g_d}^2y_b^2,\\
       \delta_{\epsilon_e}&=\frac{4}{81} \epsilon _e^2y_b^2  y_{\tau }^2-\frac{8}{729}\epsilon _e^3
   y_b^2  y_{\tau }^2-\frac{244 }{6561}\epsilon _e^4 y_b^2 y_{\tau }^2, \quad 
   \delta_{L}=4y_\tau^4+3y_b^2y_\tau^2-\Tilde{g_e}^2y_\tau^2. \\
    \end{split}
    \label{eq:paramdefcase1}
\end{equation}
 In what follows, all soft scalar mass-squared parameters are assumed to include a factor of $\Lambda^2/(4\pi)^4$ and all trilinear scalar couplings are assumed to include a factor of $\Lambda/(4\pi)^2$, where $\Lambda=F/M_{\rm mess}$.
 The nonvanishing corrections to the soft supersymmetry breaking terms are as follows: 
\begin{equation}
    \begin{split}
     \delta m_{Q_{11}}^2&=
     \delta_Q-\frac{4}{9}\epsilon_u \delta_{\epsilon_u}-\frac{4}{9}\epsilon_d \delta_{\epsilon_d} +\frac{4}{27} \epsilon _d \epsilon _e y_b^2  y_{\tau }^2-\frac{82}{729} \epsilon _d \epsilon _e^2 y_b^2 y_{\tau }^2\\&\quad+\epsilon _d^2 \left(-\frac{26}{81} y_b^2
   \tilde{g}_d^2+\frac{5}{27} y_b^2 y_t^2+\frac{26}{81} y_b^2 y_{\tau }^2+\frac{367 }{81}y_b^4\right)+\delta_{\epsilon_e},
     \\
     \delta m_{Q_{22}}^2&=
     (\delta_Q+\frac{4}{9}\epsilon_u \delta_{\epsilon_u}+\frac{4}{9}\epsilon_d \delta_{\epsilon_d} +\frac{4}{81}\epsilon _d \epsilon _e y_b^2  y_{\tau }^2-\frac{2}{243} \epsilon _d \epsilon _e^2 y_b^2  y_{\tau }^2\\&\quad+\epsilon _d^2 \left(\frac{2}{3} y_b^2
   \tilde{g}_d^2-\frac{88}{81} y_b^2 y_t^2-\frac{2}{3} y_b^2 y_{\tau }^2-\frac{289 }{81}y_b^4\right)+\delta_{\epsilon_e},
     \\
      \delta m_{Q_{23}}^2&=\delta m_{Q_{32}}^2=
      \frac{4\sqrt{2}}{9}\epsilon_u(\delta_{\epsilon_u}-2y_t^4)
        +\frac{4\sqrt{2}}{9}\epsilon_d\left(\delta_{\epsilon_{d}}-2y_b^4\right)-\frac{4\sqrt{2}}{81} \epsilon _d \epsilon _ey_b^2  y_{\tau }^2+\frac{38\sqrt{2}}{729} \epsilon _d \epsilon _e^2 y_b^2 y_{\tau }^2\\&\qquad\qquad\quad+\sqrt{2}\epsilon _d^2\left(\frac{14}{81} y_b^2 \tilde{g}_d^2 -\frac{25}{162} y_b^2 y_t^2-\frac{14}{81} y_b^2 y_{\tau }^2-\frac{59}{81} y_b^4 \right),
        \\
   \delta m_{Q_{12}}^2&=\delta m_{Q_{21}}^2= 
   \frac{4\sqrt{3}}{3} \sigma _d  \left(8 y_b^4-\tilde{g}_d^2y_b^2 +2 y_t^2y_b^2 +y_b^2 y_{\tau }^2\right)+ \frac{4\sqrt{3}}{3} \sigma _u  \left(8 y_t^4-\tilde{g}_u^2y_b^2 +2 y_t^2y_b^2 \right),
   \\
   \delta m_{Q_{33}}^2&= 
   \frac{8}{81} \epsilon _d\epsilon _e  y_b^2y_{\tau }^2-\frac{44}{729}  \epsilon _d\epsilon _e^2y_b^2 y_{\tau }^2+\epsilon _d^2 \left(-\frac{4}{9} y_b^2
   \tilde{g}_d^2+\frac{17}{81} y_b^2 y_t^2+\frac{4}{9} y_b^2 y_{\tau }^2+2 y_b^4\right),
   \\
   \delta m_{Q_{13}}^2&=  \delta m_{Q_{31}}^2= 
   \sqrt{\frac{2}{3}}\sigma _d \left(12 y_b^4-2 \tilde{g}_d^2y_b^2+2 y_t^2y_b^2+2y_b^2 y_{\tau }^2\right)+4\sqrt{\frac{2}{3}}\sigma_u \left(y_t^4+y_t^2y_b^2-\tilde{g}_u^2y_t^2 \right),
   \\
   \nonumber
   \end{split}
   \end{equation}
   \begin{equation}
\begin{split}
        \delta m_{\bar{u}_{11}}^2&=
        (2\delta_{\epsilon_u}-4y_t^4)-\frac{8}{9}\epsilon_u\delta_{\epsilon_u}-\frac{8}{27}\epsilon _d^2 y_b^2  y_t^2,
        \\
         \delta m_{\bar{u}_{12}}^2&=\delta m_{\bar{u}_{21}}^2=\frac{8}{\sqrt{3}}\sigma_u \delta_{\epsilon_u},\qquad   \delta m_{\bar{u}_{13}}^2=\delta m_{\bar{u}_{31}}^2=8\sqrt{\frac{2}{3}}\sigma_u (\delta_{\epsilon_u}-2y_t^4),
        \\
    \delta m_{\bar{u}_{22}}^2&=
    (2\delta_{\epsilon_u}-4y_t^4)+\frac{8}{9}\epsilon_u\delta_{\epsilon_u}-\frac{8}{27}\epsilon _d^2 y_b^2  y_t^2,
    \\
     \delta m_{\bar{u}_{23}}^2&=\delta m_{\bar{u}_{32}}^2=
     \frac{8\sqrt{2}\epsilon_u}{9}\left( \delta_{\epsilon_u}-2y_t^4\right),
     \\
     \delta m_{\bar{u}_{33}}^2&=
     -\frac{8}{9}\epsilon_d^2 y_t^2 y_b^2,
     \\
       \delta m_{\bar{d}_{11}}^2&=
       (2\delta_{\epsilon_d}-4y_b^4)-\frac{8\epsilon_d}{9}\delta_{\epsilon_d}+ \frac{8}{27}\epsilon _d\epsilon _e y_b^2  y_{\tau }^2-\frac{164}{729}\epsilon _d\epsilon _e^2 y_b^2  y_{\tau }^2 \\&\quad+\epsilon _d^2 \left(-\frac{52}{81} y_b^2
   \tilde{g}_d^2+\frac{52}{81} y_b^2 y_t^2+\frac{52}{81} y_b^2 y_{\tau }^2+\frac{680 }{81}y_b^4\right)+2 \delta_{\epsilon_e},
   \\
    \delta m_{\bar{d}_{22}}^2&=
    (2\delta_{\epsilon_d}-4y_b^4)+\frac{8\epsilon_d}{9}\delta_{\epsilon_d}+\frac{8}{81} \epsilon _d \epsilon _e  y_b^2y_{\tau }^2-\frac{4}{243}\epsilon _d  \epsilon _e^2 y_b^2 y_{\tau }^2\\&\quad+\epsilon _d^2 \left(\frac{3}{4} y_b^2
   \tilde{g}_d^2-\frac{116}{81} y_b^2 y_t^2-\frac{4}{3} y_b^2 y_{\tau }^2-\frac{592 }{81}y_b^4\right)+2\delta_{\epsilon_e},
   \\
    \delta m_{\bar{d}_{23}}^2&=\delta m_{\bar{d}_{32}}^2=
    \frac{8\sqrt{2}}{9}\epsilon_d\delta_{\epsilon_d}-\frac{8\sqrt{2} }{81}\epsilon _d\epsilon _e y_b^2  y_{\tau }^2+ \frac{76\sqrt{2}}{729}\epsilon _d\epsilon _e^2  y_b^2  y_{\tau }^2\\&\quad\qquad\qquad+\epsilon _d^2
   \left(\frac{28\sqrt{2}}{81}  y_b^2 \tilde{g}_d^2-\frac{28\sqrt{2}}{81}  y_b^2 y_t^2-\frac{28 \sqrt{2}}{81} y_b^2 y_{\tau }^2-\frac{128 \sqrt{2}}{81} y_b^4\right),
   \\
   \delta m_{\bar{d}_{33}}^2&= 
   \frac{16}{81}\epsilon _d \epsilon _e y_b^2  y_{\tau }^2-\frac{88}{729} \epsilon _d \epsilon _e^2 y_b^2 y_{\tau }^2+\epsilon _d^2 \left(-\frac{9}{8} y_b^2
   \tilde{g}_d^2+\frac{8}{9} y_b^2 y_t^2+\frac{8}{9} y_b^2 y_{\tau }^2+\frac{392 }{81}y_b^4\right), 
   \\
    \delta m_{\bar{d}_{12}}^2&=\delta m_{\bar{d}_{12}}^2= 
    \sigma _d \left(-\frac{3}{8\sqrt{3}}  y_b^2 \tilde{g}_d^2+\frac{8 y_b^2 y_t^2}{\sqrt{3}}+\frac{8 y_b^2 y_{\tau }^2}{\sqrt{3}}+\frac{64 y_b^4}{\sqrt{3}}\right), 
    \\
     \delta m_{\bar{d}_{13}}^2&= \delta m_{\bar{d}_{31}}^2= 
     \sigma _d \left(-\frac{1}{4} \sqrt{\frac{3}{2}} y_b^2 \tilde{g}_d^2+4 \sqrt{\frac{2}{3}} y_b^2 y_t^2+4 \sqrt{\frac{2}{3}} y_b^2 y_{\tau }^2+8 \sqrt{6} y_b^4\right), 
     \\
     \delta m_{L_{11}}^2&=
     \delta_L-\frac{4\epsilon_e}{9}\left(\delta_L+2y_\tau^4\right)+ \frac{4}{9} \epsilon _d \epsilon _e y_b^2  y_{\tau }^2-\frac{22}{81} \epsilon _d\epsilon _e^2 y_b^2  y_{\tau }^2+\frac{4}{27}\epsilon _d^2 y_b^2  y_{\tau
   }^2\\&\quad+\epsilon _e^2 \left(\frac{26}{27} y_b^2 y_{\tau
   }^2-\frac{81}{26} \tilde{g}_e^2 y_{\tau }^2+\frac{283 }{81}y_{\tau }^4\right)+\epsilon _e^3
   \left(-\frac{64}{243} y_b^2 y_{\tau }^2+\frac{729}{64} \tilde{g}_e^2 y_{\tau }^2-\frac{1778}{729} y_{\tau }^4\right)\\&\quad+\epsilon _e^4 \left(-\frac{238}{729} y_b^2 y_{\tau }^2+\frac{2187}{238} \tilde{g}_e^2 y_{\tau }^2+\frac{862 }{2187}y_{\tau }^4\right),
   \\
     \delta m_{L_{22}}^2&=
     \delta_L+\frac{4\epsilon_e}{9}\left(\delta_L+2y_\tau^4\right)+\frac{4}{27} \epsilon _d \epsilon _e y_b^2  y_{\tau }^2+\frac{10}{243} \epsilon _d \epsilon _e^2 y_b^2  y_{\tau }^2+\frac{4}{27} \epsilon _d^2 y_b^2  y_{\tau
   }^2\\&\quad+\epsilon _e^2 \left(-2 y_b^2 y_{\tau
   }^2+\frac{3}{2} \tilde{g}_e^2 y_{\tau }^2-\frac{197}{81} y_{\tau }^4\right)+\epsilon _e^3
   \left(\frac{424}{243} y_b^2 y_{\tau }^2-\frac{729}{424} \tilde{g}_e^2 y_{\tau }^2+\frac{370 }{243}y_{\tau }^4\right)\\&\quad+\epsilon _e^4 \left(-\frac{1642 }{2187}y_b^2 y_{\tau }^2+\frac{6561 }{1642}\tilde{g}_e^2 y_{\tau }^2-\frac{1832 }{6561}y_{\tau }^4\right),
   \\
     \delta m_{L_{23}}^2&=\delta m_{L_{32}}^2=
     \frac{4\sqrt{2}}{9}\epsilon_e\delta_L -\frac{4}{27} \sqrt{2}\epsilon _d \epsilon _e y_b^2  y_{\tau }^2 +\frac{14}{243} \sqrt{2} \epsilon _d \epsilon _e^2 y_b^2  y_{\tau }^2 \\\quad\qquad\qquad& +\epsilon _e^2
   \left(-\frac{14}{27} \sqrt{2} y_b^2 y_{\tau }^2+\frac{81 }{7 \sqrt{2}} \tilde{g}_e^2 y_{\tau }^2-\frac{23}{81} \sqrt{2} y_{\tau }^4\right)+\epsilon _e^3
   \left(-\frac{4}{9} \sqrt{2} y_b^2 y_{\tau }^2+\frac{27 }{2 \sqrt{2}}\tilde{g}_e^2 y_{\tau }^2-\frac{530}{729} \sqrt{2} y_{\tau }^4\right)\\\quad\qquad\qquad&+\epsilon _e^4
   \left(\frac{2402 \sqrt{2}}{2187} y_b^2 y_{\tau }^2-\frac{6561 }{1201 \sqrt{2}}\tilde{g}_e^2 y_{\tau }^2+\frac{7660 \sqrt{2} }{6561}y_{\tau }^4\right),
   \\ 
   \nonumber
   \end{split}
   \end{equation}
   
   \begin{equation}
       \begin{split}
   \delta m_{L_{33}}^2&=
   \frac{8}{27} \epsilon _d\epsilon _e y_b^2  y_{\tau }^2-\frac{76}{243} \epsilon _d\epsilon _e^2 y_b^2  y_{\tau }^2 \\&\quad +\epsilon _e^2 \left(\frac{4}{3} y_b^2 y_{\tau }^2-\frac{9}{4} \tilde{g}_e^2 y_{\tau }^2+\frac{74 }{81}y_{\tau
   }^4\right)+\epsilon _e^3 \left(-\frac{376}{243} y_b^2 y_{\tau }^2+\frac{729}{376}
   \tilde{g}_e^2 y_{\tau }^2-\frac{244 }{243}y_{\tau }^4\right) \\&\quad+\epsilon _e^4 \left(\frac{1868 }{2187}y_b^2 y_{\tau
   }^2-\frac{6561 }{1868}\tilde{g}_e^2 y_{\tau }^2+\frac{436 }{729}y_{\tau }^4\right), 
   \\
      \delta m_{L_{12}}^2&=\delta m_{L_{21}}^2=
      \sigma _e \left(4 \sqrt{3} y_b^2 y_{\tau }^2-\frac{3}{4} \sqrt{3} \tilde{g}_e^2 y_{\tau }^2+8 \sqrt{3} y_{\tau }^4\right), 
      \\
      \delta m_{L_{13}}^2&=\delta m_{L_{31}}^2=
      \sigma _e \left(2 \sqrt{6} y_b^2 y_{\tau }^2-3 \sqrt{\frac{3}{2}} \tilde{g}_e^2 y_{\tau }^2+8 \sqrt{\frac{2}{3}} y_{\tau }^4\right),
      \\
     \delta m_{\bar{e}_{11}}^2&=
     2\delta_L-\frac{8\epsilon_e}{9}\left(\delta_L+2y_\tau^4\right)+\frac{8}{9} \epsilon _d \epsilon _e y_b^2  y_{\tau }^2-\frac{44}{81}  \epsilon _d \epsilon _e^2y_b^2 y_{\tau }^2+\frac{8}{27} y_b^2 \epsilon _d^2 y_{\tau
   }^2 \\&\quad +\epsilon _e^2 \left(\frac{52}{27} y_b^2
   y_{\tau }^2-\frac{81}{52} \tilde{g}_e^2 y_{\tau }^2+\frac{512 }{81}y_{\tau }^4\right)+\epsilon _e^3
   \left(-\frac{128}{243} y_b^2 y_{\tau }^2+\frac{729}{128} \tilde{g}_e^2 y_{\tau }^2-\frac{3016 }{729}y_{\tau }^4\right) \\&\quad +\epsilon _e^4 \left(-\frac{476}{729} y_b^2 y_{\tau }^2+\frac{2187}{476} \tilde{g}_e^2 y_{\tau }^2+\frac{914 }{2187}y_{\tau }^4\right),
   \\
      \delta m_{\bar{e}_{22}}^2&=
      2\delta_L+\frac{8\epsilon_e}{9}\left(\delta_L+2y_\tau^4\right)+\frac{8}{27}\epsilon _d  \epsilon _e y_b^2 y_{\tau }^2+\frac{20}{243}\epsilon _d  \epsilon _e^2 y_b^2 y_{\tau }^2+\frac{8}{27} \epsilon _d^2 y_b^2  y_{\tau
   }^2 \\&\quad +\epsilon _e^2 \left(-4 y_b^2 y_{\tau
   }^2+\frac{3}{4} \tilde{g}_e^2 y_{\tau }^2-\frac{136 }{27}y_{\tau }^4\right)+\epsilon _e^3
   \left(\frac{848}{243} y_b^2 y_{\tau }^2-\frac{729}{848} \tilde{g}_e^2 y_{\tau }^2+\frac{776}{243} y_{\tau }^4\right)\\&\quad+\epsilon _e^4 \left(-\frac{3284 }{2187}y_b^2 y_{\tau }^2+\frac{6561 }{3284}\tilde{g}_e^2 y_{\tau }^2-\frac{1138 }{2187}y_{\tau }^4\right),
   \\
     \delta m_{\bar{e}_{23}}^2&=\delta m_{\bar{e}_{32}}^2=
     \frac{8\sqrt{2}}{9}\epsilon_e\delta_L -\frac{8}{27}\epsilon _d\epsilon _e  \sqrt{2} y_b^2 y_{\tau }^2+ \frac{28}{243} \sqrt{2} \epsilon _d\epsilon _e^2y_b^2  y_{\tau }^2 \\&\quad+\epsilon _e^2
   \left(-\frac{28}{27} \sqrt{2} y_b^2 y_{\tau }^2+\frac{81 }{14 \sqrt{2}}\tilde{g}_e^2 y_{\tau }^2-\frac{56}{81} \sqrt{2} y_{\tau }^4\right)+\epsilon _e^3
   \left(-\frac{8}{9} \sqrt{2} y_b^2 y_{\tau }^2+\frac{27 }{4 \sqrt{2}}\tilde{g}_e^2 y_{\tau }^2-\frac{1096}{729} \sqrt{2} y_{\tau }^4\right) \\&\quad+\epsilon _e^4
   \left(\frac{4804 \sqrt{2}}{2187} y_b^2 y_{\tau }^2-\frac{6561 }{2402 \sqrt{2}}\tilde{g}_e^2 y_{\tau }^2+\frac{16918 \sqrt{2} }{6561}y_{\tau }^4\right),
   \\
   \delta m_{\bar{e}_{33}}^2&=
   \frac{16}{27}\epsilon _d\epsilon _e y_b^2 y_{\tau }^2-\frac{152}{243} \epsilon _d\epsilon _e^2 y_b^2 y_{\tau }^2 +\epsilon _e^2 \left(\frac{8}{3} y_b^2 y_{\tau }^2-\frac{9}{8} \tilde{g}_e^2 y_{\tau
   }^2+\frac{8 }{3}y_{\tau }^4\right)\\&\quad +\epsilon _e^3 \left(-\frac{752}{243} y_b^2 y_{\tau
   }^2+\frac{81}{28} \tilde{g}_e^2 y_{\tau }^2-\frac{704 }{243}y_{\tau }^4\right)+ \epsilon _e^4 \left(\frac{3736 }{2187}y_b^2
   y_{\tau }^2-\frac{6561}{3736} \tilde{g}_e^2 y_{\tau }^2+\frac{10028 }{6561}y_{\tau }^4\right),
   \\
    \delta m_{\bar{e}_{12}}^2&=\delta m_{\bar{e}_{21}}^2=
    \sigma _e \left(8 \sqrt{3} y_b^2 y_{\tau }^2-\frac{3}{8} \sqrt{3} \tilde{g}_e^2 y_{\tau }^2+16 \sqrt{3} y_{\tau }^4\right), 
    \\
    \delta m_{\bar{e}_{13}}^2&=\delta m_{\bar{e}_{31}}^2=
    \sigma _e \left(4 \sqrt{6} y_b^2 y_{\tau }^2-\frac{3}{2} \sqrt{\frac{3}{2}} \tilde{g}_e^2 y_{\tau }^2+16 \sqrt{\frac{2}{3}} y_{\tau }^4\right), 
    \\
     \delta m_{H_u}^2&= 
     -\frac{4}{3}\epsilon _d^2 y_b^2  y_t^2,
     \\
     \delta m_{H_d}^2&=
     \epsilon _d^2 \left(-\frac{4}{27} y_b^2 y_t^2-\frac{40 }{9}y_b^4\right)-\frac{40}{27} \epsilon _e^2 y_{\tau }^4+\frac{368}{243} \epsilon _e^3 y_{\tau
   }^4-\frac{1376}{2187}  \epsilon _e^4 y_{\tau }^4, \nonumber
   \end{split}
   \end{equation}
   \begin{equation}
  \begin{split}
   (\Tilde{A_u})_{13}&=2\sqrt{\frac{2}{3}}\sigma_d y_t y_b^2+4\sqrt{\frac{2}{3}}\sigma_uy_t^3,\qquad\qquad\qquad\quad\;\; (\Tilde{A_u})_{31}=8\sqrt{\frac{2}{3}}\sigma_uy_t^3,\\
   (\Tilde{A_u})_{22}&= -\frac{2\epsilon_u}{9}(3y_t^3+y_ty_b^2),\qquad\qquad\qquad\qquad\qquad\;\; (\Tilde{A_u})_{33}= \frac{4}{9}\epsilon_d^2 y_t y_b^2,\\ 
   (\Tilde{A_u})_{23}&= \frac{4\sqrt{2}}{9}\epsilon_uy_t^3+\frac{4\sqrt{2}}{9}\epsilon_dy_ty_b^2-\frac{14\sqrt{2}}{81}\epsilon_d^2 y_ty_b^2,\qquad (\Tilde{A_u})_{32}=\frac{8\sqrt{2}\epsilon_u}{9}y_t^3,\\
    (\Tilde{A_d})_{13}&=2\sqrt{\frac{2}{3}}\sigma_d y_b^3+4\sqrt{\frac{2}{3}}\sigma_uy_by_t^2,\qquad\qquad\qquad\quad\; (\Tilde{A_d})_{31}= 4\sqrt{\frac{2}{3}}\sigma_d y_b^3,\\
     (\Tilde{A_d})_{22}&= -\frac{2\epsilon_d}{9}(3y_b^3+y_t^2y_b)-\frac{2\epsilon_d^2}{81}(9y_b^3-y_t^2y_b),\qquad  (\Tilde{A_d})_{23}=\frac{4\sqrt{2}\epsilon_d}{9}y_b^3+\frac{4\sqrt{2}\epsilon_u}{9}y_t^2y_b-\frac{10\sqrt{2}}{27}\epsilon_d^2 y_b^3,\\
      (\Tilde{A_d})_{32}&=\frac{8\sqrt{2}\epsilon_d}{9}y_b^3-\frac{4\sqrt{2}}{9}\epsilon_d^2 y_b^3,\qquad\qquad\qquad\qquad\;\;\; (\Tilde{A_d})_{33}=\frac{4}{3}\epsilon_d^2 y_b^3,\\
      (\Tilde{A_e})_{13}&=2\sqrt{\frac{2}{3}}\sigma_e y_\tau^3,\qquad (\Tilde{A_e})_{31}= 4\sqrt{\frac{2}{3}}\sigma_e y_\tau^3,
      \\
       (\Tilde{A_e})_{22}&=-\frac{2}{3}\epsilon_ey_\tau^3-\frac{2}{9} \epsilon _e^2 y_{\tau }^3 +\frac{178}{243} \epsilon _e^3 y_{\tau }^3 -\frac{382}{729} \epsilon _e^4 y_{\tau }^3,
      \\
       (\Tilde{A_e})_{23}&=\frac{4\sqrt{2}}{9}\epsilon_ey_\tau^3-\frac{10}{27} \sqrt{2} \epsilon _e^2 y_{\tau }^3-\frac{4}{81} \sqrt{2} \epsilon _e^3 y_{\tau }^3+\frac{3274 \sqrt{2} }{6561}\epsilon _e^4 y_{\tau }^3,
      \\
      (\Tilde{A_e})_{32}&=\frac{8\sqrt{2}}{9}\epsilon_e y_\tau^3 -\frac{4}{9} \sqrt{2} \epsilon _e^2 y_{\tau }^3-\frac{20}{81} \sqrt{2} \epsilon _e^3 y_{\tau }^3+\frac{5240 \sqrt{2} }{6561}\epsilon _e^4 y_{\tau }^3,
      \\
      (\Tilde{A_e})_{33}&=\frac{4}{3} \epsilon _e^2 y_{\tau }^3-\frac{376}{243} \epsilon _e^3 y_{\tau }^3+\frac{1868 }{2187}\epsilon _e^4 y_{\tau }^3.
      \\
    \end{split}
\end{equation}


\subsection{Case 2 models}
\noindent $\bullet$ {\it Ordering $\beta_{3i}>\beta_{2i}$}.
 As before, all soft scalar mass-squared parameters are assumed to include a factor of $\Lambda^2/(4\pi)^4$, all trilinear scalar couplings are assumed to include a factor of $\Lambda/(4\pi)^2$, and we define $g_{\tilde{u}}^2 = \tilde{g}_u^2/2$,   $g_{\tilde{d}}^2 = \tilde{g}_d^2/2$, and $g_{\tilde{l}}^2 = \tilde{g}_e^2/2$ (see Eq.~(\ref{eq:paramdefcase1})).
Including all the relevant terms up to second order in $\beta_{3u}$, the nonvanishing corrections to the soft supersymmetry breaking terms take the form:
\begin{eqnarray}
(\delta m_{\tilde{Q}}^2)_{11}&=&\frac{195}{16}(y_t^4+y_b^4)+\frac{15}{4}y_t^2 y_b^2+\frac{27}{16}y_b^2 y_\tau^2 - 3y_t^2 g_{\tilde{u}}^2-3y_b^2 g_{\tilde{d}}^2+\frac{9}{16}\beta_{3d}^2y_b^2(6y_b^2+y_t^2)+\frac{9}{16}\beta_{3l}^2y_\tau^2y_b^2 \nonumber \\
&&+ \frac{9}{16}\beta_{3u}^2 y_t^2 (6y_t^2+y_b^2), \nonumber\\
(\delta m_{\tilde{Q}}^2)_{22}&=& \beta_{3d}^2\left( -2g_{\tilde{d}}^2 y_b^2 + \frac{21}{8}y_b^4+\frac{9}{16}y_b^2y_t^2+\frac{5}{8}y_b^2y_\tau^2\right)+ \beta_{3u}^2\left(-2g_{\tilde{u}}^2y_t^2 + \frac{21}{8}y_t^4 + \frac{9}{16}y_b^2y_t^2 \right)+ \frac{3}{8}\beta_{3u}\beta_{3d}y_b^2y_t^2, \nonumber \\
(\delta m_{\tilde{Q}}^2)_{33}&=&\frac{39}{16}(y_t^4+y_b^4)+\frac{5}{4}y_t^2 y_b^2+\frac{11}{16}y_b^2 y_\tau^2 - y_t^2 g_{\tilde{u}}^2-y_b^2 g_{\tilde{d}}^2 -\frac{3}{4}\beta_{3d}^2y_b^4 + \frac{1}{16}\beta_{3l}^2y_b^2y_\tau^2-\frac{3}{4}\beta_{3u}^2y_t^4+\frac{3}{8}\beta_{3u}\beta_{3d}y_b^2y_t^2, \nonumber \\
(\delta m_{\tilde{Q}}^2)_{12}&=&(\delta m_{\tilde{Q}}^2)_{21}=-\frac{3}{4\sqrt{2}}(y_t^4\beta_{3u}+y_b^4 \beta_{3d}-y_t^2 y_b^2 (\beta_{3u}+\beta_{3d})), \nonumber \\
\end{eqnarray}
\begin{eqnarray}
(\delta m_{\tilde{u}}^2)_{22}&=&\frac{189}{8}y_t^4+\frac{9}{2}y_t^2 y_b^2 - 6y_t^2 g_{\tilde{u}}^2+\beta_{3u}^2\left( -g_{\tilde{u}}^2 y_t^2 + \frac{27}{4}y_t^4\right),
 \nonumber \\
(\delta m_{\tilde{u}}^2)_{33}&=&
\frac{45}{8}y_t^4+\frac{1}{2}y_t^2 y_b^2- 2y_t^2 g_{\tilde{u}}^2+\beta_{3u}^2\left( -3g_{\tilde{u}}^2 y_t^2 + \frac{27}{4}y_t^4\right), \nonumber \\
(\delta m_{\tilde{d}}^2)_{22}&=&\frac{189}{8}y_b^4+\frac{9}{2}y_t^2 y_b^2+\frac{27}{8}y_b^2y_\tau^2 - 6y_b^2 g_{\tilde{d}}^2 +\beta_{3d}^2y_b^2 \left(-g_{\tilde{d}}^2+\frac{27}{4}y_b^2+\frac{1}{8}y_\tau^2\right)+\frac{9}{8}\beta_{3l}^2y_b^2y_\tau^2,\nonumber \\
(\delta m_{\tilde{d}}^2)_{33} &=&
\frac{45}{8}y_b^4+\frac{1}{2}y_t^2 y_b^2+\frac{11}{8}y_b^2y_\tau^2- 2y_b^2 g_{\tilde{d}}^2+\beta_{3d}^2y_b^2 \left(-\frac{1}{3}g_{\tilde{d}}^2+\frac{27}{4}y_b^2+\frac{9}{8}y_\tau^2\right)+\frac{1}{8}\beta_{3l}^2y_b^2y_\tau^2, \nonumber \\
(\delta m_{\tilde{L}}^2)_{11}&=&\frac{141}{16}y_\tau^4+\frac{81}{16}y_b^2 y_\tau^2 - 3y_\tau^2 g_{\tilde{l}}^2 +\frac{27}{16}\beta_{3d}^2y_b^2y_\tau^2 + \frac{9}{4}\beta_{3l}^2y_\tau^4, 
\nonumber \\
(\delta m_{\tilde{L}}^2)_{22}&=& \beta_{3l}^2 \left(-2g_{\tilde{l}}^2y_\tau^2 +\frac{15}{8} y_b^2 y_\tau^2 +\frac{11}{8}y_\tau^4  \right), \nonumber\\
(\delta m_{\tilde{L}}^2)_{33}&=&
\frac{17}{16}y_\tau^4+\frac{33}{16}y_b^2 y_\tau^2 - y_\tau^2 g_{\tilde{l}}^2+\frac{3}{16}\beta_{3d}^2y_b^2y_\tau^2 - \frac{7}{8}\beta_{3l}^2y_\tau^4,  \nonumber \\
(\delta m_{\tilde{L}}^2)_{12} &=&
(\delta m_{\tilde{L}}^2)_{21}=-\frac{3}{4\sqrt{2}}(y_\tau^4\beta_{3l})+\frac{3}{8\sqrt{2}}(\beta_{3l}^3 y_\tau^4), \nonumber \\
(\delta m_{\tilde{e}}^2)_{22}&=&\frac{135}{8}y_\tau^4+\frac{81}{8}y_b^2y_\tau^2 - 6y_\tau^2 g_{\tilde{l}}^2+\frac{27}{8}\beta_{3d}^2y_b^2y_\tau^2+\beta_{3l}^2\left(-g_{\tilde{l}}^2y_\tau^2 +\frac{3}{8}y_b^2y_\tau^2 +\frac{17}{4} y_\tau^4 \right), \qquad 
 \nonumber \\
(\delta m_{\tilde{e}}^2)_{33}
&=&
\frac{23}{8}y_\tau^4+\frac{33}{8}y_b^2 y_\tau^2- 2y_\tau^2 g_{\tilde{l}}^2+\frac{3}{8}\beta_{3d}^2y_b^2y_\tau^2+\beta_{3l}^2\left(-3g_{\tilde{l}}^2y_\tau^2 +\frac{27}{8}y_b^2y_\tau^2 +\frac{17}{4} y_\tau^4 \right), \nonumber\\
\delta m_{H_u}^2&=&-\frac{9}{2}y_t^4-\frac{3}{2}y_t^2y_b^2-9\beta_{3u}^2y_t^4,\nonumber \\
\delta m_{H_d}^2&=&-\frac{9}{2}y_b^4-\frac{3}{2}y_\tau^4-\frac{3}{2}y_t^2y_b^2-9\beta_{3d}^2y_b^4-3\beta_{3l}^2y_\tau^4, \nonumber\\
(\tilde{A}_u)_{22}&=&-\frac{3}{\sqrt{2}}y_t^3\beta_{3u}, \qquad\qquad\qquad\qquad (\tilde{A}_u)_{33}=-\frac{3}{2}y_t^3-\frac{1}{2}y_t y_b^2-\frac{3}{2}\beta_{3u}^2y_t^3 , \nonumber \\
(\tilde{A}_d)_{22}&=&-\frac{3}{\sqrt{2}}y_b^3\beta_{3d}, \qquad\qquad\qquad\qquad (\tilde{A}_d)_{33}=-\frac{3}{2}y_b^3-\frac{1}{2}y_b y_t^2-\frac{3}{2}\beta_{3d}^2y_b^3, \nonumber \\
(\tilde{A}_e)_{22}&=&-\frac{3}{\sqrt{2}}y_\tau^3\beta_{3l}-\frac{3}{2\sqrt{2}}y_\tau^3\beta_{3l}^3, \qquad\;\; (\tilde{A}_e)_{33}=-\frac{3}{2}y_\tau^3-\frac{3}{2}y_\tau^3\beta_{3l}^2. 
\label{eq:deltamB2-1}
\end{eqnarray}
\clearpage
\noindent $\bullet$ {\it Ordering $\beta_{2i}>\beta_{3i}$}.
Assuming the subleading $\beta_{3i}=0$, we find the following corrections to the soft terms up to second order in $\beta_{2i}$:
\begin{eqnarray}
(\delta m_{\tilde{Q}}^2)_{22}&=&\frac{195}{16}(y_t^4+y_b^4)+\frac{15}{4}y_t^2 y_b^2+\frac{27}{16}y_b^2 y_\tau^2 - 3y_t^2 g_{\tilde{u}}^2-3y_b^2 g_{\tilde{d}}^2 + \beta_{2d}^2 y_b^2 \nonumber
\left(-\frac{1}{2}g_{\tilde{d}}^2+\frac{27}{8}y_b^2+\frac{3}{16}y_t^2+\frac{1}{16}y_\tau^2 \right)
\\&&+\frac{9}{16}\beta_{2l}^2y_b^2y_\tau^2 + \beta_{2u}^2y_t^2 \left( -\frac{1}{2}g_{\tilde{u}}^2+\frac{27}{8}y_t^2+\frac{3}{16}y_b^2\right)-\frac{3}{8}\beta_{2u}\beta_{2d}y_b^2y_t^2, \nonumber \\
(\delta m_{\tilde{Q}}^2)_{33}&=&\frac{39}{16}(y_t^4+y_b^4)+\frac{5}{4}y_t^2 y_b^2+\frac{11}{16}y_b^2 y_\tau^2 - y_t^2 g_{\tilde{u}}^2-y_b^2 g_{\tilde{d}}^2+ \beta_{2d}^2 y_b^2 \left(-\frac{3}{2}g_{\tilde{d}}^2+\frac{27}{8}y_b^2+\frac{3}{16}y_t^2+\frac{9}{16}y_\tau^2 \right) \nonumber \\
&& +\beta_{2u}^2 y_t^2 \left(-\frac{3}{2}g_{\tilde{u}}^2+\frac{27}{8}y_t^2+\frac{3}{16}y_b^2 \right)-\frac{3}{8}\beta_{2u}\beta_{2d}y_b^2y_t^2, \nonumber\\
(\delta m_{\tilde{u}}^2)_{11}&=&\frac{189}{8}y_t^4+\frac{9}{2}y_t^2 y_b^2 - 6y_t^2 g_{\tilde{u}}^2+\frac{9}{4}\beta_{2d}^2y_b^2y_t^2+\frac{45}{8}\beta_{2u}^2 y_t^4,\nonumber \\
(\delta m_{\tilde{u}}^2)_{22}&=&\beta_{2u}^2 \left (-4g_{\tilde{u}}^2y_t^2+\frac{3}{2}y_b^2y_t^2 + \frac{21}{4}y_t^4 \right),\nonumber \\
(\delta m_{\tilde{u}}^2)_{33}
&=&
\frac{45}{8}y_t^4+\frac{1}{2}y_t^2 y_b^2- 2y_t^2 g_{\tilde{u}}^2 -\frac{3}{4}\beta_{2d}^2y_b^2y_t^2-\frac{3}{8}\beta_{2u}^2y_t^4, \nonumber \\
(\delta m_{\tilde{d}}^2)_{11}&=&\frac{189}{8}y_b^4+\frac{9}{2}y_t^2 y_b^2+\frac{27}{8}y_b^2y_\tau^2 - 6y_b^2 g_{\tilde{d}}^2+\frac{45}{8}\beta_{2d}^2y_b^4+\frac{9}{8}\beta_{2l}^2y_b^2y_\tau^2 + \frac{9}{4}\beta_{2u}^2y_b^2y_t^2,\nonumber \\
(\delta m_{\tilde{d}}^2)_{22} &=& \beta_{2d}^2y_b^2 \left(-\frac{1}{4}g_{\tilde{d}}^2+\frac{21}{4}y_b^2+\frac{3}{2}y_t^2+\frac{5}{4}y_{\tau}^2 \right), \nonumber
\\
(\delta m_{\tilde{d}}^2)_{33} &=& 
\frac{45}{8}y_b^4+\frac{1}{2}y_t^2 y_b^2+\frac{11}{8}y_b^2y_\tau^2- 2y_b^2 g_{\tilde{d}}^2-\frac{3}{8}\beta_{2d}^2y_b^4+\frac{1}{8}\beta_{2l}^2y_b^2y_\tau^2- \frac{3}{4}\beta_{2u}^2y_b^2y_t^2, \nonumber \\
(\delta m_{\tilde{L}}^2)_{22}&=&\frac{141}{16}y_\tau^4+\frac{81}{16}y_b^2 y_\tau^2 - 3y_\tau^2 g_{\tilde{l}}^2 +\frac{27}{16}\beta_{2d}^2y_b^2y_\tau^2 + \beta_{2l}^2\left(-\frac{1}{2}g_{\tilde{l}}^2y_\tau^2 + \frac{3}{16}y_b^2y_\tau^2+\frac{17}{8}y_\tau^4\right), 
\nonumber \\
(\delta m_{\tilde{L}}^2)_{33}&=&
\frac{17}{16}y_\tau^4+\frac{33}{16}y_b^2 y_\tau^2 - y_\tau^2 g_{\tilde{l}}^2+\frac{3}{16}\beta_{2d}^2y_b^2y_\tau^2+ \beta_{2l}^2\left(-\frac{3}{2}g_{\tilde{l}}^2y_\tau^2 + \frac{27}{16}y_b^2y_\tau^2+\frac{17}{8}y_\tau^4\right),  \nonumber \\
(\delta m_{\tilde{e}}^2)_{11}&=&\frac{135}{8}y_\tau^4+\frac{81}{8}y_b^2y_\tau^2 - 6y_\tau^2 g_{\tilde{l}}^2+\frac{27}{8}\beta_{2d}^2y_b^2y_\tau^2+\frac{27}{8}\beta_{2l}^2y_\tau^4, 
 \nonumber \\
(\delta m_{\tilde{e}}^2)_{22}&=& \beta_{2l}^2 \left(-4g_{\tilde{l}}^2 y_\tau^2 +\frac{15}{4}y_b^2y_\tau^2 +\frac{11}{4}y_\tau^4 \right), \nonumber \\
(\delta m_{\tilde{e}}^2)_{33}&=&
\frac{23}{8}y_\tau^4+\frac{33}{8}y_b^2 y_\tau^2- 2y_\tau^2 g_{\tilde{l}}^2+\frac{3}{8}\beta_{2d}^2y_b^2y_\tau^2-\frac{5}{8}\beta_{2l}^2y_\tau^4, \nonumber \\
\delta m_{H_u}^2&=&-\frac{9}{2}y_t^4-\frac{3}{2}y_t^2y_b^2-\frac{9}{4}\beta_{2d}^2y_b^2y_t^2-\frac{9}{4}\beta_{2u}^2y_t^2\left( 2y_t^2+y_b^2\right),\nonumber\\
\delta m_{H_d}^2&=&-\frac{9}{2}y_b^4-\frac{3}{2}y_\tau^4-\frac{3}{2}y_t^2y_b^2-\frac{9}{4}\beta_{2d}^2y_b^2(2y_b^2+y_t^2)-\frac{9}{4}\beta_{2u}^2y_b^2y_t^2, \nonumber\\
(\tilde{A}_u)_{22}&=&-\frac{3}{2\sqrt{2}}\beta_{2u}(y_t^3+y_ty_b^2), \qquad (\tilde{A}_u)_{33}=-\frac{3}{2}y_t^3-\frac{1}{2}y_t y_b^2-\frac{3}{4}\beta_{2d}^2y_b^2y_t-\frac{3}{4}\beta_{2u}^2y_t^3, \nonumber \\
(\tilde{A}_d)_{22}&=&-\frac{3}{2\sqrt{2}}\beta_{2d}(y_b^3+y_by_t^2), \qquad (\tilde{A}_d)_{33}=-\frac{3}{2}y_b^3-\frac{1}{2}y_b y_t^2-\frac{3}{4}\beta_{2d}^2y_b^3-\frac{3}{4}\beta_{2u}^2y_t^2y_b, \nonumber \\
(\tilde{A}_e)_{22}&=&-\frac{3}{2\sqrt{2}}\beta_{2l}y_\tau^3-\frac{9}{4\sqrt{2}}\beta_{2l}^3y_\tau^3, \quad (\tilde{A}_e)_{33}=-\frac{3}{2}y_\tau^3-\frac{3}{4}\beta_{2l}^2y_\tau^3. 
\label{eq:deltamB2-3}
\end{eqnarray}
\clearpage


\bibliography{sample.bib}

\begin{thebibliography}{55}%
\makeatletter
\providecommand \@ifxundefined [1]{%
 \@ifx{#1\undefined}
}%
\providecommand \@ifnum [1]{%
 \ifnum #1\expandafter \@firstoftwo
 \else \expandafter \@secondoftwo
 \fi
}%
\providecommand \@ifx [1]{%
 \ifx #1\expandafter \@firstoftwo
 \else \expandafter \@secondoftwo
 \fi
}%
\providecommand \natexlab [1]{#1}%
\providecommand \enquote  [1]{``#1''}%
\providecommand \bibnamefont  [1]{#1}%
\providecommand \bibfnamefont [1]{#1}%
\providecommand \citenamefont [1]{#1}%
\providecommand \href@noop [0]{\@secondoftwo}%
\providecommand \href [0]{\begingroup \@sanitize@url \@href}%
\providecommand \@href[1]{\@@startlink{#1}\@@href}%
\providecommand \@@href[1]{\endgroup#1\@@endlink}%
\providecommand \@sanitize@url [0]{\catcode `\\12\catcode `\$12\catcode
  `\&12\catcode `\#12\catcode `\^12\catcode `\_12\catcode `\%12\relax}%
\providecommand \@@startlink[1]{}%
\providecommand \@@endlink[0]{}%
\providecommand \url  [0]{\begingroup\@sanitize@url \@url }%
\providecommand \@url [1]{\endgroup\@href {#1}{\urlprefix }}%
\providecommand \urlprefix  [0]{URL }%
\providecommand \Eprint [0]{\href }%
\providecommand \doibase [0]{https://doi.org/}%
\providecommand \selectlanguage [0]{\@gobble}%
\providecommand \bibinfo  [0]{\@secondoftwo}%
\providecommand \bibfield  [0]{\@secondoftwo}%
\providecommand \translation [1]{[#1]}%
\providecommand \BibitemOpen [0]{}%
\providecommand \bibitemStop [0]{}%
\providecommand \bibitemNoStop [0]{.\EOS\space}%
\providecommand \EOS [0]{\spacefactor3000\relax}%
\providecommand \BibitemShut  [1]{\csname bibitem#1\endcsname}%
\let\auto@bib@innerbib\@empty
\bibitem [{\citenamefont {Dine}\ \emph {et~al.}(1981)\citenamefont {Dine},
  \citenamefont {Fischler},\ and\ \citenamefont {Srednicki}}]{Dine:1981za}%
  \BibitemOpen
  \bibfield  {author} {\bibinfo {author} {\bibfnamefont {M.}~\bibnamefont
  {Dine}}, \bibinfo {author} {\bibfnamefont {W.}~\bibnamefont {Fischler}},\
  and\ \bibinfo {author} {\bibfnamefont {M.}~\bibnamefont {Srednicki}},\
  }\bibfield  {title} {\bibinfo {title} {{Supersymmetric Technicolor}},\ }\href
  {https://doi.org/10.1016/0550-3213(81)90582-4} {\bibfield  {journal}
  {\bibinfo  {journal} {Nucl. Phys. B}\ }\textbf {\bibinfo {volume} {189}},\
  \bibinfo {pages} {575} (\bibinfo {year} {1981})}\BibitemShut {NoStop}%
\bibitem [{\citenamefont {Alvarez-Gaumé}\ \emph {et~al.}(1982)\citenamefont
  {Alvarez-Gaumé}, \citenamefont {Claudson},\ and\ \citenamefont
  {Wise}}]{ALVAREZGAUME198296}%
  \BibitemOpen
  \bibfield  {author} {\bibinfo {author} {\bibfnamefont {L.}~\bibnamefont
  {Alvarez-Gaumé}}, \bibinfo {author} {\bibfnamefont {M.}~\bibnamefont
  {Claudson}},\ and\ \bibinfo {author} {\bibfnamefont {M.~B.}\ \bibnamefont
  {Wise}},\ }\bibfield  {title} {\bibinfo {title} {Low-energy supersymmetry},\
  }\href {https://doi.org/https://doi.org/10.1016/0550-3213(82)90138-9}
  {\bibfield  {journal} {\bibinfo  {journal} {Nuclear Physics B}\ }\textbf
  {\bibinfo {volume} {207}},\ \bibinfo {pages} {96} (\bibinfo {year}
  {1982})}\BibitemShut {NoStop}%
\bibitem [{\citenamefont {Dimopoulos}\ and\ \citenamefont
  {Raby}(1981)}]{DIMOPOULOS1981353}%
  \BibitemOpen
  \bibfield  {author} {\bibinfo {author} {\bibfnamefont {S.}~\bibnamefont
  {Dimopoulos}}\ and\ \bibinfo {author} {\bibfnamefont {S.}~\bibnamefont
  {Raby}},\ }\bibfield  {title} {\bibinfo {title} {Supercolor},\ }\href
  {https://doi.org/https://doi.org/10.1016/0550-3213(81)90430-2} {\bibfield
  {journal} {\bibinfo  {journal} {Nuclear Physics B}\ }\textbf {\bibinfo
  {volume} {192}},\ \bibinfo {pages} {353} (\bibinfo {year}
  {1981})}\BibitemShut {NoStop}%
\bibitem [{\citenamefont {Dimopoulos}\ and\ \citenamefont
  {Raby}(1983)}]{DIMOPOULOS1983479}%
  \BibitemOpen
  \bibfield  {author} {\bibinfo {author} {\bibfnamefont {S.}~\bibnamefont
  {Dimopoulos}}\ and\ \bibinfo {author} {\bibfnamefont {S.}~\bibnamefont
  {Raby}},\ }\bibfield  {title} {\bibinfo {title} {Geometric hierarchy},\
  }\href {https://doi.org/https://doi.org/10.1016/0550-3213(83)90652-1}
  {\bibfield  {journal} {\bibinfo  {journal} {Nuclear Physics B}\ }\textbf
  {\bibinfo {volume} {219}},\ \bibinfo {pages} {479} (\bibinfo {year}
  {1983})}\BibitemShut {NoStop}%
\bibitem [{\citenamefont {Dine}\ and\ \citenamefont
  {Fischler}(1982)}]{DINE1982227}%
  \BibitemOpen
  \bibfield  {author} {\bibinfo {author} {\bibfnamefont {M.}~\bibnamefont
  {Dine}}\ and\ \bibinfo {author} {\bibfnamefont {W.}~\bibnamefont
  {Fischler}},\ }\bibfield  {title} {\bibinfo {title} {A phenomenological model
  of particle physics based on supersymmetry},\ }\href
  {https://doi.org/https://doi.org/10.1016/0370-2693(82)91241-2} {\bibfield
  {journal} {\bibinfo  {journal} {Physics Letters B}\ }\textbf {\bibinfo
  {volume} {110}},\ \bibinfo {pages} {227} (\bibinfo {year}
  {1982})}\BibitemShut {NoStop}%
\bibitem [{\citenamefont {Nappi}\ and\ \citenamefont
  {Ovrut}(1982)}]{Nappi:1982hm}%
  \BibitemOpen
  \bibfield  {author} {\bibinfo {author} {\bibfnamefont {C.~R.}\ \bibnamefont
  {Nappi}}\ and\ \bibinfo {author} {\bibfnamefont {B.~A.}\ \bibnamefont
  {Ovrut}},\ }\bibfield  {title} {\bibinfo {title} {{Supersymmetric Extension
  of the SU(3) x SU(2) x U(1) Model}},\ }\href
  {https://doi.org/10.1016/0370-2693(82)90418-X} {\bibfield  {journal}
  {\bibinfo  {journal} {Phys. Lett. B}\ }\textbf {\bibinfo {volume} {113}},\
  \bibinfo {pages} {175} (\bibinfo {year} {1982})}\BibitemShut {NoStop}%
\bibitem [{\citenamefont {Dine}\ and\ \citenamefont
  {Nelson}(1993)}]{Dine_1993}%
  \BibitemOpen
  \bibfield  {author} {\bibinfo {author} {\bibfnamefont {M.}~\bibnamefont
  {Dine}}\ and\ \bibinfo {author} {\bibfnamefont {A.~E.}\ \bibnamefont
  {Nelson}},\ }\bibfield  {title} {\bibinfo {title} {Dynamical supersymmetry
  breaking at low energies},\ }\href {https://doi.org/10.1103/physrevd.48.1277}
  {\bibfield  {journal} {\bibinfo  {journal} {Physical Review D}\ }\textbf
  {\bibinfo {volume} {48}},\ \bibinfo {pages} {1277–1287} (\bibinfo {year}
  {1993})}\BibitemShut {NoStop}%
\bibitem [{\citenamefont {Dine}\ \emph {et~al.}(1995)\citenamefont {Dine},
  \citenamefont {Nelson},\ and\ \citenamefont {Shirman}}]{Dine_1995}%
  \BibitemOpen
  \bibfield  {author} {\bibinfo {author} {\bibfnamefont {M.}~\bibnamefont
  {Dine}}, \bibinfo {author} {\bibfnamefont {A.~E.}\ \bibnamefont {Nelson}},\
  and\ \bibinfo {author} {\bibfnamefont {Y.}~\bibnamefont {Shirman}},\
  }\bibfield  {title} {\bibinfo {title} {Low energy dynamical supersymmetry
  breaking simplified},\ }\href {https://doi.org/10.1103/physrevd.51.1362}
  {\bibfield  {journal} {\bibinfo  {journal} {Physical Review D}\ }\textbf
  {\bibinfo {volume} {51}},\ \bibinfo {pages} {1362–1370} (\bibinfo {year}
  {1995})}\BibitemShut {NoStop}%
\bibitem [{\citenamefont {Dine}\ \emph {et~al.}(1996)\citenamefont {Dine},
  \citenamefont {Nelson}, \citenamefont {Nir},\ and\ \citenamefont
  {Shirman}}]{Dine_1996}%
  \BibitemOpen
  \bibfield  {author} {\bibinfo {author} {\bibfnamefont {M.}~\bibnamefont
  {Dine}}, \bibinfo {author} {\bibfnamefont {A.~E.}\ \bibnamefont {Nelson}},
  \bibinfo {author} {\bibfnamefont {Y.}~\bibnamefont {Nir}},\ and\ \bibinfo
  {author} {\bibfnamefont {Y.}~\bibnamefont {Shirman}},\ }\bibfield  {title}
  {\bibinfo {title} {New tools for low energy dynamical supersymmetry
  breaking},\ }\href {https://doi.org/10.1103/physrevd.53.2658} {\bibfield
  {journal} {\bibinfo  {journal} {Physical Review D}\ }\textbf {\bibinfo
  {volume} {53}},\ \bibinfo {pages} {2658–2669} (\bibinfo {year}
  {1996})}\BibitemShut {NoStop}%
\bibitem [{\citenamefont {Dine}\ \emph {et~al.}(1997)\citenamefont {Dine},
  \citenamefont {Nir},\ and\ \citenamefont {Shirman}}]{Dine_1997}%
  \BibitemOpen
  \bibfield  {author} {\bibinfo {author} {\bibfnamefont {M.}~\bibnamefont
  {Dine}}, \bibinfo {author} {\bibfnamefont {Y.}~\bibnamefont {Nir}},\ and\
  \bibinfo {author} {\bibfnamefont {Y.}~\bibnamefont {Shirman}},\ }\bibfield
  {title} {\bibinfo {title} {Variations on minimal gauge-mediated supersymmetry
  breaking},\ }\href {https://doi.org/10.1103/physrevd.55.1501} {\bibfield
  {journal} {\bibinfo  {journal} {Physical Review D}\ }\textbf {\bibinfo
  {volume} {55}},\ \bibinfo {pages} {1501–1508} (\bibinfo {year}
  {1997})}\BibitemShut {NoStop}%
\bibitem [{\citenamefont {Giudice}\ and\ \citenamefont
  {Rattazzi}(1999)}]{Giudice_1999}%
  \BibitemOpen
  \bibfield  {author} {\bibinfo {author} {\bibfnamefont {G.}~\bibnamefont
  {Giudice}}\ and\ \bibinfo {author} {\bibfnamefont {R.}~\bibnamefont
  {Rattazzi}},\ }\bibfield  {title} {\bibinfo {title} {Theories with
  gauge-mediated supersymmetry breaking},\ }\href
  {https://doi.org/10.1016/s0370-1573(99)00042-3} {\bibfield  {journal}
  {\bibinfo  {journal} {Physics Reports}\ }\textbf {\bibinfo {volume} {322}},\
  \bibinfo {pages} {419–499} (\bibinfo {year} {1999})}\BibitemShut {NoStop}%
\bibitem [{\citenamefont {Draper}\ \emph {et~al.}(2012)\citenamefont {Draper},
  \citenamefont {Meade}, \citenamefont {Reece},\ and\ \citenamefont
  {Shih}}]{Draper_2012}%
  \BibitemOpen
  \bibfield  {author} {\bibinfo {author} {\bibfnamefont {P.}~\bibnamefont
  {Draper}}, \bibinfo {author} {\bibfnamefont {P.}~\bibnamefont {Meade}},
  \bibinfo {author} {\bibfnamefont {M.}~\bibnamefont {Reece}},\ and\ \bibinfo
  {author} {\bibfnamefont {D.}~\bibnamefont {Shih}},\ }\bibfield  {title}
  {\bibinfo {title} {Implications of a 125 gev higgs boson for the mssm and
  low-scale supersymmetry breaking},\ }\bibfield  {journal} {\bibinfo
  {journal} {Physical Review D}\ }\textbf {\bibinfo {volume} {85}},\ \href
  {https://doi.org/10.1103/physrevd.85.095007} {10.1103/physrevd.85.095007}
  (\bibinfo {year} {2012})\BibitemShut {NoStop}%
\bibitem [{\citenamefont {Arbey}\ \emph {et~al.}(2012)\citenamefont {Arbey},
  \citenamefont {Battaglia}, \citenamefont {Djouadi}, \citenamefont
  {Mahmoudi},\ and\ \citenamefont {Quevillon}}]{Arbey_2012}%
  \BibitemOpen
  \bibfield  {author} {\bibinfo {author} {\bibfnamefont {A.}~\bibnamefont
  {Arbey}}, \bibinfo {author} {\bibfnamefont {M.}~\bibnamefont {Battaglia}},
  \bibinfo {author} {\bibfnamefont {A.}~\bibnamefont {Djouadi}}, \bibinfo
  {author} {\bibfnamefont {F.}~\bibnamefont {Mahmoudi}},\ and\ \bibinfo
  {author} {\bibfnamefont {J.}~\bibnamefont {Quevillon}},\ }\bibfield  {title}
  {\bibinfo {title} {Implications of a 125 gev higgs for supersymmetric
  models},\ }\href {https://doi.org/10.1016/j.physletb.2012.01.053} {\bibfield
  {journal} {\bibinfo  {journal} {Physics Letters B}\ }\textbf {\bibinfo
  {volume} {708}},\ \bibinfo {pages} {162–169} (\bibinfo {year}
  {2012})}\BibitemShut {NoStop}%
\bibitem [{\citenamefont {Adeel~Ajaib}\ \emph {et~al.}(2012)\citenamefont
  {Adeel~Ajaib}, \citenamefont {Gogoladze}, \citenamefont {Nasir},\ and\
  \citenamefont {Shafi}}]{Adeel_Ajaib_2012}%
  \BibitemOpen
  \bibfield  {author} {\bibinfo {author} {\bibfnamefont {M.}~\bibnamefont
  {Adeel~Ajaib}}, \bibinfo {author} {\bibfnamefont {I.}~\bibnamefont
  {Gogoladze}}, \bibinfo {author} {\bibfnamefont {F.}~\bibnamefont {Nasir}},\
  and\ \bibinfo {author} {\bibfnamefont {Q.}~\bibnamefont {Shafi}},\ }\bibfield
   {title} {\bibinfo {title} {Revisiting mgmsb in light of a 125 gev higgs},\
  }\href {https://doi.org/10.1016/j.physletb.2012.06.036} {\bibfield  {journal}
  {\bibinfo  {journal} {Physics Letters B}\ }\textbf {\bibinfo {volume}
  {713}},\ \bibinfo {pages} {462–468} (\bibinfo {year} {2012})}\BibitemShut
  {NoStop}%
\bibitem [{\citenamefont {Fischler}\ and\ \citenamefont
  {Tangarife}(2014)}]{Fischler_2014}%
  \BibitemOpen
  \bibfield  {author} {\bibinfo {author} {\bibfnamefont {W.}~\bibnamefont
  {Fischler}}\ and\ \bibinfo {author} {\bibfnamefont {W.}~\bibnamefont
  {Tangarife}},\ }\bibfield  {title} {\bibinfo {title} {Vector-like fields,
  messenger mixing and the higgs mass in gauge mediation},\ }\bibfield
  {journal} {\bibinfo  {journal} {Journal of High Energy Physics}\ }\textbf
  {\bibinfo {volume} {2014}},\ \href {https://doi.org/10.1007/jhep05(2014)151}
  {10.1007/jhep05(2014)151} (\bibinfo {year} {2014})\BibitemShut {NoStop}%
\bibitem [{\citenamefont {Calibbi}\ \emph {et~al.}(2014)\citenamefont
  {Calibbi}, \citenamefont {Paradisi},\ and\ \citenamefont
  {Ziegler}}]{Calibbi_2014}%
  \BibitemOpen
  \bibfield  {author} {\bibinfo {author} {\bibfnamefont {L.}~\bibnamefont
  {Calibbi}}, \bibinfo {author} {\bibfnamefont {P.}~\bibnamefont {Paradisi}},\
  and\ \bibinfo {author} {\bibfnamefont {R.}~\bibnamefont {Ziegler}},\
  }\bibfield  {title} {\bibinfo {title} {Lepton flavor violation in flavored
  gauge mediation},\ }\bibfield  {journal} {\bibinfo  {journal} {The European
  Physical Journal C}\ }\textbf {\bibinfo {volume} {74}},\ \href
  {https://doi.org/10.1140/epjc/s10052-014-3211-x}
  {10.1140/epjc/s10052-014-3211-x} (\bibinfo {year} {2014})\BibitemShut
  {NoStop}%
\bibitem [{\citenamefont {Meade}\ \emph {et~al.}(2009)\citenamefont {Meade},
  \citenamefont {Seiberg},\ and\ \citenamefont {Shih}}]{Meade:2008wd}%
  \BibitemOpen
  \bibfield  {author} {\bibinfo {author} {\bibfnamefont {P.}~\bibnamefont
  {Meade}}, \bibinfo {author} {\bibfnamefont {N.}~\bibnamefont {Seiberg}},\
  and\ \bibinfo {author} {\bibfnamefont {D.}~\bibnamefont {Shih}},\ }\bibfield
  {title} {\bibinfo {title} {{General Gauge Mediation}},\ }\href
  {https://doi.org/10.1143/PTPS.177.143} {\bibfield  {journal} {\bibinfo
  {journal} {Prog. Theor. Phys. Suppl.}\ }\textbf {\bibinfo {volume} {177}},\
  \bibinfo {pages} {143} (\bibinfo {year} {2009})},\ \Eprint
  {https://arxiv.org/abs/0801.3278} {arXiv:0801.3278 [hep-ph]} \BibitemShut
  {NoStop}%
\bibitem [{\citenamefont {Chacko}\ and\ \citenamefont
  {Pontón}(2002)}]{Chacko_2002}%
  \BibitemOpen
  \bibfield  {author} {\bibinfo {author} {\bibfnamefont {Z.}~\bibnamefont
  {Chacko}}\ and\ \bibinfo {author} {\bibfnamefont {E.}~\bibnamefont
  {Pontón}},\ }\bibfield  {title} {\bibinfo {title} {Yukawa deflected gauge
  mediation},\ }\bibfield  {journal} {\bibinfo  {journal} {Physical Review D}\
  }\textbf {\bibinfo {volume} {66}},\ \href
  {https://doi.org/10.1103/physrevd.66.095004} {10.1103/physrevd.66.095004}
  (\bibinfo {year} {2002})\BibitemShut {NoStop}%
\bibitem [{\citenamefont {Shadmi}\ and\ \citenamefont
  {Szabo}(2012)}]{Shadmi_2012}%
  \BibitemOpen
  \bibfield  {author} {\bibinfo {author} {\bibfnamefont {Y.}~\bibnamefont
  {Shadmi}}\ and\ \bibinfo {author} {\bibfnamefont {P.~Z.}\ \bibnamefont
  {Szabo}},\ }\bibfield  {title} {\bibinfo {title} {Flavored gauge-mediation},\
  }\bibfield  {journal} {\bibinfo  {journal} {Journal of High Energy Physics}\
  }\textbf {\bibinfo {volume} {2012}},\ \href
  {https://doi.org/10.1007/jhep06(2012)124} {10.1007/jhep06(2012)124} (\bibinfo
  {year} {2012})\BibitemShut {NoStop}%
\bibitem [{\citenamefont {Evans}\ \emph
  {et~al.}(2011{\natexlab{a}})\citenamefont {Evans}, \citenamefont {Ibe},\ and\
  \citenamefont {Yanagida}}]{Evans_2011}%
  \BibitemOpen
  \bibfield  {author} {\bibinfo {author} {\bibfnamefont {J.~L.}\ \bibnamefont
  {Evans}}, \bibinfo {author} {\bibfnamefont {M.}~\bibnamefont {Ibe}},\ and\
  \bibinfo {author} {\bibfnamefont {T.~T.}\ \bibnamefont {Yanagida}},\
  }\bibfield  {title} {\bibinfo {title} {Relatively heavy higgs boson in more
  generic gauge mediation},\ }\href
  {https://doi.org/10.1016/j.physletb.2011.10.031} {\bibfield  {journal}
  {\bibinfo  {journal} {Physics Letters B}\ }\textbf {\bibinfo {volume}
  {705}},\ \bibinfo {pages} {342–348} (\bibinfo {year}
  {2011}{\natexlab{a}})}\BibitemShut {NoStop}%
\bibitem [{\citenamefont {Evans}\ \emph
  {et~al.}(2011{\natexlab{b}})\citenamefont {Evans}, \citenamefont {Ibe},\ and\
  \citenamefont {Yanagida}}]{evans2011probing}%
  \BibitemOpen
  \bibfield  {author} {\bibinfo {author} {\bibfnamefont {J.~L.}\ \bibnamefont
  {Evans}}, \bibinfo {author} {\bibfnamefont {M.}~\bibnamefont {Ibe}},\ and\
  \bibinfo {author} {\bibfnamefont {T.~T.}\ \bibnamefont {Yanagida}},\
  }\href@noop {} {\bibinfo {title} {Probing extra matter in gauge mediation
  through the lightest higgs boson mass}} (\bibinfo {year}
  {2011}{\natexlab{b}}),\ \Eprint {https://arxiv.org/abs/1108.3437}
  {arXiv:1108.3437 [hep-ph]} \BibitemShut {NoStop}%
\bibitem [{\citenamefont {Evans}\ \emph {et~al.}(2012)\citenamefont {Evans},
  \citenamefont {Ibe}, \citenamefont {Shirai},\ and\ \citenamefont
  {Yanagida}}]{Evans_2012}%
  \BibitemOpen
  \bibfield  {author} {\bibinfo {author} {\bibfnamefont {J.~L.}\ \bibnamefont
  {Evans}}, \bibinfo {author} {\bibfnamefont {M.}~\bibnamefont {Ibe}}, \bibinfo
  {author} {\bibfnamefont {S.}~\bibnamefont {Shirai}},\ and\ \bibinfo {author}
  {\bibfnamefont {T.~T.}\ \bibnamefont {Yanagida}},\ }\bibfield  {title}
  {\bibinfo {title} {A 125 gev higgs boson and muong−2in more generic gauge
  mediation},\ }\bibfield  {journal} {\bibinfo  {journal} {Physical Review D}\
  }\textbf {\bibinfo {volume} {85}},\ \href
  {https://doi.org/10.1103/physrevd.85.095004} {10.1103/physrevd.85.095004}
  (\bibinfo {year} {2012})\BibitemShut {NoStop}%
\bibitem [{\citenamefont {Kang}\ \emph {et~al.}(2012)\citenamefont {Kang},
  \citenamefont {Li}, \citenamefont {Liu}, \citenamefont {Tong},\ and\
  \citenamefont {Yang}}]{Kang_2012}%
  \BibitemOpen
  \bibfield  {author} {\bibinfo {author} {\bibfnamefont {Z.}~\bibnamefont
  {Kang}}, \bibinfo {author} {\bibfnamefont {T.}~\bibnamefont {Li}}, \bibinfo
  {author} {\bibfnamefont {T.}~\bibnamefont {Liu}}, \bibinfo {author}
  {\bibfnamefont {C.}~\bibnamefont {Tong}},\ and\ \bibinfo {author}
  {\bibfnamefont {J.~M.}\ \bibnamefont {Yang}},\ }\bibfield  {title} {\bibinfo
  {title} {Heavy standard model-like higgs boson and a light stop from
  yukawa-deflected gauge mediation},\ }\bibfield  {journal} {\bibinfo
  {journal} {Physical Review D}\ }\textbf {\bibinfo {volume} {86}},\ \href
  {https://doi.org/10.1103/physrevd.86.095020} {10.1103/physrevd.86.095020}
  (\bibinfo {year} {2012})\BibitemShut {NoStop}%
\bibitem [{\citenamefont {Craig}\ \emph {et~al.}(2013)\citenamefont {Craig},
  \citenamefont {Knapen}, \citenamefont {Shih},\ and\ \citenamefont
  {Zhao}}]{Craig_2013}%
  \BibitemOpen
  \bibfield  {author} {\bibinfo {author} {\bibfnamefont {N.}~\bibnamefont
  {Craig}}, \bibinfo {author} {\bibfnamefont {S.}~\bibnamefont {Knapen}},
  \bibinfo {author} {\bibfnamefont {D.}~\bibnamefont {Shih}},\ and\ \bibinfo
  {author} {\bibfnamefont {Y.}~\bibnamefont {Zhao}},\ }\bibfield  {title}
  {\bibinfo {title} {A complete model of low-scale gauge mediation},\
  }\bibfield  {journal} {\bibinfo  {journal} {Journal of High Energy Physics}\
  }\textbf {\bibinfo {volume} {2013}},\ \href
  {https://doi.org/10.1007/jhep03(2013)154} {10.1007/jhep03(2013)154} (\bibinfo
  {year} {2013})\BibitemShut {NoStop}%
\bibitem [{\citenamefont {Albaid}\ and\ \citenamefont
  {Babu}(2013)}]{Albaid_2013}%
  \BibitemOpen
  \bibfield  {author} {\bibinfo {author} {\bibfnamefont {A.}~\bibnamefont
  {Albaid}}\ and\ \bibinfo {author} {\bibfnamefont {K.~S.}\ \bibnamefont
  {Babu}},\ }\bibfield  {title} {\bibinfo {title} {Higgs boson of mass 125 gev
  in gauge mediated supersymmetry breaking models with matter-messenger
  mixing},\ }\bibfield  {journal} {\bibinfo  {journal} {Physical Review D}\
  }\textbf {\bibinfo {volume} {88}},\ \href
  {https://doi.org/10.1103/physrevd.88.055007} {10.1103/physrevd.88.055007}
  (\bibinfo {year} {2013})\BibitemShut {NoStop}%
\bibitem [{\citenamefont {Abdullah}\ \emph {et~al.}(2013)\citenamefont
  {Abdullah}, \citenamefont {Galon}, \citenamefont {Shadmi},\ and\
  \citenamefont {Shirman}}]{Abdullah_2013}%
  \BibitemOpen
  \bibfield  {author} {\bibinfo {author} {\bibfnamefont {M.}~\bibnamefont
  {Abdullah}}, \bibinfo {author} {\bibfnamefont {I.}~\bibnamefont {Galon}},
  \bibinfo {author} {\bibfnamefont {Y.}~\bibnamefont {Shadmi}},\ and\ \bibinfo
  {author} {\bibfnamefont {Y.}~\bibnamefont {Shirman}},\ }\bibfield  {title}
  {\bibinfo {title} {Flavored gauge mediation, a heavy higgs, and
  supersymmetric alignment},\ }\bibfield  {journal} {\bibinfo  {journal}
  {Journal of High Energy Physics}\ }\textbf {\bibinfo {volume} {2013}},\ \href
  {https://doi.org/10.1007/jhep06(2013)057} {10.1007/jhep06(2013)057} (\bibinfo
  {year} {2013})\BibitemShut {NoStop}%
\bibitem [{\citenamefont {Pérez}\ \emph {et~al.}(2013)\citenamefont {Pérez},
  \citenamefont {Ramond},\ and\ \citenamefont {Zhang}}]{P_rez_2013}%
  \BibitemOpen
  \bibfield  {author} {\bibinfo {author} {\bibfnamefont {M.~J.}\ \bibnamefont
  {Pérez}}, \bibinfo {author} {\bibfnamefont {P.}~\bibnamefont {Ramond}},\
  and\ \bibinfo {author} {\bibfnamefont {J.}~\bibnamefont {Zhang}},\ }\bibfield
   {title} {\bibinfo {title} {Mixing supersymmetry and family symmetry
  breakings},\ }\bibfield  {journal} {\bibinfo  {journal} {Physical Review D}\
  }\textbf {\bibinfo {volume} {87}},\ \href
  {https://doi.org/10.1103/physrevd.87.035021} {10.1103/physrevd.87.035021}
  (\bibinfo {year} {2013})\BibitemShut {NoStop}%
\bibitem [{\citenamefont {Byakti}\ and\ \citenamefont
  {Ray}(2013)}]{Byakti_2013}%
  \BibitemOpen
  \bibfield  {author} {\bibinfo {author} {\bibfnamefont {P.}~\bibnamefont
  {Byakti}}\ and\ \bibinfo {author} {\bibfnamefont {T.~S.}\ \bibnamefont
  {Ray}},\ }\bibfield  {title} {\bibinfo {title} {Burgeoning the higgs mass to
  125 gev through messenger-matter interactions in gmsb models},\ }\bibfield
  {journal} {\bibinfo  {journal} {Journal of High Energy Physics}\ }\textbf
  {\bibinfo {volume} {2013}},\ \href {https://doi.org/10.1007/jhep05(2013)055}
  {10.1007/jhep05(2013)055} (\bibinfo {year} {2013})\BibitemShut {NoStop}%
\bibitem [{\citenamefont {Evans}\ and\ \citenamefont
  {Shih}(2013)}]{Evans_2013}%
  \BibitemOpen
  \bibfield  {author} {\bibinfo {author} {\bibfnamefont {J.~A.}\ \bibnamefont
  {Evans}}\ and\ \bibinfo {author} {\bibfnamefont {D.}~\bibnamefont {Shih}},\
  }\bibfield  {title} {\bibinfo {title} {Surveying extended gmsb models with m
  h = 125 gev},\ }\bibfield  {journal} {\bibinfo  {journal} {Journal of High
  Energy Physics}\ }\textbf {\bibinfo {volume} {2013}},\ \href
  {https://doi.org/10.1007/jhep08(2013)093} {10.1007/jhep08(2013)093} (\bibinfo
  {year} {2013})\BibitemShut {NoStop}%
\bibitem [{\citenamefont {Calibbi}\ \emph {et~al.}(2013)\citenamefont
  {Calibbi}, \citenamefont {Paradisi},\ and\ \citenamefont
  {Ziegler}}]{Calibbi_2013}%
  \BibitemOpen
  \bibfield  {author} {\bibinfo {author} {\bibfnamefont {L.}~\bibnamefont
  {Calibbi}}, \bibinfo {author} {\bibfnamefont {P.}~\bibnamefont {Paradisi}},\
  and\ \bibinfo {author} {\bibfnamefont {R.}~\bibnamefont {Ziegler}},\
  }\bibfield  {title} {\bibinfo {title} {Gauge mediation beyond minimal flavor
  violation},\ }\bibfield  {journal} {\bibinfo  {journal} {Journal of High
  Energy Physics}\ }\textbf {\bibinfo {volume} {2013}},\ \href
  {https://doi.org/10.1007/jhep06(2013)052} {10.1007/jhep06(2013)052} (\bibinfo
  {year} {2013})\BibitemShut {NoStop}%
\bibitem [{\citenamefont {Evans}\ \emph {et~al.}(2015)\citenamefont {Evans},
  \citenamefont {Shih},\ and\ \citenamefont {Thalapillil}}]{evans2015chiral}%
  \BibitemOpen
  \bibfield  {author} {\bibinfo {author} {\bibfnamefont {J.~A.}\ \bibnamefont
  {Evans}}, \bibinfo {author} {\bibfnamefont {D.}~\bibnamefont {Shih}},\ and\
  \bibinfo {author} {\bibfnamefont {A.}~\bibnamefont {Thalapillil}},\
  }\href@noop {} {\bibinfo {title} {Chiral flavor violation from extended gauge
  mediation}} (\bibinfo {year} {2015}),\ \Eprint
  {https://arxiv.org/abs/1504.00930} {arXiv:1504.00930 [hep-ph]} \BibitemShut
  {NoStop}%
\bibitem [{\citenamefont {Galon}\ \emph {et~al.}(2013)\citenamefont {Galon},
  \citenamefont {Perez},\ and\ \citenamefont {Shadmi}}]{Galon_2013}%
  \BibitemOpen
  \bibfield  {author} {\bibinfo {author} {\bibfnamefont {I.}~\bibnamefont
  {Galon}}, \bibinfo {author} {\bibfnamefont {G.}~\bibnamefont {Perez}},\ and\
  \bibinfo {author} {\bibfnamefont {Y.}~\bibnamefont {Shadmi}},\ }\bibfield
  {title} {\bibinfo {title} {Non-degenerate squarks from flavored gauge
  mediation},\ }\bibfield  {journal} {\bibinfo  {journal} {Journal of High
  Energy Physics}\ }\textbf {\bibinfo {volume} {2013}},\ \href
  {https://doi.org/10.1007/jhep09(2013)117} {10.1007/jhep09(2013)117} (\bibinfo
  {year} {2013})\BibitemShut {NoStop}%
\bibitem [{\citenamefont {Joaquim}\ and\ \citenamefont
  {Rossi}(2007)}]{Joaquim:2006mn}%
  \BibitemOpen
  \bibfield  {author} {\bibinfo {author} {\bibfnamefont {F.~R.}\ \bibnamefont
  {Joaquim}}\ and\ \bibinfo {author} {\bibfnamefont {A.}~\bibnamefont
  {Rossi}},\ }\bibfield  {title} {\bibinfo {title} {{Phenomenology of the
  triplet seesaw mechanism with Gauge and Yukawa mediation of SUSY breaking}},\
  }\href {https://doi.org/10.1016/j.nuclphysb.2006.11.030} {\bibfield
  {journal} {\bibinfo  {journal} {Nucl. Phys. B}\ }\textbf {\bibinfo {volume}
  {765}},\ \bibinfo {pages} {71} (\bibinfo {year} {2007})},\ \Eprint
  {https://arxiv.org/abs/hep-ph/0607298} {arXiv:hep-ph/0607298} \BibitemShut
  {NoStop}%
\bibitem [{\citenamefont {Joaquim}\ and\ \citenamefont
  {Rossi}(2006)}]{Joaquim:2006uz}%
  \BibitemOpen
  \bibfield  {author} {\bibinfo {author} {\bibfnamefont {F.~R.}\ \bibnamefont
  {Joaquim}}\ and\ \bibinfo {author} {\bibfnamefont {A.}~\bibnamefont
  {Rossi}},\ }\bibfield  {title} {\bibinfo {title} {{Gauge and Yukawa mediated
  supersymmetry breaking in the triplet seesaw scenario}},\ }\href
  {https://doi.org/10.1103/PhysRevLett.97.181801} {\bibfield  {journal}
  {\bibinfo  {journal} {Phys. Rev. Lett.}\ }\textbf {\bibinfo {volume} {97}},\
  \bibinfo {pages} {181801} (\bibinfo {year} {2006})},\ \Eprint
  {https://arxiv.org/abs/hep-ph/0604083} {arXiv:hep-ph/0604083} \BibitemShut
  {NoStop}%
\bibitem [{\citenamefont {Ierushalmi}\ \emph {et~al.}(2016)\citenamefont
  {Ierushalmi}, \citenamefont {Iwamoto}, \citenamefont {Lee}, \citenamefont
  {Nepomnyashy},\ and\ \citenamefont {Shadmi}}]{Ierushalmi_2016}%
  \BibitemOpen
  \bibfield  {author} {\bibinfo {author} {\bibfnamefont {N.}~\bibnamefont
  {Ierushalmi}}, \bibinfo {author} {\bibfnamefont {S.}~\bibnamefont {Iwamoto}},
  \bibinfo {author} {\bibfnamefont {G.}~\bibnamefont {Lee}}, \bibinfo {author}
  {\bibfnamefont {V.}~\bibnamefont {Nepomnyashy}},\ and\ \bibinfo {author}
  {\bibfnamefont {Y.}~\bibnamefont {Shadmi}},\ }\bibfield  {title} {\bibinfo
  {title} {Lhc benchmarks from flavored gauge mediation},\ }\bibfield
  {journal} {\bibinfo  {journal} {Journal of High Energy Physics}\ }\textbf
  {\bibinfo {volume} {2016}},\ \href {https://doi.org/10.1007/jhep07(2016)058}
  {10.1007/jhep07(2016)058} (\bibinfo {year} {2016})\BibitemShut {NoStop}%
\bibitem [{\citenamefont {Jeliński}\ and\ \citenamefont
  {Gluza}(2015)}]{Jeli_ski_2015}%
  \BibitemOpen
  \bibfield  {author} {\bibinfo {author} {\bibfnamefont {T.}~\bibnamefont
  {Jeliński}}\ and\ \bibinfo {author} {\bibfnamefont {J.}~\bibnamefont
  {Gluza}},\ }\bibfield  {title} {\bibinfo {title} {Analytical two-loop soft
  mass terms of sfermions in extended gmsb models},\ }\href
  {https://doi.org/10.1016/j.physletb.2015.11.003} {\bibfield  {journal}
  {\bibinfo  {journal} {Physics Letters B}\ }\textbf {\bibinfo {volume}
  {751}},\ \bibinfo {pages} {541–547} (\bibinfo {year} {2015})}\BibitemShut
  {NoStop}%
\bibitem [{\citenamefont {Everett}\ and\ \citenamefont
  {Garon}(2018)}]{Everett_2018}%
  \BibitemOpen
  \bibfield  {author} {\bibinfo {author} {\bibfnamefont {L.~L.}\ \bibnamefont
  {Everett}}\ and\ \bibinfo {author} {\bibfnamefont {T.~S.}\ \bibnamefont
  {Garon}},\ }\bibfield  {title} {\bibinfo {title} {Flavored gauge mediation
  with discrete non-abelian symmetries},\ }\bibfield  {journal} {\bibinfo
  {journal} {Physical Review D}\ }\textbf {\bibinfo {volume} {97}},\ \href
  {https://doi.org/10.1103/physrevd.97.095028} {10.1103/physrevd.97.095028}
  (\bibinfo {year} {2018})\BibitemShut {NoStop}%
\bibitem [{\citenamefont {Everett}\ \emph {et~al.}(2019)\citenamefont
  {Everett}, \citenamefont {Garon},\ and\ \citenamefont {Rock}}]{Everett_2019}%
  \BibitemOpen
  \bibfield  {author} {\bibinfo {author} {\bibfnamefont {L.~L.}\ \bibnamefont
  {Everett}}, \bibinfo {author} {\bibfnamefont {T.~S.}\ \bibnamefont {Garon}},\
  and\ \bibinfo {author} {\bibfnamefont {A.~B.}\ \bibnamefont {Rock}},\
  }\bibfield  {title} {\bibinfo {title} {Sizable stop mixing in flavored gauge
  mediation models with discrete non-abelian symmetries},\ }\bibfield
  {journal} {\bibinfo  {journal} {Physical Review D}\ }\textbf {\bibinfo
  {volume} {100}},\ \href {https://doi.org/10.1103/physrevd.100.015039}
  {10.1103/physrevd.100.015039} (\bibinfo {year} {2019})\BibitemShut {NoStop}%
\bibitem [{\citenamefont {Ahmed}\ \emph {et~al.}(2017)\citenamefont {Ahmed},
  \citenamefont {Calibbi}, \citenamefont {Li}, \citenamefont {Mustafayev},\
  and\ \citenamefont {Raza}}]{Ahmed_2017}%
  \BibitemOpen
  \bibfield  {author} {\bibinfo {author} {\bibfnamefont {W.}~\bibnamefont
  {Ahmed}}, \bibinfo {author} {\bibfnamefont {L.}~\bibnamefont {Calibbi}},
  \bibinfo {author} {\bibfnamefont {T.}~\bibnamefont {Li}}, \bibinfo {author}
  {\bibfnamefont {A.}~\bibnamefont {Mustafayev}},\ and\ \bibinfo {author}
  {\bibfnamefont {S.}~\bibnamefont {Raza}},\ }\bibfield  {title} {\bibinfo
  {title} {Low fine-tuning in yukawa-deflected gauge mediation},\ }\bibfield
  {journal} {\bibinfo  {journal} {Physical Review D}\ }\textbf {\bibinfo
  {volume} {95}},\ \href {https://doi.org/10.1103/physrevd.95.095031}
  {10.1103/physrevd.95.095031} (\bibinfo {year} {2017})\BibitemShut {NoStop}%
\bibitem [{\citenamefont {Xing}(1997)}]{Xing_1997}%
  \BibitemOpen
  \bibfield  {author} {\bibinfo {author} {\bibfnamefont {Z.-z.}\ \bibnamefont
  {Xing}},\ }\bibfield  {title} {\bibinfo {title} {Implications of the quark
  mass hierarchy on flavour mixings},\ }\href
  {https://doi.org/10.1088/0954-3899/23/11/006} {\bibfield  {journal} {\bibinfo
   {journal} {Journal of Physics G: Nuclear and Particle Physics}\ }\textbf
  {\bibinfo {volume} {23}},\ \bibinfo {pages} {1563–1578} (\bibinfo {year}
  {1997})}\BibitemShut {NoStop}%
\bibitem [{\citenamefont {Fritzsch}\ and\ \citenamefont
  {Holtmannspötter}(1994)}]{Fritzsch_1994}%
  \BibitemOpen
  \bibfield  {author} {\bibinfo {author} {\bibfnamefont {H.}~\bibnamefont
  {Fritzsch}}\ and\ \bibinfo {author} {\bibfnamefont {D.}~\bibnamefont
  {Holtmannspötter}},\ }\bibfield  {title} {\bibinfo {title} {The breaking of
  subnuclear democracy as the origin of flavour mixing},\ }\href
  {https://doi.org/10.1016/0370-2693(94)91380-3} {\bibfield  {journal}
  {\bibinfo  {journal} {Physics Letters B}\ }\textbf {\bibinfo {volume}
  {338}},\ \bibinfo {pages} {290–294} (\bibinfo {year} {1994})}\BibitemShut
  {NoStop}%
\bibitem [{\citenamefont {Fritzsch}\ and\ \citenamefont
  {Xing}(2000)}]{Fritzsch_2000}%
  \BibitemOpen
  \bibfield  {author} {\bibinfo {author} {\bibfnamefont {H.}~\bibnamefont
  {Fritzsch}}\ and\ \bibinfo {author} {\bibfnamefont {Z.-Z.}\ \bibnamefont
  {Xing}},\ }\bibfield  {title} {\bibinfo {title} {Mass and flavor mixing
  schemes of quarks and leptons},\ }\href
  {https://doi.org/10.1016/s0146-6410(00)00102-2} {\bibfield  {journal}
  {\bibinfo  {journal} {Progress in Particle and Nuclear Physics}\ }\textbf
  {\bibinfo {volume} {45}},\ \bibinfo {pages} {1–81} (\bibinfo {year}
  {2000})}\BibitemShut {NoStop}%
\bibitem [{\citenamefont {Everett}\ \emph {et~al.}(2020)\citenamefont
  {Everett}, \citenamefont {Garon},\ and\ \citenamefont {Rock}}]{Everett_2020}%
  \BibitemOpen
  \bibfield  {author} {\bibinfo {author} {\bibfnamefont {L.~L.}\ \bibnamefont
  {Everett}}, \bibinfo {author} {\bibfnamefont {T.~S.}\ \bibnamefont {Garon}},\
  and\ \bibinfo {author} {\bibfnamefont {A.~B.}\ \bibnamefont {Rock}},\
  }\bibfield  {title} {\bibinfo {title} {Generating the cabibbo angle in
  flavored gauge mediation models with discrete non-abelian symmetries},\
  }\bibfield  {journal} {\bibinfo  {journal} {Physical Review D}\ }\textbf
  {\bibinfo {volume} {101}},\ \href
  {https://doi.org/10.1103/physrevd.101.115003} {10.1103/physrevd.101.115003}
  (\bibinfo {year} {2020})\BibitemShut {NoStop}%
\bibitem [{Note1()}]{Note1}%
  \BibitemOpen
  \bibinfo {note} {The triplet messengers and the $X_T$ field are chosen for
  simplicity to be singlets with respect to the $\protect \mathcal {S}_3$
  Higgs-messenger symmetry. Note that this choice is not consistent with a full
  embedding of this scenario into a grand unified theory. This would require
  more extended model-building that would also need to address the well-known
  doublet-triplet splitting issue in grand unified models).}\BibitemShut
  {Stop}%
\bibitem [{Note2()}]{Note2}%
  \BibitemOpen
  \bibinfo {note} {Note that if we neglect the subleading $\sigma $
  perturbations, this scheme is equivalent to replacing $\beta _{2u,3u,4u} =
  1+\epsilon _u$ in the general form for the superpotential couplings. As such,
  the mixing matrices are easily obtained using the results of Eq.~(\ref
  {eq:case1diagmatrices}) with the appropriate substitutions.}\BibitemShut
  {Stop}%
\bibitem [{\citenamefont {Allanach}(2002)}]{Allanach:2001kg}%
  \BibitemOpen
  \bibfield  {author} {\bibinfo {author} {\bibfnamefont {B.~C.}\ \bibnamefont
  {Allanach}},\ }\bibfield  {title} {\bibinfo {title} {{SOFTSUSY: a program for
  calculating supersymmetric spectra}},\ }\href
  {https://doi.org/10.1016/S0010-4655(01)00460-X} {\bibfield  {journal}
  {\bibinfo  {journal} {Comput. Phys. Commun.}\ }\textbf {\bibinfo {volume}
  {143}},\ \bibinfo {pages} {305} (\bibinfo {year} {2002})},\ \Eprint
  {https://arxiv.org/abs/hep-ph/0104145} {arXiv:hep-ph/0104145 [hep-ph]}
  \BibitemShut {NoStop}%
\bibitem [{\citenamefont {Zyla}\ \emph {et~al.}(2020)\citenamefont {Zyla} \emph
  {et~al.}}]{Zyla:2020zbs}%
  \BibitemOpen
  \bibfield  {author} {\bibinfo {author} {\bibfnamefont {P.~A.}\ \bibnamefont
  {Zyla}} \emph {et~al.} (\bibinfo {collaboration} {Particle Data Group}),\
  }\bibfield  {title} {\bibinfo {title} {{Review of Particle Physics}},\ }\href
  {https://doi.org/10.1093/ptep/ptaa104} {\bibfield  {journal} {\bibinfo
  {journal} {PTEP}\ }\textbf {\bibinfo {volume} {2020}},\ \bibinfo {pages}
  {083C01} (\bibinfo {year} {2020})},\ \bibinfo {note} {and 2021
  updates}\BibitemShut {NoStop}%
\bibitem [{Note3()}]{Note3}%
  \BibitemOpen
  \bibinfo {note} {Note that MIA is a good approximation in this scenario
  although we have non-degenerate squark masses, since the squark masses are
  not strongly hierarchical~\cite {Raz_2002}.}\BibitemShut {Stop}%
\bibitem [{\citenamefont {Misiak}\ \emph {et~al.}(1998)\citenamefont {Misiak},
  \citenamefont {Pokorski},\ and\ \citenamefont {Rosiek}}]{Misiak_1998}%
  \BibitemOpen
  \bibfield  {author} {\bibinfo {author} {\bibfnamefont {M.}~\bibnamefont
  {Misiak}}, \bibinfo {author} {\bibfnamefont {S.}~\bibnamefont {Pokorski}},\
  and\ \bibinfo {author} {\bibfnamefont {J.}~\bibnamefont {Rosiek}},\
  }\bibfield  {title} {\bibinfo {title} {Supersymmetry and fcnc effects},\
  }\href {https://doi.org/10.1142/9789812812667_0012} {\bibfield  {journal}
  {\bibinfo  {journal} {Advanced Series on Directions in High Energy Physics}\
  ,\ \bibinfo {pages} {795–828}} (\bibinfo {year} {1998})}\BibitemShut
  {NoStop}%
\bibitem [{\citenamefont {Abi}\ \emph {et~al.}(2021)\citenamefont {Abi} \emph
  {et~al.}}]{PhysRevLett.126.141801}%
  \BibitemOpen
  \bibfield  {author} {\bibinfo {author} {\bibfnamefont {B.}~\bibnamefont
  {Abi}} \emph {et~al.} (\bibinfo {collaboration} {Muon $g\ensuremath{-}2$
  Collaboration}),\ }\bibfield  {title} {\bibinfo {title} {Measurement of the
  positive muon anomalous magnetic moment to 0.46 ppm},\ }\href
  {https://doi.org/10.1103/PhysRevLett.126.141801} {\bibfield  {journal}
  {\bibinfo  {journal} {Phys. Rev. Lett.}\ }\textbf {\bibinfo {volume} {126}},\
  \bibinfo {pages} {141801} (\bibinfo {year} {2021})}\BibitemShut {NoStop}%
\bibitem [{\citenamefont {Sirunyan}\ \emph {et~al.}(2019)\citenamefont
  {Sirunyan} \emph {et~al.}}]{CMS:2018eqb}%
  \BibitemOpen
  \bibfield  {author} {\bibinfo {author} {\bibfnamefont {A.~M.}\ \bibnamefont
  {Sirunyan}} \emph {et~al.} (\bibinfo {collaboration} {CMS}),\ }\bibfield
  {title} {\bibinfo {title} {{Search for supersymmetric partners of electrons
  and muons in proton-proton collisions at $\sqrt{s}=$ 13 TeV}},\ }\href
  {https://doi.org/10.1016/j.physletb.2019.01.005} {\bibfield  {journal}
  {\bibinfo  {journal} {Phys. Lett. B}\ }\textbf {\bibinfo {volume} {790}},\
  \bibinfo {pages} {140} (\bibinfo {year} {2019})},\ \Eprint
  {https://arxiv.org/abs/1806.05264} {arXiv:1806.05264 [hep-ex]} \BibitemShut
  {NoStop}%
\bibitem [{\citenamefont {Steffen}(2006)}]{Steffen:2006hw}%
  \BibitemOpen
  \bibfield  {author} {\bibinfo {author} {\bibfnamefont {F.~D.}\ \bibnamefont
  {Steffen}},\ }\bibfield  {title} {\bibinfo {title} {{Gravitino dark matter
  and cosmological constraints}},\ }\href
  {https://doi.org/10.1088/1475-7516/2006/09/001} {\bibfield  {journal}
  {\bibinfo  {journal} {JCAP}\ }\textbf {\bibinfo {volume} {09}},\ \bibinfo
  {pages} {001}},\ \Eprint {https://arxiv.org/abs/hep-ph/0605306}
  {arXiv:hep-ph/0605306} \BibitemShut {NoStop}%
\bibitem [{\citenamefont {Feng}\ \emph {et~al.}(2004)\citenamefont {Feng},
  \citenamefont {Su},\ and\ \citenamefont {Takayama}}]{Feng:2004zu}%
  \BibitemOpen
  \bibfield  {author} {\bibinfo {author} {\bibfnamefont {J.~L.}\ \bibnamefont
  {Feng}}, \bibinfo {author} {\bibfnamefont {S.-f.}\ \bibnamefont {Su}},\ and\
  \bibinfo {author} {\bibfnamefont {F.}~\bibnamefont {Takayama}},\ }\bibfield
  {title} {\bibinfo {title} {{SuperWIMP gravitino dark matter from slepton and
  sneutrino decays}},\ }\href {https://doi.org/10.1103/PhysRevD.70.063514}
  {\bibfield  {journal} {\bibinfo  {journal} {Phys. Rev. D}\ }\textbf {\bibinfo
  {volume} {70}},\ \bibinfo {pages} {063514} (\bibinfo {year} {2004})},\
  \Eprint {https://arxiv.org/abs/hep-ph/0404198} {arXiv:hep-ph/0404198}
  \BibitemShut {NoStop}%
\bibitem [{\citenamefont {Asaka}\ \emph {et~al.}(2000)\citenamefont {Asaka},
  \citenamefont {Hamaguchi},\ and\ \citenamefont {Suzuki}}]{ASAKA2000136}%
  \BibitemOpen
  \bibfield  {author} {\bibinfo {author} {\bibfnamefont {T.}~\bibnamefont
  {Asaka}}, \bibinfo {author} {\bibfnamefont {K.}~\bibnamefont {Hamaguchi}},\
  and\ \bibinfo {author} {\bibfnamefont {K.}~\bibnamefont {Suzuki}},\
  }\bibfield  {title} {\bibinfo {title} {Cosmological gravitino problem in
  gauge-mediated supersymmetry breaking models},\ }\href
  {https://doi.org/https://doi.org/10.1016/S0370-2693(00)00959-X} {\bibfield
  {journal} {\bibinfo  {journal} {Physics Letters B}\ }\textbf {\bibinfo
  {volume} {490}},\ \bibinfo {pages} {136} (\bibinfo {year}
  {2000})}\BibitemShut {NoStop}%
\bibitem [{\citenamefont {Raz}(2002)}]{Raz_2002}%
  \BibitemOpen
  \bibfield  {author} {\bibinfo {author} {\bibfnamefont {G.}~\bibnamefont
  {Raz}},\ }\bibfield  {title} {\bibinfo {title} {Mass insertion approximation
  without squark degeneracy},\ }\bibfield  {journal} {\bibinfo  {journal}
  {Physical Review D}\ }\textbf {\bibinfo {volume} {66}},\ \href
  {https://doi.org/10.1103/physrevd.66.037701} {10.1103/physrevd.66.037701}
  (\bibinfo {year} {2002})\BibitemShut {NoStop}%
\end{thebibliography}%


%
\end{document}